\documentclass[a4paper,11pt]{article}
\usepackage{jcappub} 
\usepackage{lineno}

\usepackage[T1]{fontenc}
\usepackage{graphicx}
\usepackage{newtxtext,newtxmath}
\usepackage{amssymb}
\usepackage{eurosym}
\usepackage{nicefrac}
\usepackage{float}
\usepackage{array}
\usepackage{glossaries}
\usepackage{pdfpages}
\usepackage{multirow}
\usepackage{longtable}

\newcommand{\krakow}{Krak\'ow{}}
\newcommand{\dcam}{DigiCam}
\newcommand{\degc}{$^{\circ}$C}
\title{The SST-1M imaging atmospheric Cherenkov telescope for gamma-ray astrophysics\note{Web page: https://sst-1m.space}}

\author[a]{C.~Alispach}
\author[i, q, s]{, A.~Araudo}
\author[a]{, M.~Balbo}
\author[o]{, V.~Beshley}
\author[c]{, A.~Biland}
\author[i]{, J.~Bla\v{z}ek}
\author[h]{, J.~Borkowski}
\author[e]{, T.~Bulik}
\author[a]{, F.~Cadoux}
\author[d]{, S.~Casanova}
\author[i]{, A.~Christov}
\author[i]{, J.~Chudoba}
\author[i]{, L.~Chytka}
\author[i]{, P.~D\v{e}di\v{c}}
\author[a]{, D.~della Volpe}
\author[a]{, Y.~Favre}
\author[w]{, M.~Garczarczyk}
\author[t]{, L.~Gibaud}
\author[d]{, T.~Gieras}
\author[l]{, P.~Hamal}
\author[a]{, M.~Heller}
\author[l]{, M.~Hrabovsk\'y}
\author[i]{, P.~Jane\v{c}ek}
\author[r]{, M.~Jel\'inek}
\author[l]{, V.~J\'ilek}
\author[i]{, J.~Jury\v{s}ek}
\author[q]{, V.~Karas}
\author[a]{, B.~Lacave}
\author[b]{, E.~Lyard}
\author[d]{, E.~Mach}
\author[i]{, D.~Mand\'at}
\author[d]{, W.~Marek}
\author[l]{, S.~Michal}
\author[d]{, J.~Micha{\l}owski}
\author[h]{, R.~Moderski}
\author[a]{, T.~Montaruli}
\author[h]{, A.~Muraczewski}
\author[i]{, S.~Muthyala}
\author[i, q]{, A.~L.~Müller}
\author[a]{, A.~Nagai}
\author[d]{, K.~Nalewajski}
\author[c]{, D.~Neise}
\author[d]{, J.~Niemiec}
\author[t]{, M.~Niko{\l}ajuk}
\author[i, u]{, V.~Novotn\'y}
\author[f]{, M.~Ostrowski}
\author[i]{, M.~Palatka}
\author[i]{, M.~Pech}
\author[i]{, M.~Prouza}
\author[k]{, P.~Rajda}
\author[i]{, P.~Schovanek}
\author[g]{, K.~Seweryn}
\author[b]{, V.~Sliusar}
\author[f]{, {\L}.~Stawarz}
\author[w]{, R.~Sternberger}
\author[a]{, M.~Stodulska}
\author[d]{, J.~\'{S}wierblewski}
\author[d]{, P.~\'{S}wierk}
\author[r]{, J.~\v{S}trobl}
\author[i]{, T.~Tavernier}
\author[i]{, P.~Tr\'avn\'i\v{c}ek}
\author[a]{, I.~Troyano Pujadas}
\author[i]{, J.~V\'icha}
\author[i, l]{, M.~Vacula}
\author[b]{, R.~Walter}
\author[f]{, K.~Zi{\c e}tara}

\affiliation[a]{D\'epartement de Physique Nucl\'eaire, Facult\'e de Sciences, Universit\'e de Gen\`eve, Quai Ernest Ansermet 24, CH-1205 Gen\`eve, Switzerland}
\affiliation[b]{D\'epartement d'Astronomie, Facult\'e de Science, Universit\'e de Gen\`eve, Chemin d'Ecogia 16, CH-1290 Versoix, Switzerland}
\affiliation[c]{Institute for Particle Physics and Astrophysics, ETH Zurich, Otto-Stern-Weg 5, 8093 Zurich, Switzerland}
\affiliation[d]{Institute of Nuclear Physics Polish Academy of Sciences,  ul. Radzikowskiego 152, 31-342 \krakow, Poland}
\affiliation[e]{Astronomical Observatory, University of Warsaw, Al. Ujazdowskie 4, 00-478 Warsaw, Poland}
\affiliation[f]{Astronomical Observatory, Jagiellonian University, ul. Orla 171, 30-244 \krakow, Poland}
\affiliation[g]{Centrum Bada{\'n} Kosmicznych Polskiej Akademii Nauk,  18a Bartycka str., 00-716 Warsaw, Poland}
\affiliation[h]{Nicolaus Copernicus Astronomical Center, Polish Academy of Sciences,  ul. Bartycka 18, 00-716 Warsaw, Poland}
\affiliation[i]{Institute of Physics of the Czech Academy of Sciences, FZU, Na Slovance 1999/2, Prague 8, Czech Republic}
\affiliation[k]{AGH University of Science and Technology, al. Mickiewicza 30, 30-059 \krakow, Poland}
\affiliation[l]{Faculty of Science, Palack\'y University Olomouc, 17. listopadu 50, Olomouc, Czech Republic}
\affiliation[o]{Pidstryhach Institute for Applied Problems of Mechanics and Mathematics, National Academy of Sciences of Ukraine, 3-b Naukova St., 79060, Lviv, Ukraine}
\affiliation[q]{Astronomical Institute of the Czech Academy of Sciences, Bo\v{c}n\'i~II 1401, CZ-14100 Prague, Czech Republic}
\affiliation[r]{Astronomical Institute of the Czech Academy of Sciences, Fri\v{c}ova~298, CZ-25165 Ond\v{r}ejov, Czech Republic}
\affiliation[s]{ELI Beamlines Facility, The ExtremeLight Infrastructure ERIC, Za Radnic\'i 835, CZ-25241,  Doln\'i B\v{r}e\v{z}any, Czech Republic}
\affiliation[t]{Faculty of Physics, University of Bia{\l}ystok, ul. K. Cio{\l}kowskiego 1L, 15-245 Bia{\l}ystok, Poland}
\affiliation[u]{Institute of Particle and Nuclear Physics, Faculty of Mathematics and Physics, Charles University, V Hole\v{s}ovi\v{c}k\'{a}ch 2, Prague, Czech Republic}
\affiliation[w]{Deutsches Elektronen-Synchrotron (DESY) Platanenallee 6, D-15738 Zeuthen, Germany}

\emailAdd{cyril.alispach@unige.ch}
\emailAdd{araudo@fzu.cz}
\emailAdd{matteo.balbo@unige.ch}
\emailAdd{beshley.vasyl@gmail.com}
\emailAdd{biland@phys.ethz.ch}
\emailAdd{blazekj@fzu.cz}
\emailAdd{jubork@ncac.torun.pl}
\emailAdd{tb@astrouw.edu.pl}
\emailAdd{franck.cadoux@unige.ch}
\emailAdd{sabrina.casanova@ifj.edu.pl}
\emailAdd{christov@fzu.cz}
\emailAdd{chudoba@fzu.cz}
\emailAdd{chytka@fzu.cz}
\emailAdd{premysl.dedic@cvut.cz}
\emailAdd{domenico.dellavolpe@unige.ch}
\emailAdd{yannick.favre@unige.ch}
\emailAdd{markus.garczarczyk@desy.de}
\emailAdd{l.gibaud@uwb.edu.pl}
\emailAdd{tomasz.gieras@ifj.edu.pl}
\emailAdd{p.hamal@upol.cz}
\emailAdd{matthieu.heller@unige.ch}
\emailAdd{miroslav.hrabovsky@upol.cz}
\emailAdd{janecekp@fzu.cz}
\emailAdd{martin.jelinek@asu.cas.cz}
\emailAdd{vlastimil.jilek@upol.cz}
\emailAdd{jurysek@fzu.cz}
\emailAdd{vladimir.karas@asu.cas.cz}
\emailAdd{bastien.lacave@unige.ch}
\emailAdd{etienne.lyard@unige.ch}
\emailAdd{emilmach@gmail.com}
\emailAdd{mandat@fzu.cz}
\emailAdd{wojciech.marek@ifj.edu.pl}
\emailAdd{stanislav.michal@slo.upol.cz}
\emailAdd{jerzy.michalowski@ifj.edu.pl}
\emailAdd{moderski@camk.edu.pl}
\emailAdd{teresa.montaruli@unige.ch}
\emailAdd{murak@ncac.torun.pl}
\emailAdd{muthyalasrijareddy@gmail.com}
\emailAdd{mulleral@fzu.cz}
\emailAdd{a.o.nagai@gmail.com}
\emailAdd{krzysztof.nalewajski@ifj.edu.pl}
\emailAdd{dominik.neise@cta-consortium.org}
\emailAdd{jacek.niemiec@ifj.edu.pl}
\emailAdd{m.nikolajuk@uwb.edu.pl}
\emailAdd{vladimir.novotny@matfyz.cuni.cz}
\emailAdd{michal.ostrowski@uj.edu.pl}
\emailAdd{miroslav.palatka@upol.cz}
\emailAdd{miroslav.pech@upol.cz}
\emailAdd{prouza@fzu.cz}
\emailAdd{pjrajda@agh.edu.pl}
\emailAdd{petr.schovanek@upol.cz}
\emailAdd{kseweryn@cbk.waw.pl}
\emailAdd{vitalii.sliusar@unige.ch}
\emailAdd{lukasz.1.stawarz@uj.edu.pl}
\emailAdd{ronny.sternberger@desy.de}
\emailAdd{magdalena.stodulska@unige.ch}
\emailAdd{jacek.swierblewski@ifj.edu.pl}
\emailAdd{pawel.swierk@ifj.edu.pl}
\emailAdd{strobl@asu.cas.cz}
\emailAdd{tavernier@fzu.cz}
\emailAdd{petr.travnicek@fzu.cz}
\emailAdd{isaac.troyano@unige.ch}
\emailAdd{vicha@fzu.cz}
\emailAdd{martin.vacula@jointlab.upol.cz}
\emailAdd{roland.walter@unige.ch}
\emailAdd{krzysztof.zietara@uj.edu.pl}

\abstract{The SST-1M is a Small-Sized Telescope (SST) designed to provide a cost-effective and high-performance solution for gamma-ray astrophysics, particularly for energies beyond a few TeV.
The goal is to integrate this telescope into an array of similar instruments, leveraging its lightweight design, earthquake resistance, and established Davies-Cotton configuration. Additionally, its optical system is designed to function without a protective dome, allowing it to withstand the harsh atmospheric conditions typical of mountain environments above 2000 m a.s.l.
The SST-1M utilizes a fully digitizing camera system based on silicon photomultipliers (SiPMs). This camera is capable of digitizing all signals from the UV-optical light detectors, allowing for the implementation of various triggers and data analysis methods. We detail the process of designing, prototyping, and validating this system, ensuring that it meets the stringent requirements for gamma-ray detection and performance.
An SST-1M stereo system is currently operational and collecting data at the Ondřejov observatory in the Czech Republic, situated at 500~m a.s.l. Preliminary results from this system are promising.
A forthcoming paper will provide a comprehensive analysis of the telescope’s performance in detecting gamma rays and operating under real-world conditions}

\begin{document}
\maketitle
\flushbottom


\section{Introduction}
\label{sec:introduction}
The vast majority of today's ground-based observations of gamma-ray astrophysical sources with emission in the energy range from tens of GeV to hundreds of TeV are performed by  Imaging Atmospheric Cherenkov Telescope (IACT) arrays. The Single-Mirror Small-Sized Telescope (SST-1M) is an IACT developed by a collaboration of institutions from Switzerland, Poland, and the Czech Republic.
The SST-1M features a Davies-Cotton (D-C) optical system which concentrates most of the innovation in its camera \cite{CameraPaperHeller2017} by adopting SiPM and fully digitising electronics. It was originally designed to be part of an array of 70 similar SSTs \cite{Montaruli:2015xya}, providing a cost-effective and easy-to-maintain solution for the Cherenkov Telescope Array Observatory (CTAO) and satisfying its sensitivity requirement to a flux of gamma-rays from astrophysical sources beyond energies of 3~TeV\footnote{At the time when the SST-1M was conceived, the sensitivity curves for point-sources of the CTAO Southern observatory in Paranal, for different off-axis angles and 50~h of observations, were shown in Fig.~5 in Ref.~\cite{2017ICRC...35..846M}. The region beyond a few TeV is dominated by an array of 70 SSTs. The SST-1M array sensitivity was below the 1\% of the Crab flux above 10~TeV.}. In Sec.~\ref{sec:design_principles}, we describe the general drivers of the SST-1M design and its concept. We then examine each subsystem in detail concerning their design definition and prototype realisation: the mechanical structure in Sec.~\ref{subsec:struct_design}
(including the drive system in Sec.~\ref{sec:drive}), the optical system in Sec.~\ref{subsec:optics_design}, the camera in Sec.~\ref{sec:camera_design}, and the control system in Sec.~\ref{subsec:telcontrol_design}. The validation of the three systems is described in Sec.~\ref{sec:design_valid}.



\section{The SST-1M design general principles and drivers}
\label{sec:design_principles}

\subsection{Optical designs for IACTs}
 
\begin{figure}[hbt]
	\centering\includegraphics[width=0.6\textwidth]{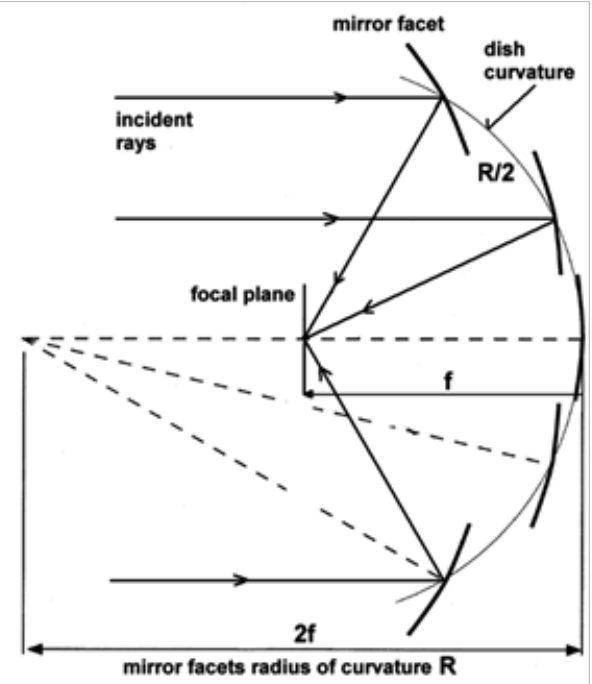}
	\caption{Sketch of the Davies-Cotton optical design from \cite{2011ExA....32..193A}.}
	\label{fig:DC_design}
\end{figure}

IACT optical designs are typically based on the Davies-Cotton or parabolic designs. Originally, the D-C design was invented for the optics of a solar concentrator \cite{1957SoEn....1...16D} as this system reduces the size of off-axis optical aberrations for field of views (FoVs) of several degrees. These aberrations have a lower value in comparison to a parabolic mirror with identical aperture, defined as $f/D$, where $D$ is the optical surface diameter and $f$ the focal length. 
The reason why the D-C is adopted in IACTs is related to cost and ease of maintenance in the field, as they function without a protective dome. The reflector of a D-C telescope is tessellated into identical elements with the same radius of curvature $R$, shape and size arranged on a spherical surface (dish) with a radius of $\frac{R}{2}$, which also defines the focal distance from the detection plane, namely the camera (see Fig.~\ref{fig:DC_design}). The orientation of the mirror segments is such that the light from the point-like source at infinity is reflected to the focal plane of the D-C setup. 
 Identical segments with a spherical reflective surface are easy to manufacture and can be easily adjusted or replaced when necessary. In contrast, solutions, such as parabolic mirrors, cannot use identical segments for the entire surface. 

\begin{figure}[hbt]
	\centering
\includegraphics[width=0.75\textwidth]{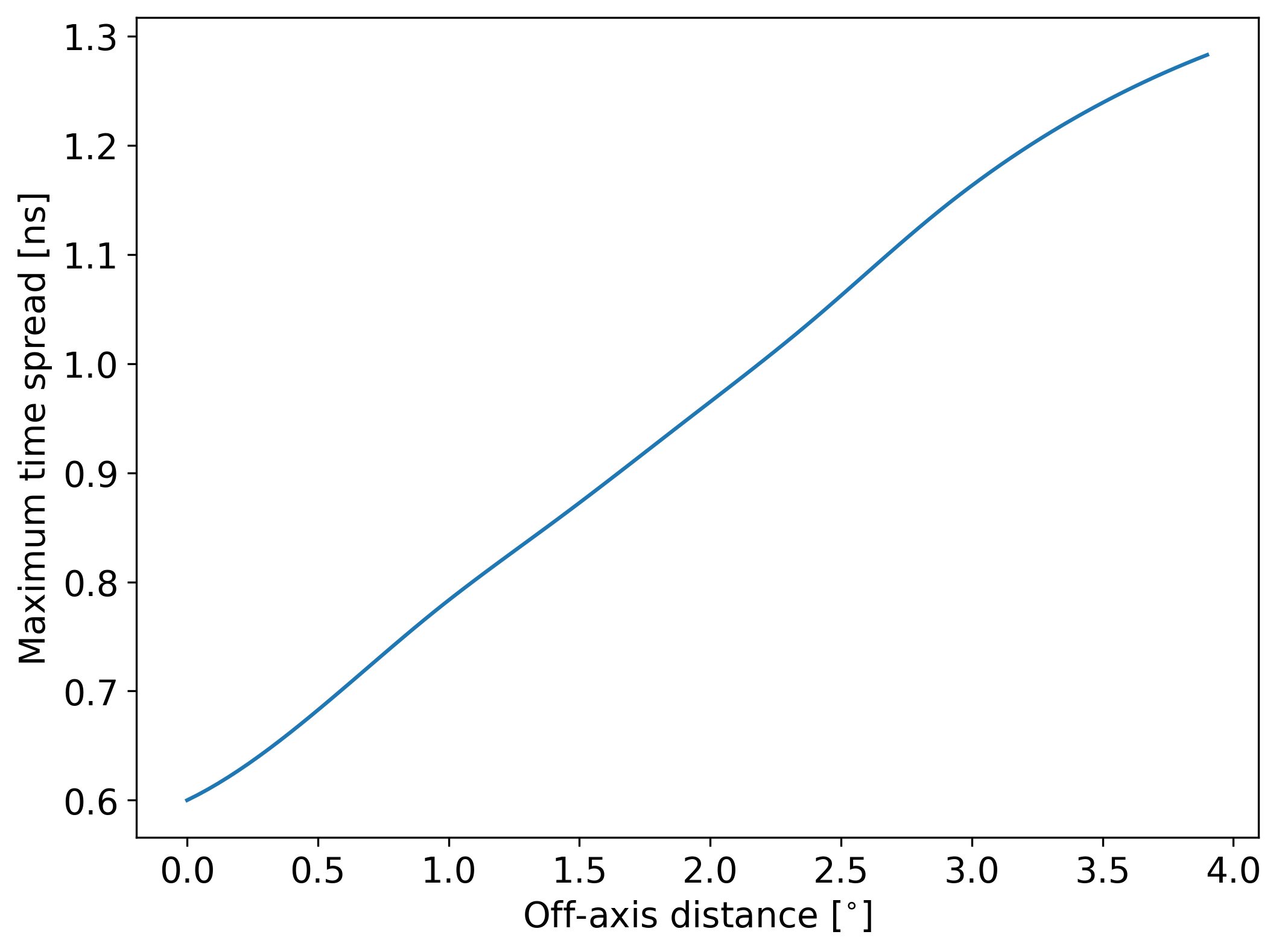}
	\caption{The root-mean-square of the time spread of signals between two pixels at different positions across the camera from simulation.}
	\label{fig:timespread}
\end{figure}

The reflective dish of a D-C design is not isochronous, unlike a parabolic dish. This is why large telescopes, such as MAGIC \cite{MAGICweb}, H.E.S.S. II \cite{HESSweb} and the CTAO Large-Sized Telescopes (LSTs) \cite{CTAOweb}, adopt a parabolic design. In the D-C case, the time extension of light pulses grows more than in parabolic designs with increasing focal length and decreasing aperture, which depends on the telescope size. For instance, the first D-C telescope deployed in 1968, Whipple in Arizona at 1750 m a.s.l., with a reflector diameter of 10~m and a focal length of 7.3~m, exhibits a pulse time extension of approximately 6.5~ns, often larger than the gamma-ray Cherenkov pulses \cite{2007APh....28..182K}.
For the SST-1M, the time difference between two pixels located at the maximum distance in the camera plane is less than 2~ns (see Fig.~\ref{fig:timespread}). This was determined using a full simulation of the telescope based on sim\_telarray \cite{simtelarray}. The size of the individual mirror segments influences the point spread function (PSF) of the telescope as a function of the FoV. Smaller segments result in a smaller PSF linear size on-axis, and the difference in the PSF for different size of segments decreases off-axis. 

\subsection{The SST-1M main features}
\label{sec:features}

\begin{figure}[hbt]
\centering
\includegraphics[width=\textwidth]{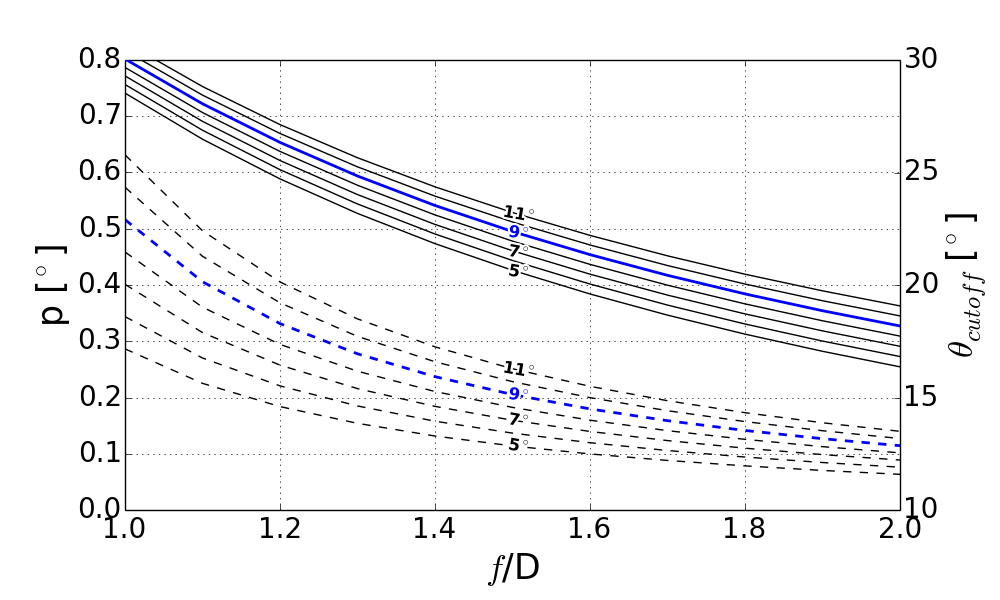}
\caption{The pixel angular size on the left axis and the cut-off angle of light-guides on the right axes vs f/D for the Davies-Cotton design for various fields of view. Figure from \cite{Aguilar:WinstonCones_2014}.}
\label{fig:pixel}
\end{figure}

\begin{figure}[hbt]
  \centering
\includegraphics[width=\textwidth]{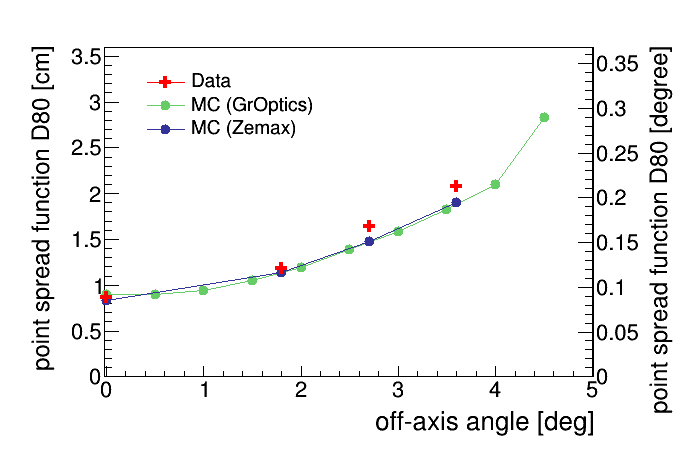}
\caption{The PSF vs. off-axis angle from measurements (see Sec.~\ref{sec:optval}) and simulations obtained in the design phase with Zemax and then with another ray-tracing and geometry simulation GrOptics (left).} 
\label{fig:PSF1D}
\end{figure}

The design of the SST-1M telescope was guided by the relationship between the Root Mean Square (RMS) of the Point Spread Function (PSF) in two orthogonal directions on the focal plane and the cut-off angle formula. The resulting dependence on the aperture of the pixels size (left y-axis) and the cut-off angle (right y-axis) is shown for different field of views in Fig.~\ref{fig:pixel} from \cite{Aguilar:WinstonCones_2014}. 
The SST-1M has a focal length of $f = 5.6$~m, the dish diameter of $D=4$~m, and thus a focal-to-dish diameter ratio of $f/D = 1.4$. Hence, the resulting pixel size is about $0.24^\circ$ (dashed lines) and the cut-off angle is $24^\circ$ for a FoV of about $9^\circ$. For the SST-1M, the FoV is about $8.9^\circ$, despite the PSF degrading with off-axis angle (see Fig.~\ref{fig:PSF1D}). 
The geometrical size of the optical surface is $9.4$~m$^2$. When reduced by shadowing from the mechanical structure and corrected for the average mirror reflectivity between $300$~m and $550$~m before operation, the effective area of the reflecting surface is about $6.5$~m$^2$. These parameters and other main characteristics of the SST-1M are summarised in the table in Fig.~\ref{fig:sst1m_table}.

The physics objectives of the SSTs span the energy range from $500$~GeV to beyond $100$~TeV. The sensitivity at the highest energies, which is critical for exploring the Galaxy, depends on the extension of the light pool on the ground and on the number of synchronised telescopes forming the array to sample it and on their mutual distance. These physics goals include large-scale surveys, high-precision spectral measurements in the high-energy range to understand cosmic ray acceleration to PeV energies (the energy of the `knee' in the cosmic ray spectrum), and the search for cosmic ray sources. 
To effectively collect the sparse gamma-ray data above 10~TeV, it is essential to deploy arrays of numerous telescopes spanning the largest possible area. Each telescope should also maximize its exposure time and field of view (FoV). This approach enables better aperture overlaps between telescopes and facilitates observations of the many intriguing high-energy, extended galactic sources.

The optical system and structure of the SST-1M follow a traditional design, albeit with novel solutions for selected elements of the structure and drive system. The most innovative part is the SiPM-based camera with fully digitising electronics.
The SiPM technology has been pioneered in the field of gamma-ray astronomy by the First G-APD Cherenkov Telescope (FACT) collaboration \cite{FACT}. It benefits from the low-cost and mass-producible semiconductor photosensing technology and its robustness to continuous light (the night sky background or NSB, produced by starlight, reflections of moonlight and by human activity). The FACT SiPM camera can safely operate with the Moon in the field of view without risk of damage \cite{2013ICRC...33.1132K}. This robustness enables longer exposure times, as it experiences minimal ageing and requires no human intervention to manage high levels of NSB. 
The adoption of SiPMs also allows for a cost-effective and easy-maintenance system.
The SST-1M and FACT allow for a deep understanding of how to operate and calibrate SiPMs for long-term operation. 

The SiPM sensors are lightweight, consume roughly an order of magnitude less power than photomultiplier tubes (PMTs), and offer high photodetection efficiency (PDE). However, they have the drawback of higher noise, which is influenced by their signal duration and capacitance, which is proportional to size. SiPM provide a PDE that is 10-20~\% higher than that of comparable PMTs. Additionally, their sensitivity surpasses that of the typical PMTs beyond 600~nm and extends to wavelengths above $800$~nm, unlike PMTs. While their higher dark count rate than PMTs at room temperature is not a major concern as NSB dominates, their higher sensitivity in the red and near-infrared regions requires the adoption of filters on the camera entrance window, or lightguides, or sensors. As the NSB increases in this wavelength region, the SST-1M Borofloat entrance window is coated with a dichroic filter cutting wavelengths longer than 550~nm (see Sec.~\ref{sec_PDP} and \cite{Alispach:2020}). Interestingly, the SST-1M has a DC-coupled preamplifier camera electronics, unlike FACT which is AC-coupled. This configuration allows for the monitoring of the NSB through changes of the baseline.  

The camera pixel size, derived by the D-C formulas in Fig.~\ref{fig:pixel}, is constructed with a monolithic hexagonal silicon photomultiplier, developed by Hamamatsu in cooperation with the University of Geneva group. The SiPM has an approximate area of $1$~cm$^2$ \cite{NAGAI2019162796}, and is coupled to a lightguide that extends the linear dimension of the pixel to $2.4$~cm. The sensor has four channels, and their outputs are pre-amplified and then summed.

\begin{figure}
	\centering
		\includegraphics[width=\textwidth]{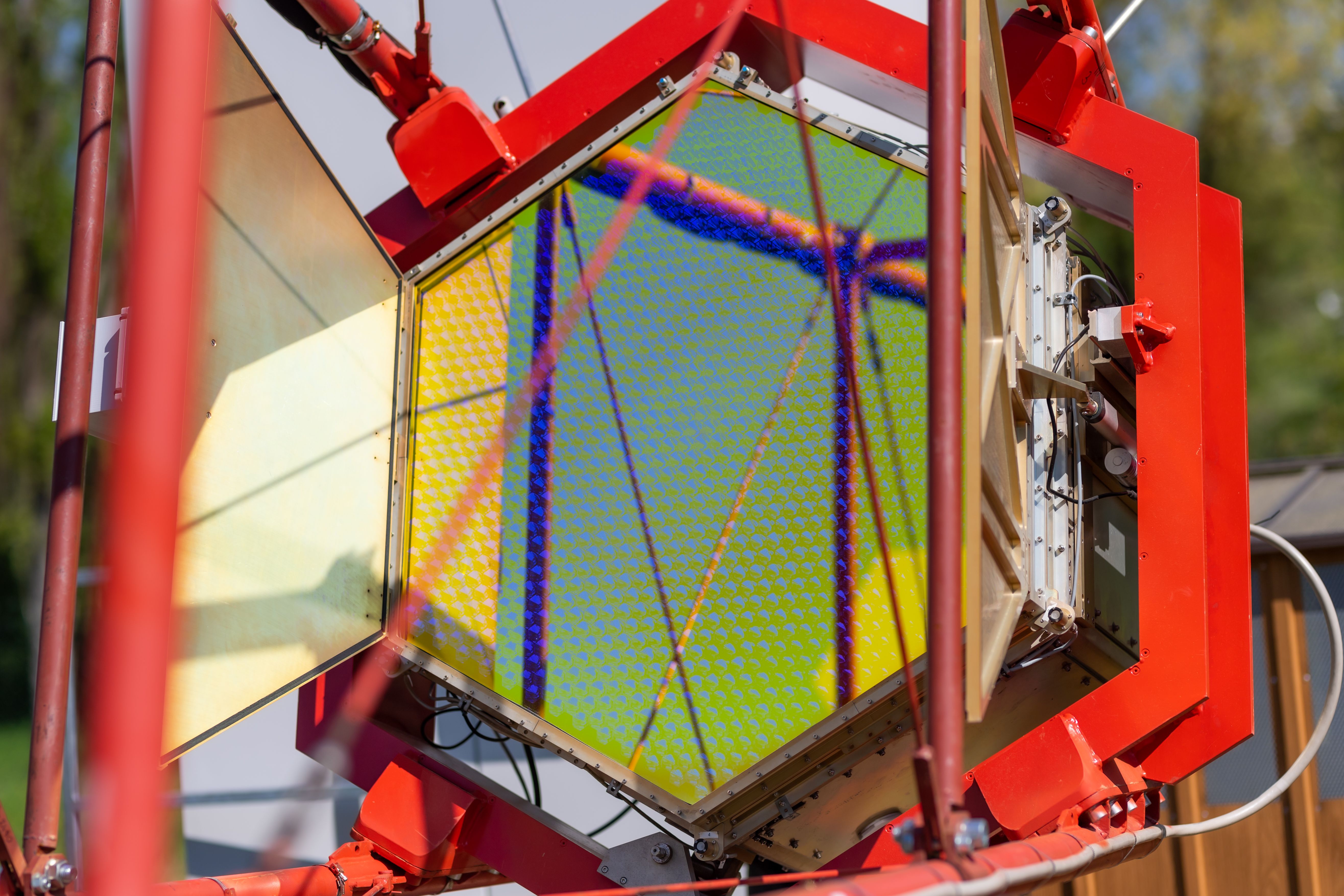}
	\caption{The camera of an SST-1M at the Ond\v{r}ejov site. The reflectance in the yellow-red of the narrow-band filter on the camera window is visible.}
	\label{fig:camera}
\end{figure}

\begin{figure}[hbt]
\includegraphics[width=\textwidth]{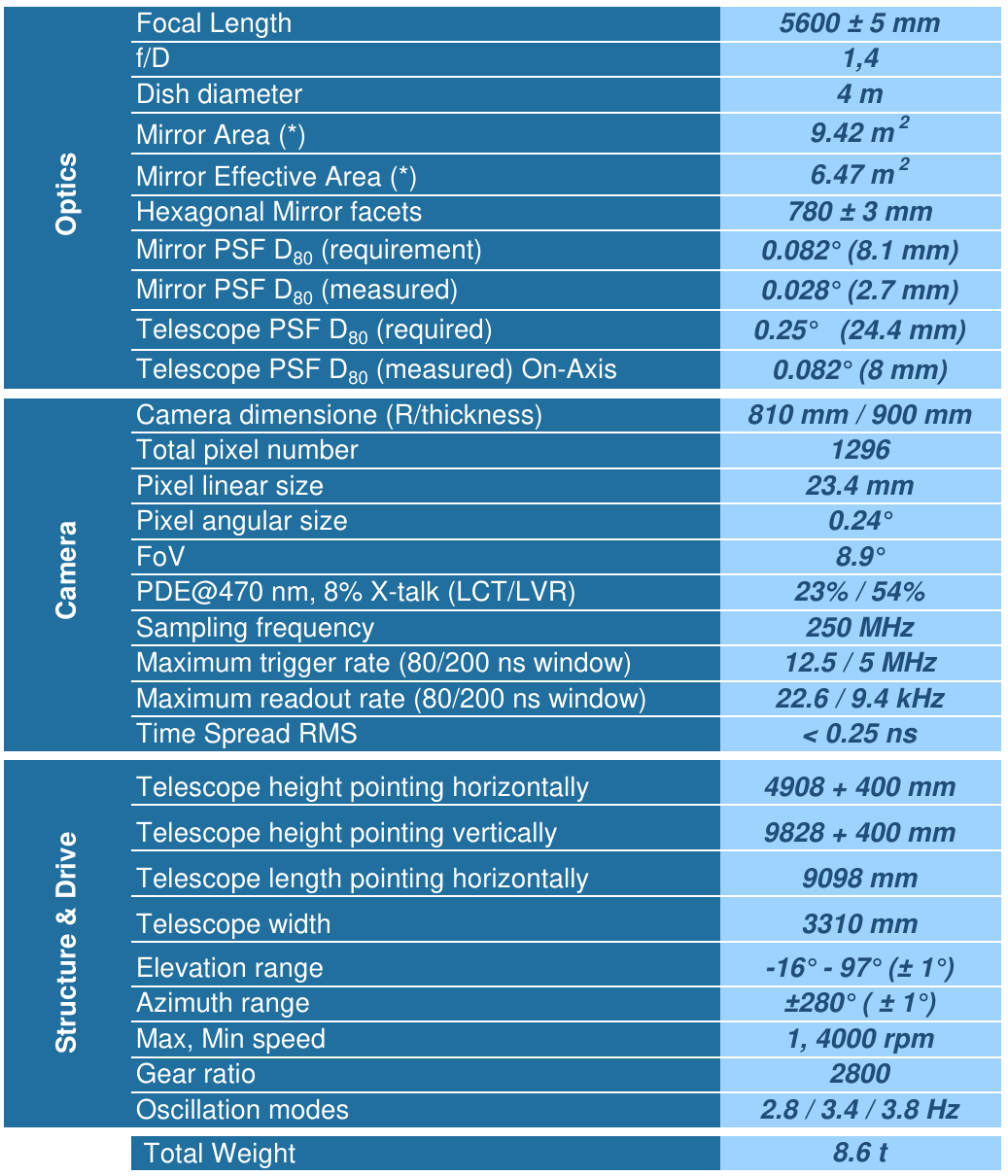}
{\footnotesize *) The mirror effective area takes the reflectivity of mirrors and the telescope structure shadowing into account.}
\caption{Table of main SST-1M parameters.}
\label{fig:sst1m_table}
\end{figure}

The camera, also described in \cite{CameraPaperHeller2017} and shown in Fig.~\ref{fig:camera}, contains a total of 1296 pixels organised into three sectors, with each sector divided into modules of 12 sensors. Analogue signals from the pixels are routed to the crate-based electronics called \dcam{} \cite{Zietara:SPIE2014}, where they are digitised using independent 250~MSPS/12bit Flash Analog-to-Digital Converters (FADCs). Further processing, including camera triggering and readout, is performed digitally in the FPGA chips. Such a fully digital approach offers numerous advantages, such as dead-time-free operation, flexible triggering, full waveform readout, and high performance.  

\begin{figure}
\centering
\includegraphics[width=\textwidth]{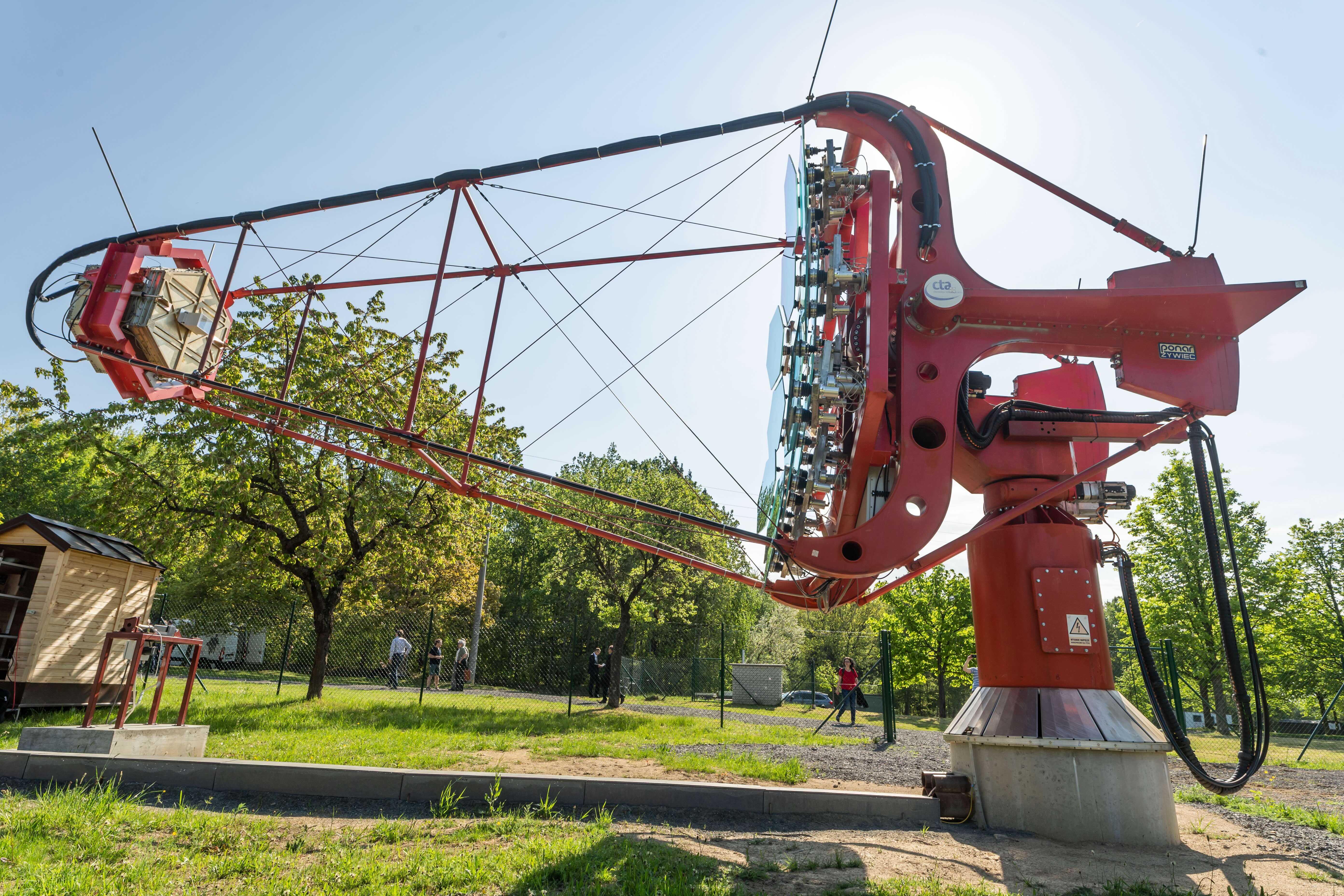}
\includegraphics[width=\textwidth]{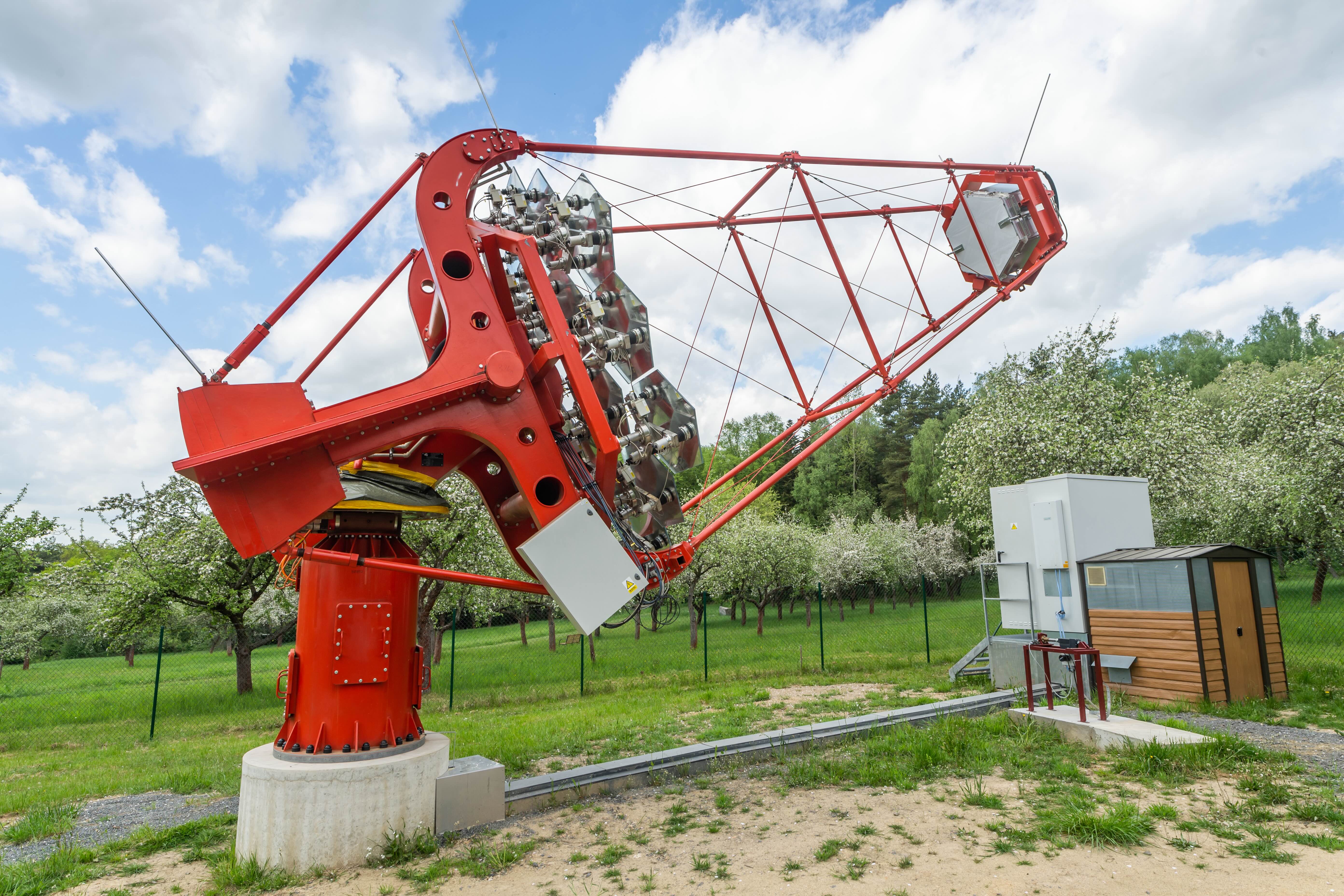}
\caption{The first (top) and second (bottom) SST-1M installed at Ond\v{r}ejov.}
\label{fig:SST-1M}
\end{figure}

The adoption of SiPMs allowed for a smaller mirror size compared to a D-C design using PMTs, which would require a mirror diameter of about 7~m \cite{2013ExA....36..223R}. 
The prototype telescope, which validated the design specifications, and the second telescope, which improved aspects concerning easier maintenance and weatherproofing of the mechanics, are shown in Fig.~\ref{fig:SST-1M}. 

\section{The telescope design and prototyping}
\label{sec:designproto}

In the following, we discuss the project design of the  main systems composing the SST-1M: the mechanical structure with the drive system and control, the optical system, the camera, and the control software.


\subsection{The telescope structure and drive system}
\label{subsec:struct_design}

The SST-1M features a simple, compact, and lightweight mechanical structure with a total weight of approximately 8.6 tons. The size and dimensions of the telescope are listed in Fig.~\ref{fig:sst1m_table}. Designed to be cost-effective and easy to build, transport, install, and maintain, the structure is predominantly made from steel, using off-the-shelf steel profiles and tubes from the industry. The telescope frame fits into a standard 12 m long container ($2.3 \times 2.6 \times 12~\rm{m}^3$). Its compact design simplifies maintenance, as the distance between the camera and the ground is only about 1~m when the telescope is parked. Despite its simplicity, the design is robust and solid, suitable for sites at altitudes above 2 km and capable of withstanding the environmental conditions at the southern array of CTAO in Chile.

In this section, we describe the design of the telescope structure and drive system. This description is based on the design of the second telescope prototype, which incorporated improvements and modifications made after constructing the first prototype.

\subsubsection{The telescope frame}
\label{sec:structure}

\begin{figure}
  \centering
\includegraphics[width=\textwidth]{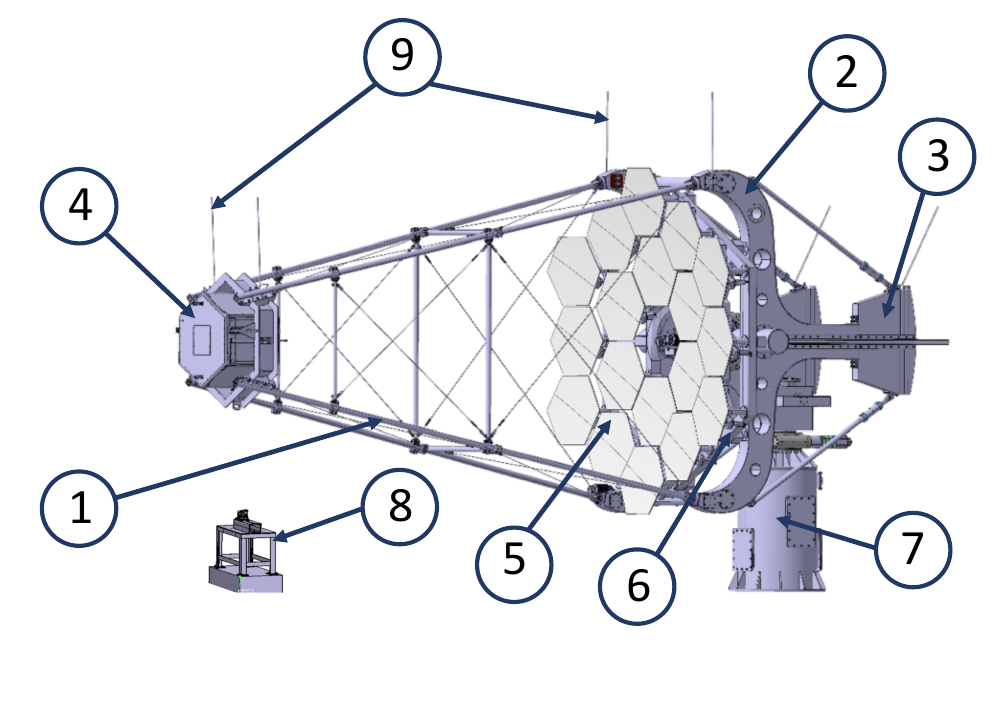}
\caption{SST-1M structure with its sub-systems. Numbered labels are referred to in the text.} 
\label{fig:Structure}
\end{figure}

The CAD model of the telescope structure is shown in Fig.~\ref{fig:Structure}. The structure is composed of several sub-systems. The mast (1) is directly connected to the dish support structure (2) to which the counterweight (3) is also attached. This solution decouples the deformations of the mast from
the deformations of the dish (6).
The mast positions the camera (4) from the reflecting mirror surface (5) at the focal distance $f$. The mirrors are mounted on the dish. The dish support structure is mounted on the telescope support (7). 
The docking station (8) locks the telescope in the parking position. The telescope is also equipped with lightning protection rods (9).
The movement of the telescope around the elevation and the azimuth axes is realised using the IMO \cite{IMO} slew drive and one roller bearing for each axis. The slew drive is a compact system that combines a twin worm gear with a motor and a roller bearing. Each axis is driven by two servo motors. The slew drives and servo motors of both the azimuth and elevation drive systems are identical.
The concept of the SST-1M drive system with IMO slew drives and roller bearings, is developed in synergy with the solution used in the Medium-Sized Telescopes (MSTs \cite{2015ICRC...34..959G}) of CTAO.

\begin{figure}
\centering
\includegraphics[width=0.6\textwidth]{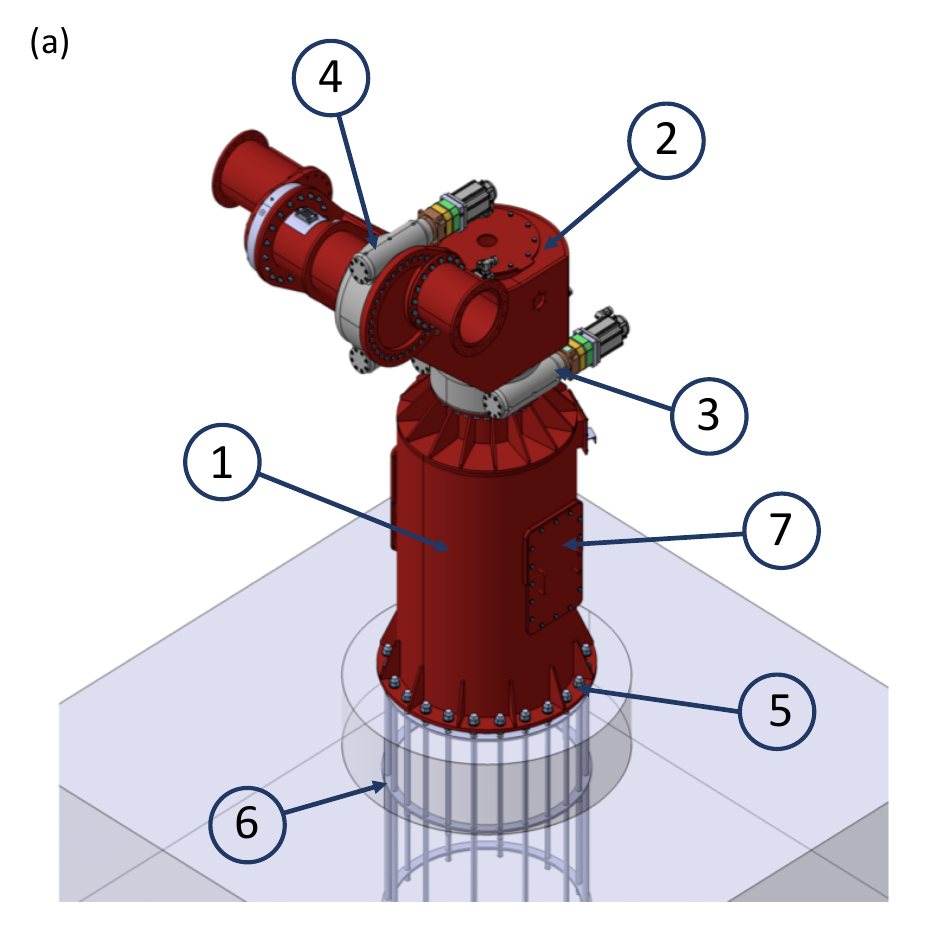}
\includegraphics[width=0.6\textwidth]{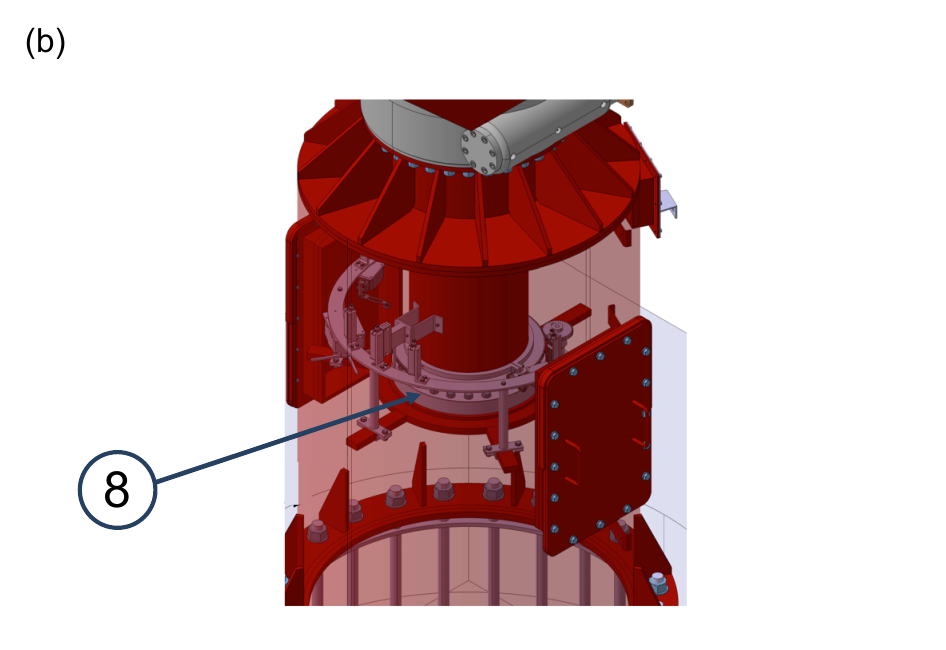}
\includegraphics[width=0.6\textwidth]{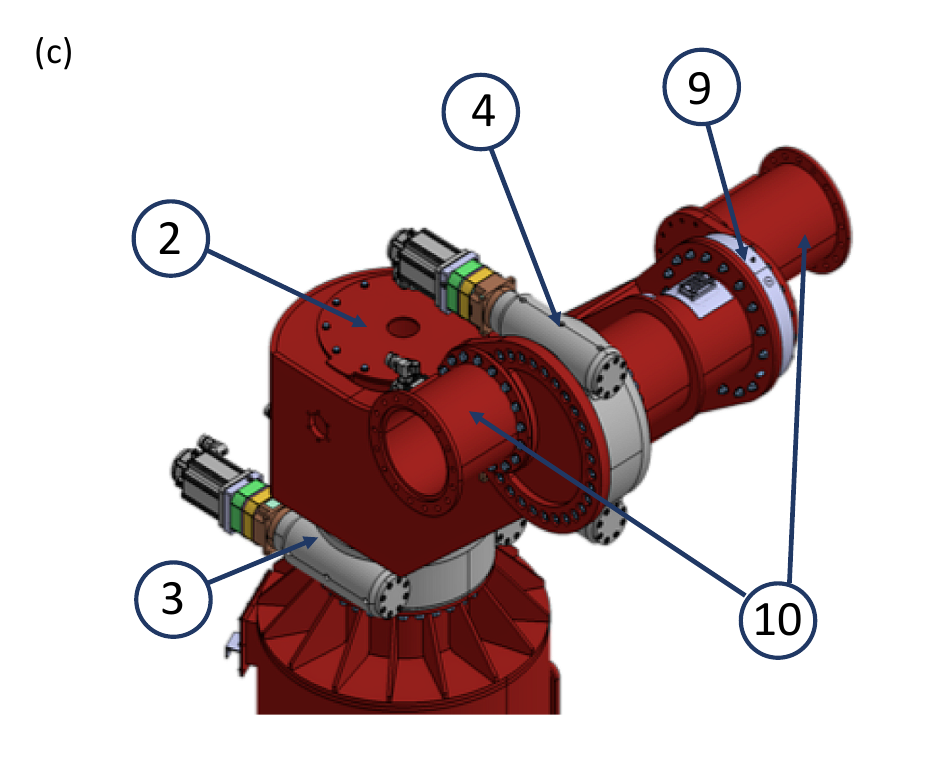}
\caption{Design of the telescope support and foundation (a), inside view of the tower (b), and design of the telescope head (c). Numbered labels are referred to in the text.}
\label{fig:support} 
\end{figure}

The SST-1M support has two main components shown in Fig.~\ref{fig:support}: the tower (1) and the head (2). The elements of the azimuth drive system (3) are incorporated into the tower and those of the elevation drive system (4) are integrated into the head. As shown in Fig.~\ref{fig:support}~(a), the tower is bolted to a reinforced concrete foundation (5). The part of the foundation above the ground level has a diameter of 1.4~m and a height of 0.4~m.
The elevation of the tower above ground level protects it from water ingress.
Inside the part below ground level, there is a steel anchor grid structure (6), with two reinforcement rings manufactured for easy and precise installation and removal of the support from the foundation. 

\begin{figure}
\centering
\includegraphics[width=\textwidth]{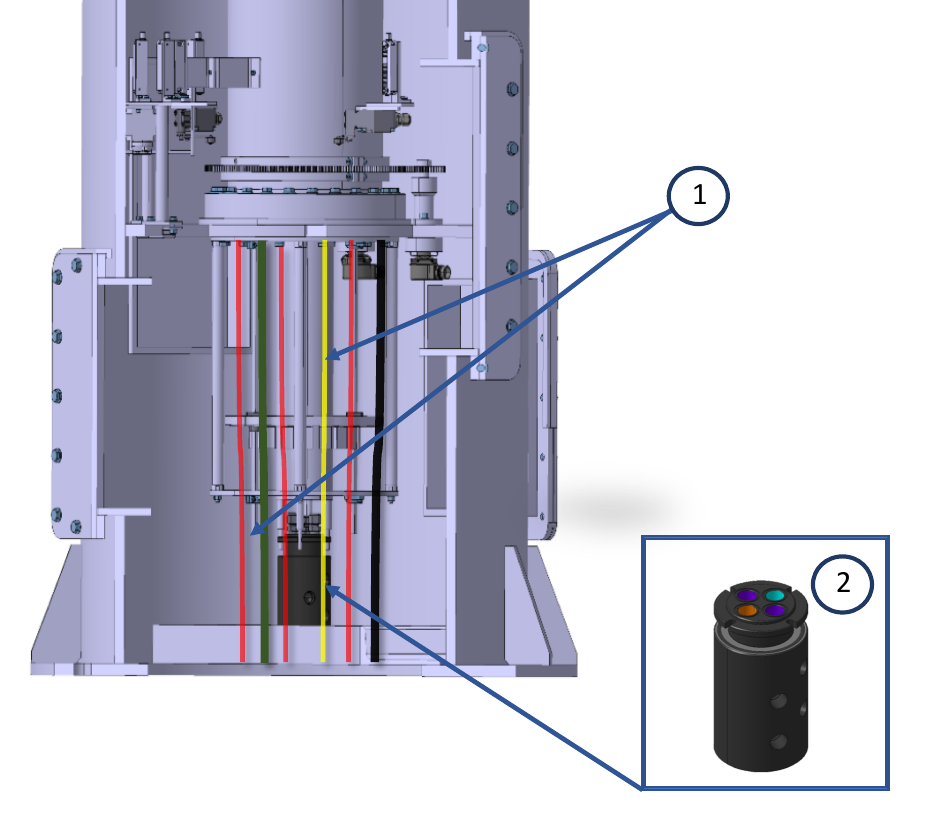}
\caption{Design of the cable rooting system in the telescope tower.}
\label{fig:twist} 
\end{figure}

The slew drive (3) for the azimuth rotation of the telescope is fixed to the
conical part of the tower (the tower cap). Inside the tower, there is a roller bearing (8) connected to the
azimuth slew drive via a tube that stabilises the entire structure, Fig.~\ref{fig:support}~(b). The housing of the azimuth bearing is built into the tower. 
The roller bearing and other components of the drive system can be accessed via two openings (7) for installation and maintenance. The head is rotated by the azimuth slew drive, Figs.~\ref{fig:support} (a, c). 
The elevation drive system, including the slew drive (4) and the roller bearing (9), are bolted to the pads of the circular tubes (10) incorporated into the forks of the dish support structure, Fig.~\ref{fig:support} (c).

The telescope design incorporates a novel solution for routing cables inside the tower and mast, enhancing their protection. This system allows cables and pipes to rotate within the tower, as shown in Fig.~\ref{fig:twist}. Special flexible power and optical cables (1) have been selected for this purpose, and a rotary union (2) is used for the pipes carrying cooling fluid to the camera.

\begin{figure}
\centering
\includegraphics[width=\textwidth]{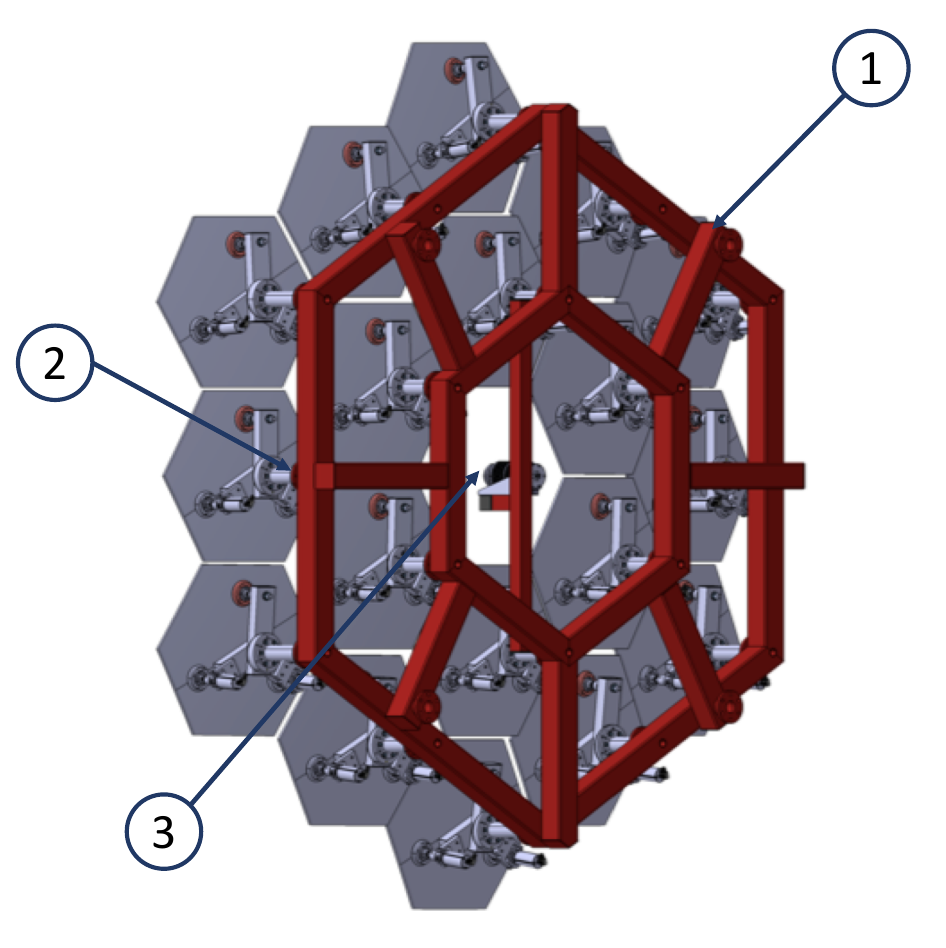}
\caption{Design of the mirror dish and the mirror layout.}
\label{fig:Mirror_Dish}  
\end{figure}

The mirror dish, Fig.~\ref{fig:Mirror_Dish}, is a welded structure made of square steel profiles of size $120~$mm$\times 120~$mm with a wall thickness of 4~mm. Thicker profiles of 6~mm are used at four points (1) where the dish is connected to the dish support structure. The required spherical shape of the dish is achieved by welding together two hexagonal sections using eight straight profiles in a star shape. On the dish, 18 pads (2) are welded. Precise machining of the pads is required after welding. Once the pads are levelled, the holes for pins and screws are drilled.
The centre of the dish hosts the CCD camera (3) for the mirror alignment and telescope pointing systems.

\begin{figure}
\centering
\includegraphics[width=\textwidth]{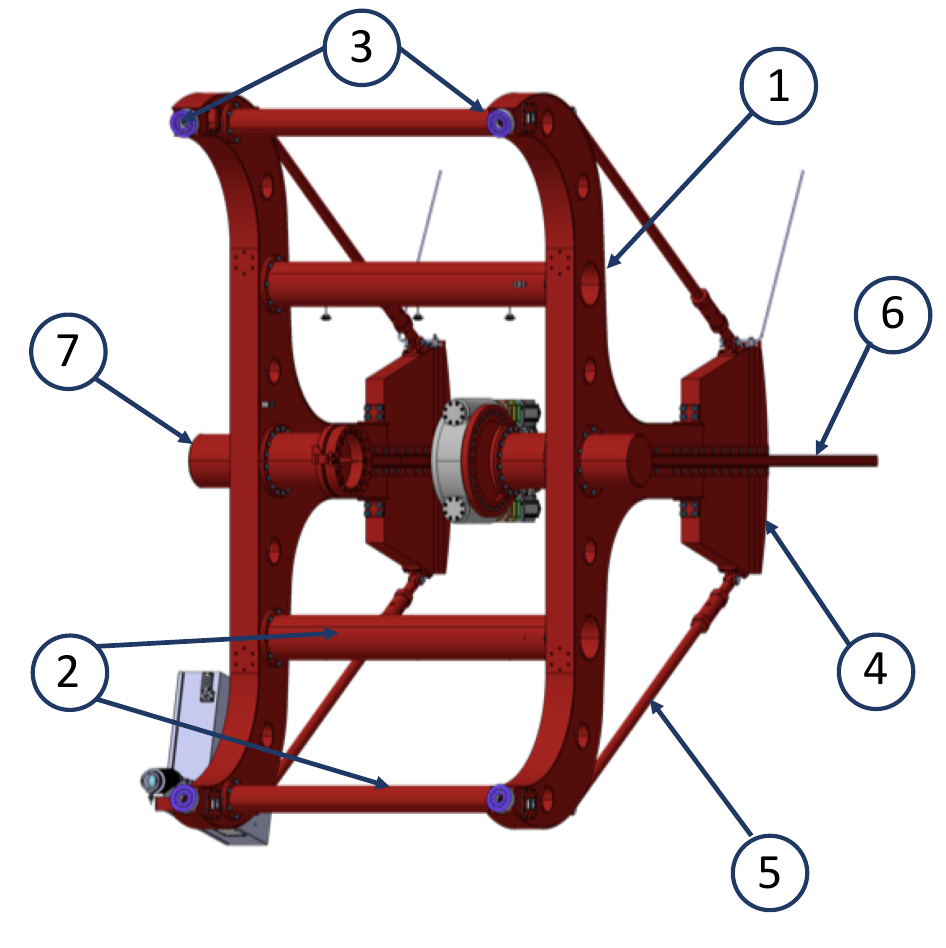}
\caption{Design of the dish support structure.}
\label{fig:Dish_support}  
\end{figure}

The dish support structure is shown in Fig.~\ref{fig:Dish_support}. It is composed of two Y-shaped fork pieces (1), which are box structures welded from steel sheets and tubes. The Y-shaped forks are separated by four steel tubes (2). As mentioned, the long tubes of the mast are bolted to the four ends of the forks (3). Counter-weights (4), made of cast iron, are fixed to the opposite sides of the fork ends. Four struts (5) and steel sheets (wings) (6) ensure the necessary stiffness of the structure, with the wings also stabilising it against wind gusts. Two circular tubes with pads (7) are also incorporated into the forks. The pads serve as interfaces for bolting the elevation drive system components, including the slew drive and the roller bearing mounted on the sides of the telescope support head.

\begin{figure}
\centering
\includegraphics[width=0.49\textwidth]{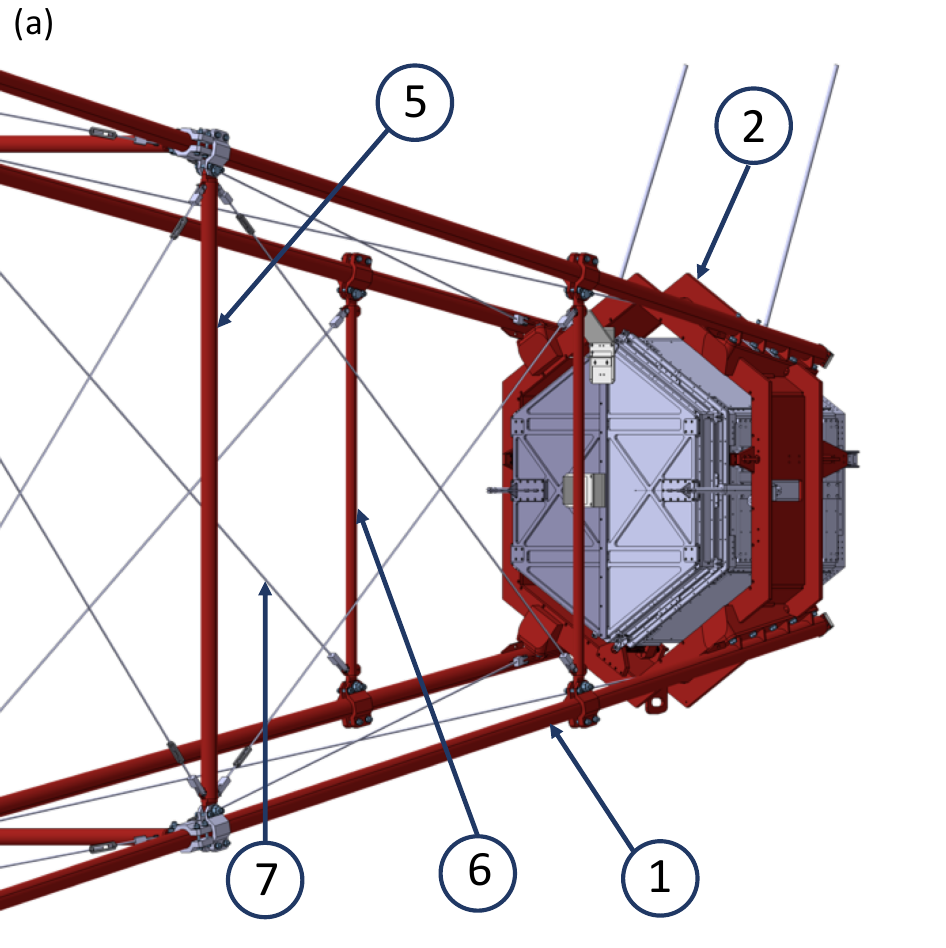}
\includegraphics[width=0.49\textwidth]{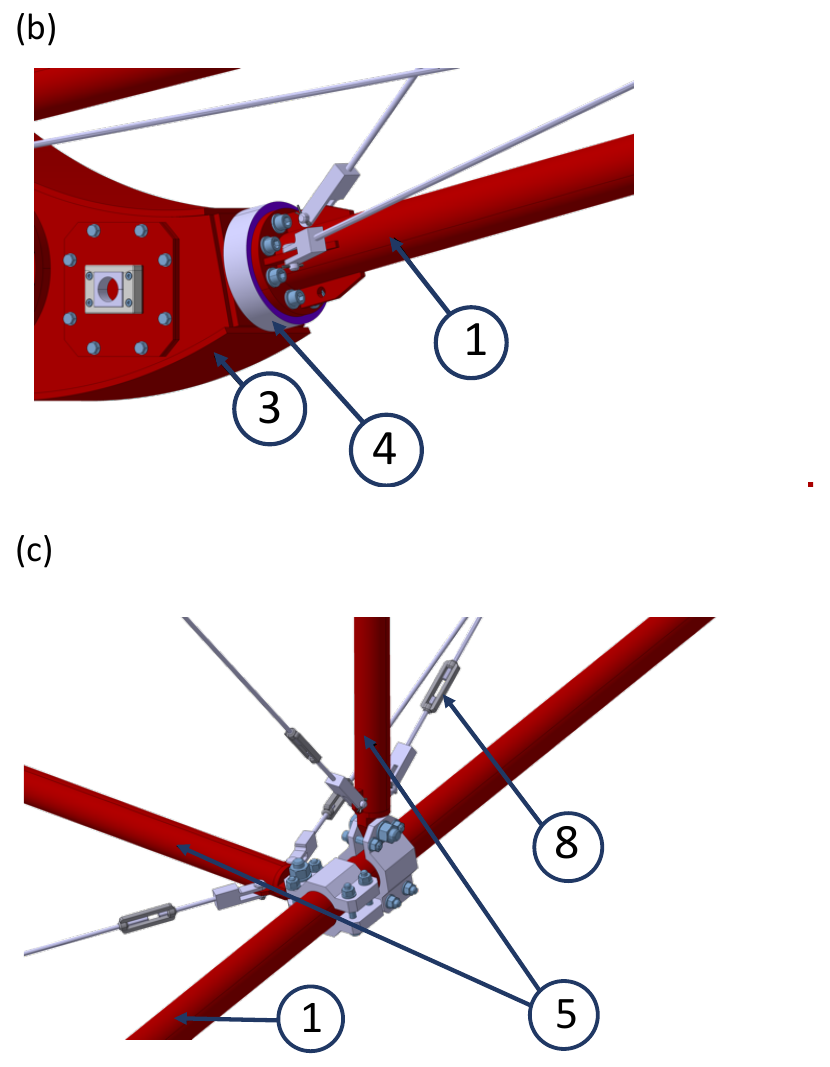}
\caption{Details of the mast design:  extremity to the camera (a), to the dish structure (b), and an intermediate part of the mast (c).}
\label{fig:mast}  
\end{figure}

The detailed design of the mast is shown in Fig.~\ref{fig:mast}. Its main components are four steel pipes (1) with a length of 6.15~m, an outer diameter of 70~mm and a wall thickness of 6.3~mm. The four long pipes without divisions allow for routing services, such as compressed air pipes, water cooling pipes, power cables, and optical fibres, inside them. 
At one extremity, Fig.~\ref{fig:mast} (a), the four tubes are connected to the camera interface (2) made of aluminium sheet. At the other end, Fig.~\ref{fig:mast} (b), the tubes are bolted to the dish support structure (3) using precisely machined setting shims
(4) to achieve the proper angle to the camera basket.
A rectangular frame (5), made of circular steel tubes with a diameter of 60.3~mm and a thickness of 6.3~mm, connects the tubes using clamps halfway along their length to increase the stiffness of the mast, Figs.~\ref{fig:mast} (c) 
and~\ref{fig:Structure}. 
For similar purposes, two vertical tubes (6) are added closer to the camera basket.
Sixteen rods (7) with a diameter of 8~mm, pre-stressed with a force of 5100~N, are fixed within the mast frame with turnbuckles (8) to further increase the rigidity of the mast.  

The camera interface is designed to be rigid and compact and facilitate easy mounting of the entire camera system. This is already equipped with the shutter and its motors. This design ensures the camera can be a self-contained system ready for installation. The shutter protects it also during shipments. Once the camera is inserted into the basket, it is fixed in place and cabled to become operational. 
From simulations, it has been determined that minimal shadowing of the shutter occurs when it is open at $110^{\circ}$ with respect to the camera plane. The arrangement of the tubes and tensioning rods in the mast was optimised to achieve this configuration. 

\subsubsection{The drive system}
\label{sec:drive}

\begin{figure}
\centering
\includegraphics[width=\textwidth]{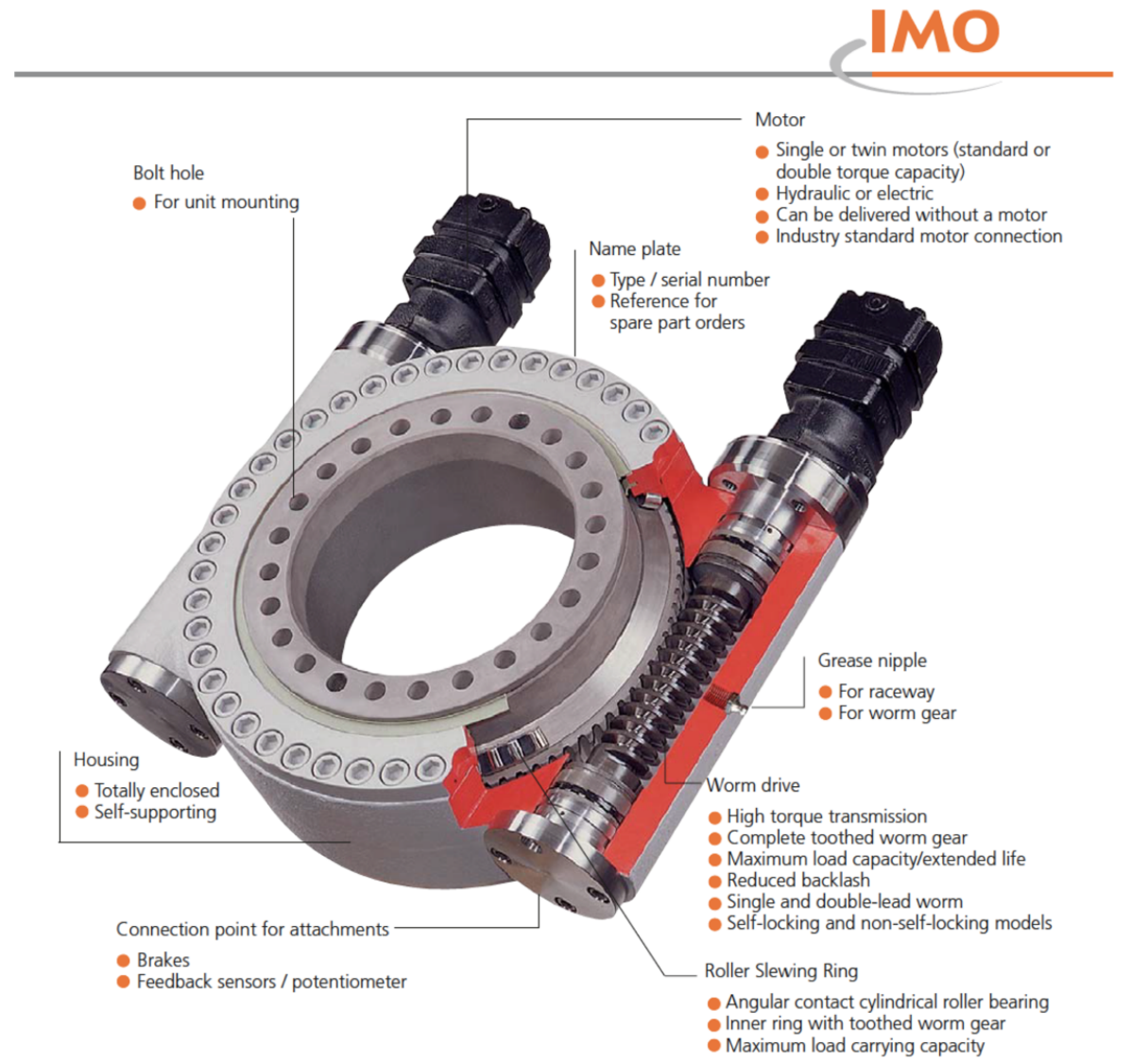}
\caption{The IMO slew drive model WD-H. Reproduced with permission of IMO Holding GmbH. This figure is not publicly available online and should be requested on the manufacturer website: \href{https://www.imo.de/en/products/slew-drives}{https://www.imo.de/en/products/slew-drives}.}
\label{fig:driveIMO}  
\end{figure}

\begin{figure}
\centering
\includegraphics[width=\textwidth]{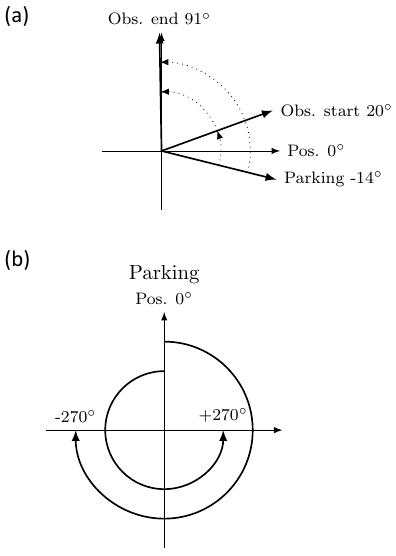}
\caption{Telescope movement range in elevation (a) and azimuth (b).}
\label{fig:drive_coordinates}  
\end{figure}

The telescope positioning and tracking system uses two independent drive axes:  azimuth and elevation. Each axis employs an IMO slew drive, combining twin worm gears with motors and a roller slewing ring to facilitate the transmission of both radial and axial forces. The slew drive has a fully enclosed and self-supporting housing. Rotating telescope components are fixed to this housing with bolts.
The SST-1M design adopts an IMO slew drive of the WD-H type, the same type
as for the MSTs of CTAO.
The general features of this IMO slew drive are illustrated in Fig.~\ref{fig:driveIMO}. 
The inclusion of twin worm gears with two servo motors in each slew drive enhances torque capacity and eliminates backlash, ensuring higher pointing accuracy. The redundancy of motors increases the reliability of the system 
and provides flexibility for high-acceleration maneuvers, with both motors capable of delivering torque in the same direction for rapid slewing.

The elevation and azimuth drive systems enable rotations of the telescope around the horizontal and vertical axes between \( -14^\circ \) (parking position) and \( 91^\circ \) and \(\pm 270^\circ \), respectively, Fig.~\ref{fig:drive_coordinates}.
This range is needed for telescope positioning during observations, technical tests, and stowing.
It is also assumed that in operation mode, the telescope can fast slew to any defined point in the sky within 1~min, including 6~s for both the acceleration at a constant rate from the start position and the deceleration. 
The resulting maximum angular velocities are $\omega_{max,a}=6^\circ$/s and $\omega_{max,e}  = 2^\circ$/s with respect to the azimuth and the elevation axes, respectively. The pointing accuracy requirement is within 7 arcseconds, and the tracking accuracy should not exceed 5 arcminutes. The full range of telescope movement in elevation and azimuth is shown in Fig.~\ref{fig:drive_coordinates}. 

To meet these requirements, the drive system components have been chosen after calculating mechanical torques and power needs. In these calculations, the
telescope site is assumed to be at an altitude of 2500~m, with operational temperatures ranging between -20\degc{} and +30\degc. The observation mode assumes a maximum average wind velocity of 50~km/h over a 10 minutes, while the emergency mode allows for safe parking of the telescope in conditions with wind velocity reaching up to 100~km/h, including gusts.
The maximum calculated total torques about the elevation and the azimuth axes are 3426~Nm and 6828~Nm, respectively, for a wind velocity of 50~km/h, and 6064~Nm and 11842~Nm for a wind velocity of 100~km/h. The total torques are dominated by the wind torques. In the calculations for the 50 km/h wind, the wind gust coefficient is assumed to take the value of 2.2, which is unrealistic under normal operating conditions. Thus, the parameters of the drive system components selected based on these calculations include a fair amount of safe margin.

In the process of selecting the drive system, the data obtained at a 50 km/h wind velocity were used to select the servo motors, which need to operate at their nominal speed during the observation mode. 
The data obtained for the wind velocity of 100 km/h were used to define the parameters of the gearboxes, which must withstand the loads during the safety-mode telescope parking. 
The final selection of the drive system components was based on the aforementioned calculations, with additional corrections to account for decreased motor efficiency at high altitudes, increased resistance in the gears at negative temperatures, and increased resistance during emergency mode operation. 
For both the azimuth and elevation axes, the worm drive WD-H 0373/3-00028 produced by IMO was selected, with a gear ratio of 1:56. 
The synchronous servo motors chosen for both axes are Bosch-Rexroth MKS70C-0450-NN, paired with the mechanically compatible planetary gear GTM 140-NN2-050, which has a gear ratio of 50, resulting in a total gear ratio of 2800. The motor speed limit is set to 4000 rpm, but the Bosch-Rexroth motor can operate at speeds below 1~rpm. Using the same components for both drives simplifies the system. 

\subsection{The SST-1M optical system}
\label{subsec:optics_design}

The optical system of the SST-1M provides the optical path for the Cherenkov light emitted by showers in the atmosphere and concentrates it on the camera to be recorded. It consists of 18 hexagonal mirror facets forming the telescope's reflective dish, the mirror alignment system for proper orientation of the mirror facets with respect to the focal plane by enabling mirror reorientation (see Sec.~\ref{subsec:optics_align}), and the pointing system (see Sec.~\ref{sec:point}). 

\subsubsection{The mirror facets}
The 18 hexagonal mirror facets have a radius of curvature of 11.2~m and size of 780~mm flat-to-flat. They are arranged in two concentric rings (see, e.g., Fig.~\ref{fig:Structure}) and form a reflecting spherical dish with a diameter of 4~m, with a 2~cm space between the tiles. The central facet would be obscured by the camera and therefore it is not installed. The mirrors are connected to the telescope dish through the fixing set of the mirror alignment system (see the next section). 

The mirror substrate is a 15~mm thick borosilicate glass. 
The thickness of the mirror substrate was selected based on an extensive Finite Analysis Method (FEM) calculations, taking into account factors such as wind and gravity loads. 
The front surface of the substrate is coated with aluminium and protective SiO$_2$ thin layers. 
The thickness of the coatings is optimised to reflect the characteristic spectrum of the Cherenkov light emitted by atmospheric showers in the wavelength interval $300 - 550$~nm. The required spectral on-focus reflectivity of the mirror is larger than $85\%$ over these entire wavelength range. The micro-roughness of the mirror surface is controlled during manufacturing to reach a limit of less than 1~nm, verified with the CASI instrument \cite{Nozka:11}, to minimise the scattering component of the mirror. 
The quality of the mirror surface shape is determined by the size of the Point Spread Function (PSF) in the focal plane at $f=5.6$~m, defined to be the angular diameter of a circle containing 80\% of photons emitted by a point source, denoted by $D_{80}$. For all of the SST-1M mirrors $D_{80}<8.1$~mm. 

\subsubsection{Mirror alignment system}
\label{subsec:optics_align}

\begin{figure}
\centering
    \includegraphics[width=0.8\textwidth]{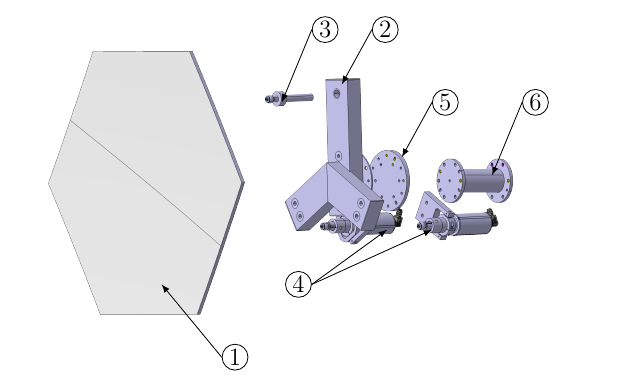}
    \includegraphics[width=0.8\textwidth]{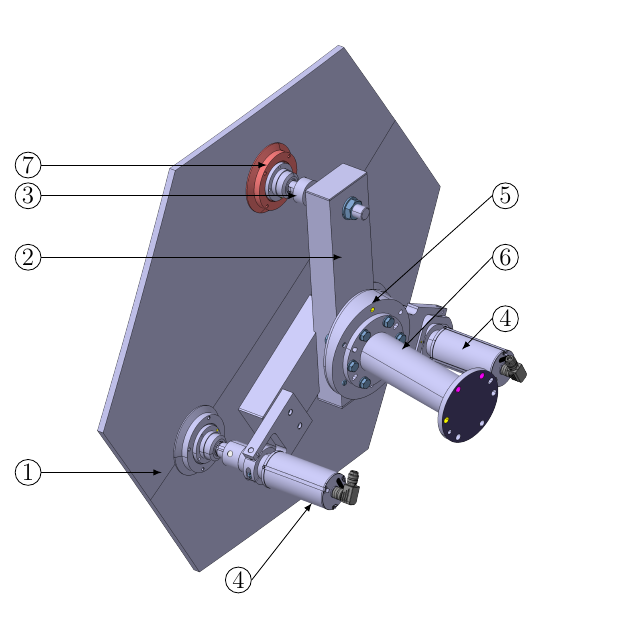}
    \caption{The mirror alignment components: mirror facet (1), mechanical interface (triangle) (2),  fix-point (3), actuators (4), setting shim (5), tube (6), mirror flange (7).}
\label{fig:Fixation}
\end{figure}

The mirror alignment system is based on a set of two actuators and one fix-point. They reliably fix the mirror facets to the dish and prevent the transfer of thermoelastic stresses from the structure. The actuators also serve as the executive elements for aligning the mirrors. They enable precise control of the orientation of each facet's optical axis with respect to the camera, which has to be stable at the arc minute level. Fig.~\ref{fig:Fixation} illustrates the mechanical components of the single mirror alignment set.

\begin{figure}
    \centering
    \includegraphics[width=\textwidth]{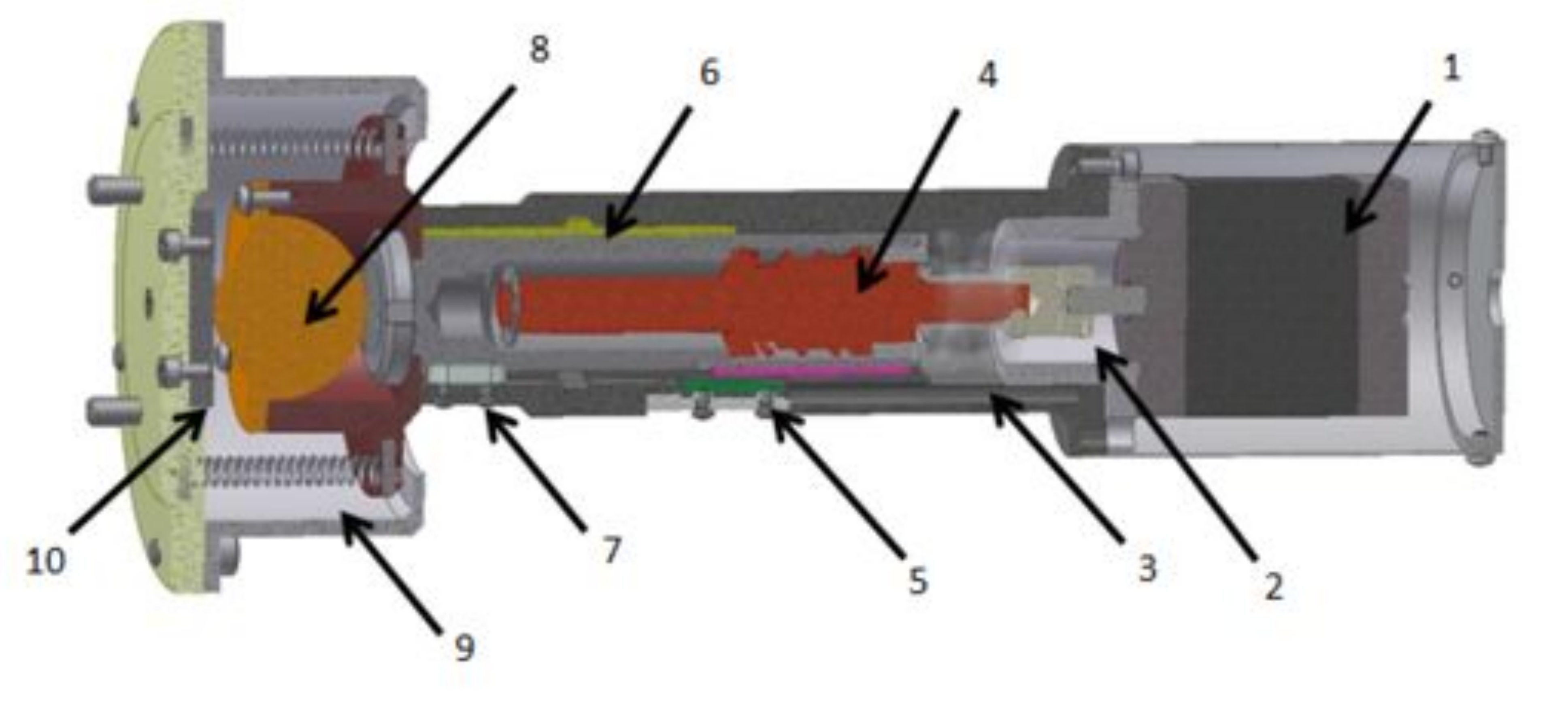}
    \caption{Design and components of the actuator used in SST-1M: stepper motor (1), Oldham clutch (2), ball bearings (3), lead screw unit (4), magnetic encoder (5), linear bearing (6), sliding key (7), ball transfer unit (8), springs (90), plane (10).}
\label{fig:actuator}
\end{figure}

The mechanical structure of the alignment set is composed of a tube (6), a setting shim (5) and a triangle (2). These elements are identical for all telescope facets. Each mirror is fixed to the interface via a metallic flange (7) glued to its rear surface. 
Two types of actuators with the desired accuracy, one with 4 degrees of freedom (DoF) and another with 5 DoF, are used to achieve the desired performance. Together with the fix-point, they allow the isostatic mounting of the mirror.
The operation principle of the actuator is based on the stepper motor, which is connected to the screw lead unit through an Oldham clutch, Fig.~\ref{fig:actuator}. The position of the actuator is controlled by a magnetic encoder comprising a magnetic strip and an integrated Hall sensor. The system is connected and operated via a wireless Zigbee network.  

\subsubsection{Pointing and PSF system}
\label{sec:point}

The telescope pointing system uses astrometry of stars projected onto the telescope's lid observed by a CCD camera installed in the centre of the dish. The field of view of the CCD camera covers the entire lid and the pointing LEDs for the telescope structure bending monitoring. The second CCD camera mounted next to the pointing CCD provides an image of the telescope PSF and is used during the mirror alignment. As the focal plane of the telescope is located behind the lid inside the Cherenkov camera, a dedicated PSF screen was designed with 45$^\circ$ mirrors and a screen located at the centre of the LID. The screen is located at a focal distance outside the camera and the CCD camera monitors the PSF using the $45^\circ$ mirror.

\subsection{Telescope control system}
\label{subsec:telcontrol_design}

\subsubsection{The control software}

To control and monitor all telescope subsystems, onsite or remotely, a web-based engineering graphical user interface (GUI) has been developed. 
The GUI enables multiple operators or subsystem experts to connect simultaneously, command and monitor the telescope hardware and software. 
During data acquisition using ZeroMQ\footnote{https://github.com/zeromq/libzmq}, a monitoring stream from Zfitswriter \cite{CTASST-1MProject:2017atf} continuously displays randomly selected events in the GUI. During a shift, the operators can receive real-time alerts for inclement weather via the GUI, which also prompts the telescope master to initiate parking procedures. 

\begin{figure}
	\centering
		\includegraphics[width=0.7\textwidth]{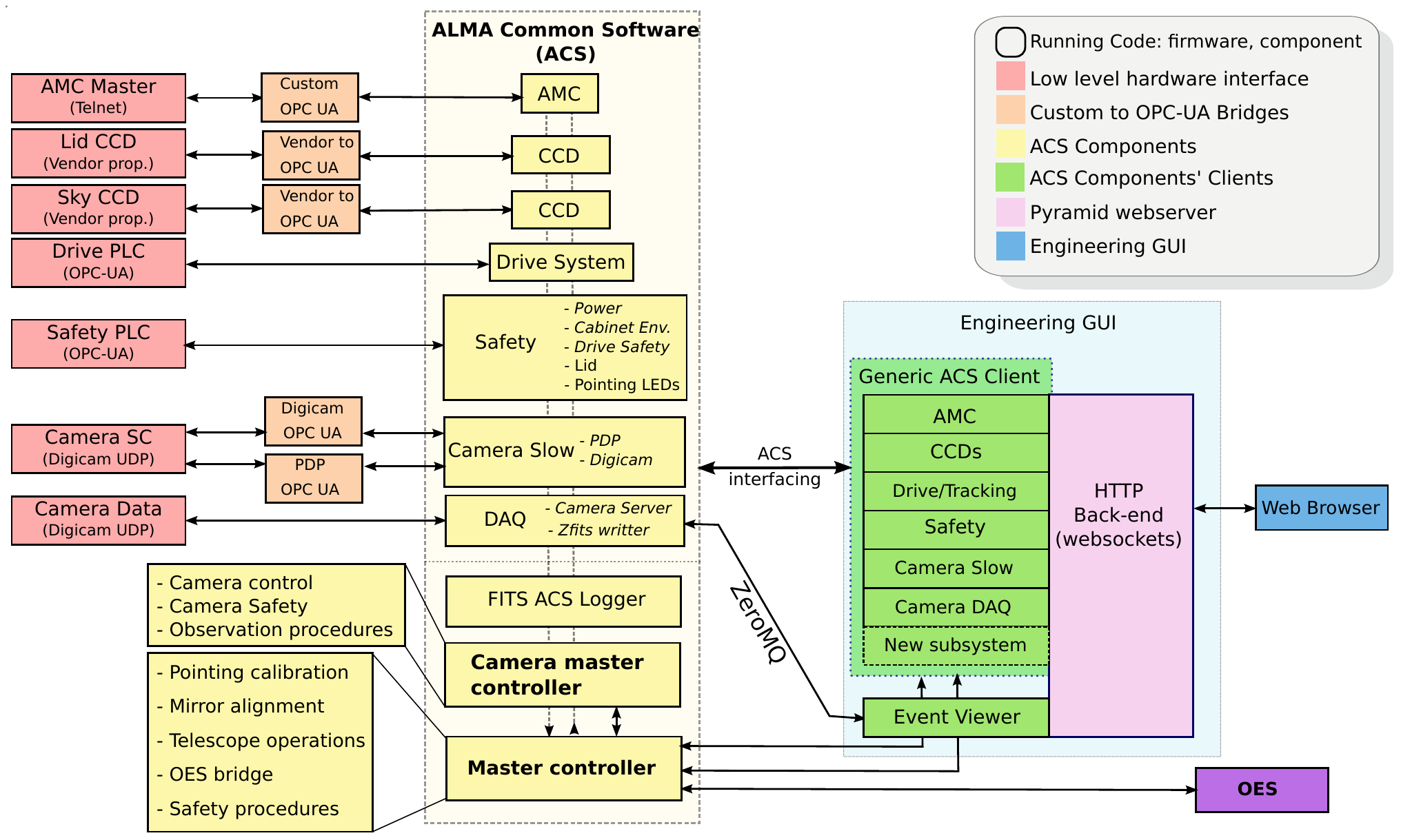}
	\caption{SST-1M control software diagram.}
	\label{fig:concept}
\end{figure}

The control software concept diagram is shown in Fig.~\ref{fig:concept}.
The SST-1M control software is built around the ALMA Common Software (ACS) infrastructure, originally developed for the Atacama Large Millimeter Array. Each subsystem within ACS is represented by a separate ACS component. 
The SST-1M integrates the following subsystems: active mirror control, drive system, safety system, photo-detection plane, CCD cameras, \dcam{} control, data acquisition system (including the \dcam{} readout done in the camera server and the raw data writer with Zfitswriter), logs writer, telescope master, and camera master. 
Observations are performed in fully automatic schedule-driven mode. Operators create the schedule during the day, and the telescope master software executes all the required actions, such as telescope un-parking and parking, movement, tracking, pointing to an astrophysical source and moving to another, and camera control through the camera master. The camera master conducts dark runs, trigger rate scans, data acquisition control, and safety checks.

\begin{figure}
	\centering
		\includegraphics[width=0.7\textwidth]{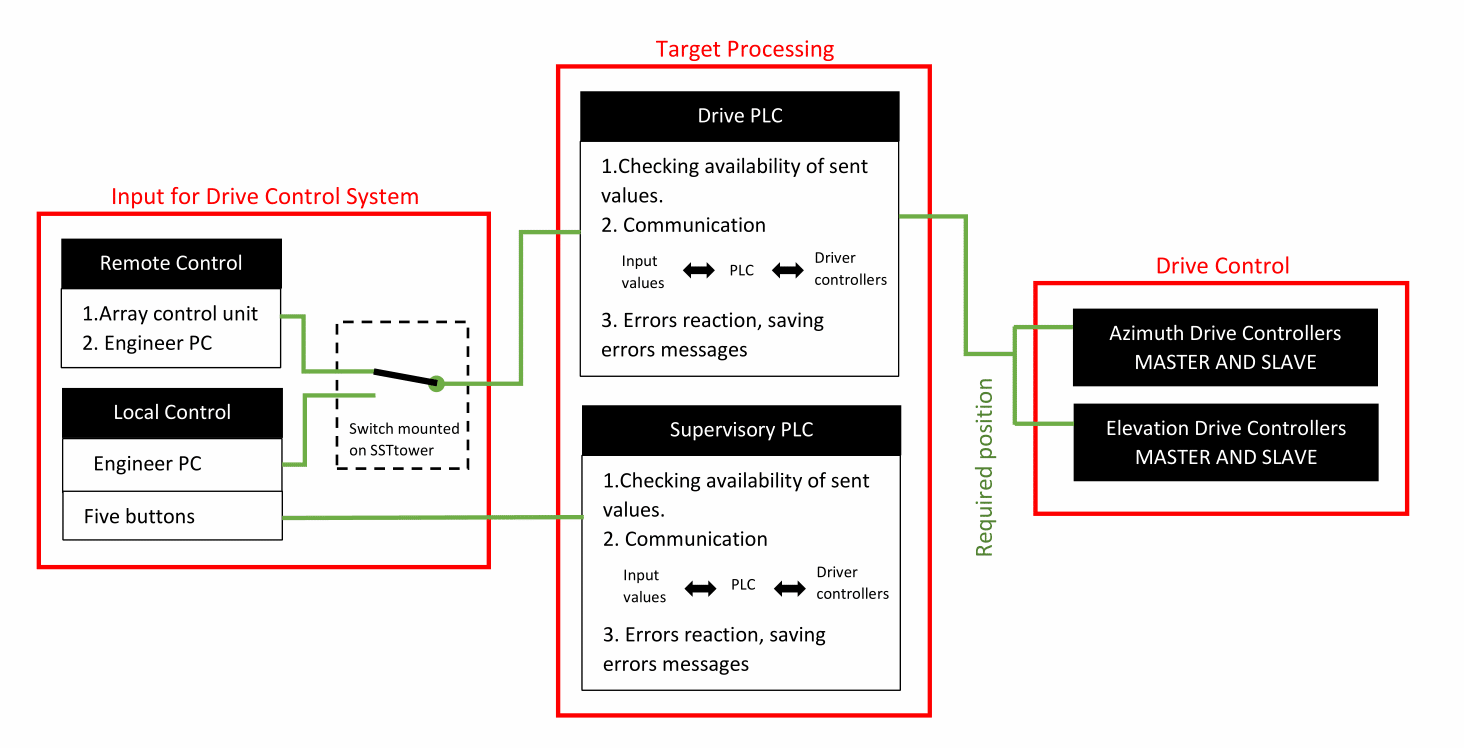}
	\caption{Simplified control structure in normal operation.}
	\label{fig:general control}
\end{figure}

\begin{figure}
	\centering
		\includegraphics[width=0.7\textwidth]{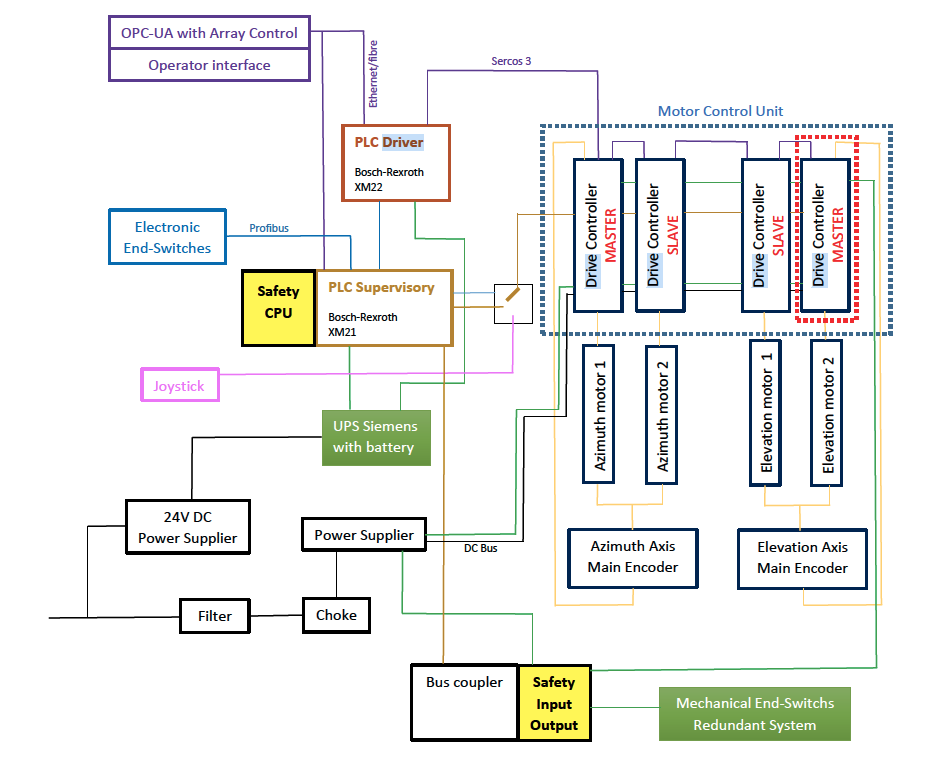}
	\caption{Simplified diagram of the SST-1M drive control system.}
	\label{fig:drive control}
\end{figure}

\subsubsection{Drive control and safety system}

The drive control and safety system of the SST-1M is similar to that used in the MST. It is managed by two Bosch-Rexroth Programmable Logic Controllers\footnote{https://www.boschrexroth.com/en/dc/} (PLCs): the Drive PLC oversees the movement of the drive system, while the Supervisory PLC monitors the system's condition and coordinates all operations. The Supervisory PLC takes on a critical role in ensuring safety, especially during emergencies such as strong winds, loss of communication with the Observation Execution System (OES), or power failures. It is equipped with a Safety CPU module to enhance the safety functionality and manage emergency procedures. Additionally, the Supervisory PLC monitors the camera's environmental conditions through continuous data from sensors, even when the camera is inactive. Additionally, it undertakes corrective actions as activation of heaters when moisture levels approaches the dew point to prevent condensation. It also oversees the start up and shut down procedures of the camera, ensuring optimal conditions before the electronic systems are powered.

The general structure of the control system is shown in Fig.~\ref{fig:general control}. 
The control system contains three main elements: the input, target processing, and drive control sections. 
Inputs for drive control, such as new target positions and/or velocities, can be delivered in three ways: locally via a manual control mode using switches and buttons in the operator panel (joystick) connected to the PLC, remotely through the array control software in remote control mode, or via an engineer PC in local mode. Due to safety considerations, switching between these control modes requires a dedicated hardware switch on the control panel. The PLC processes all target data to check their validity before transmitting them as reference position values to the individual drive controls. 

The telescope can move within its operating zone as shown in Fig.~\ref{fig:drive_coordinates}. At the start/end of observations, it moves from/to a parking position at $-14^{\circ}$ in elevation to/from the start position at $0^{\circ}$ in elevation and azimuth. This start position is the centre of the normal operating zone, where all telescope movements are allowed. 

Outside this zone, namely in the end zone, movement limits enforce an automatic stop, allowing movement only back into the allowed zone.
Three safety levels are implemented in the end zone. The first level is the `software limit' in the PLC driver, allowing for movements within $\pm270^\circ$ azimuth and $0^\circ$ to $91^\circ$ elevation. If a new target position is outside this range, a warning is generated, and the telescope stops. The second level uses electronic safety sensors as limit switches, providing diagnostic feedback to the control code. The third level involves mechanical limit switches, which disconnect the power supply if the telescope moves in an uncontrolled way. Since the movement range of the azimuth axis is greater than $\pm 270^\circ$, the system also includes a mechanical cross-switch to enable clockwise and anti-clock-wise rotations of the telescope.

\begin{figure*}
  \centering\includegraphics[width=\textwidth]{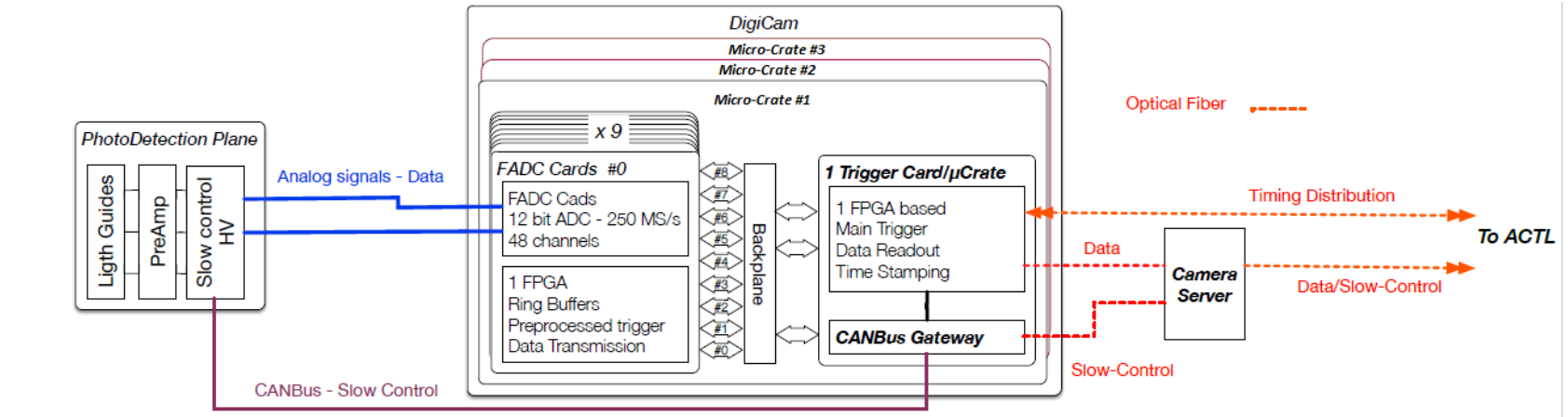}
  \caption{Block diagram of the camera architecture. The content of just one of the 3 micro-crates needed to read out 432 pixel each is shown.}\label{fig:CameraBlockDiagram}
\end{figure*}


A simplified concept of the drive control system for the SST-1M is shown in Fig.~\ref{fig:drive control}. The drive system is powered with a high-voltage 400~V three-phase maximum 64~A current. The main HV Power Supply provides 750~V DC power to the four motion controllers via a DC bus. The controllers convert the DC to 400~V AC supplied directly to the servo motors of the azimuth and elevation drive units. The power source is connected through a filter and a choke system. The main electric power source also supplies a low-voltage 24~V DC power for the slow control components of the drive system, including the main PLC driver. For  redundancy, the 24~V power supply is connected with the PLC driver through a battery module composed of a Siemens UPS and a 24~V DC battery. This ensures power to the PLC driver even if the main 24~V DC source is lost. In such a case, the UPS switches to battery power. Instead of hard wires, DC bus splints are used to connect the HV power supply to the four motion controllers. Thin copper plates connect the power supply and motion controllers on both the high and low-voltage sides. They minimise unnecessary cabling, enhance safety against electrical shocks, and optimise the control cabinet space.

The main HV power supply is equipped with several extra input slots for auxiliary components, including mechanical end-switches. If a signal from any mechanical limit switch is detected, the main HV power supply cuts the power to the motion controllers. The connection between the motor control units and servo motors is provided through two types of shielded cables: one for the motor supply and the other for the encoder built into the motor shaft. These cables are custom-designed for the SST-1M by Bosch Rexroth. The main PLC and the Supervisory PLC are connected to the OPC-UA server of the OES, allowing for remote control. PLCs communicate via Profibus interface, where the Supervisory PLC sends commands, and the main PLC controls the motors via Sercos 3 interface. In the event of a main PLC failure, the Supervisory PLC recognises it and switches the mode using the digital outputs (I/O Control) of the drives. The safety algorithm uses feedback from electronic limit switches.

Each axis of the telescope is driven by two servo motors in a MASTER/SLAVE configuration with external encoders for each axis sending signals to the MASTER controllers. This setup eliminates backlash in bearings and gears, ensuring precise tracking. Independent microprocessors cross-check the encoder signals, allowing movement only if both acknowledge a correct position, enhancing accuracy. The synchronous operation, with torque differences fed to a PID controller, allows precise SLAVE motor control. For high acceleration, both drives can supply torque in the same direction, reducing costs for motors, gears, and converters. Built-in safety-on-board functionality suspends movement commands and initiates a movement to the park position if discrepancies are detected, ensuring reliability and safety.

\subsection{The \dcam{} SiPM-based camera design}
\label{sec:camera_design}

The camera adopts an ``horizontal'' architecture similar to the MST camera based on photomultipliers and named FlashCam \cite{flashcam}. Its building blocks are: the photo-detection plane (PDP), the readout electronics (\dcam{}) with its trigger system, 
and the data acquisition system in the camera server (DAQ) (see Fig.~\ref{fig:CameraBlockDiagram}). 
\dcam{} is a further developed version of FlashCam to achieve a compact and lightweight design with the SiPM technology instead of PMTs adopted by the MST camera. The design of the digital readout and trigger systems (\dcam) foresees compact and high performance electronics, which at the same time has to be as standard as possible to be easily mass-producible. 
As a comparison, if FlashCam was used for the SST-1M, 9 mini-crates would have been needed with a total dimension of 1.2 $\times$ 0.65 $\times$ 0.3~m$^3$ not fitting the camera chassis. In addition, it would have implied 324 cables routing analogue signals along the telescope mast to reach the electronics, which would have travelled over 12~m long cables. This implied a significant risk of signal degradation (cross-talk, ElectroMagnetic Compatibility or EMC, etc...). 

The lower trigger rate and lower number of camera pixels expected for the SST-1M compared to the MST, due to the larger diameter (about 4~m for the SST-1M and 12~m for the MST) reduce the required total throughput of the camera data. More channels can be served by a single board for the SST-1M making its design more compact than FlashCam.
Moreover, the reduced number of boards also simplifies the interconnections between the micro-crates that are needed to transfer the data for triggering across the system. From the original 192 cables needed for triggering, only three Infiniband cables of 10~Gb/s are needed for SST-1M.
Additionally, more powerful FPGAs have been selected in order to provide further flexibility to the trigger algorithms.

Besides these advantages, improving the reliability, maintainability,  operational stability, and physics performance were among the main design prerequisites of \dcam{}.
The approach used for \dcam{} of having fully digital electronics is also very innovative in the gamma astronomy field.
The signals routed via analogue cables from the PDP are digitised in \dcam{}, by FADC converters and stored in digital ring buffers for their subsequent elaboration.
The temporarily stored data is used by the FPGAs of the system to produce trigger decisions and send them to the data processing and archiving.
This approach is highly flexible as the trigger logic, the amount and the type of transferred information are completely defined by the FPGA logic that can evolve over time.
Moreover, the size of the DDR3 SDRAM memory (SODIMM) on the trigger cards, used for event buffering, can be changed according to the requirements of different trigger algorithms.

\begin{figure*}
  \centering
  \includegraphics[width=\textwidth]{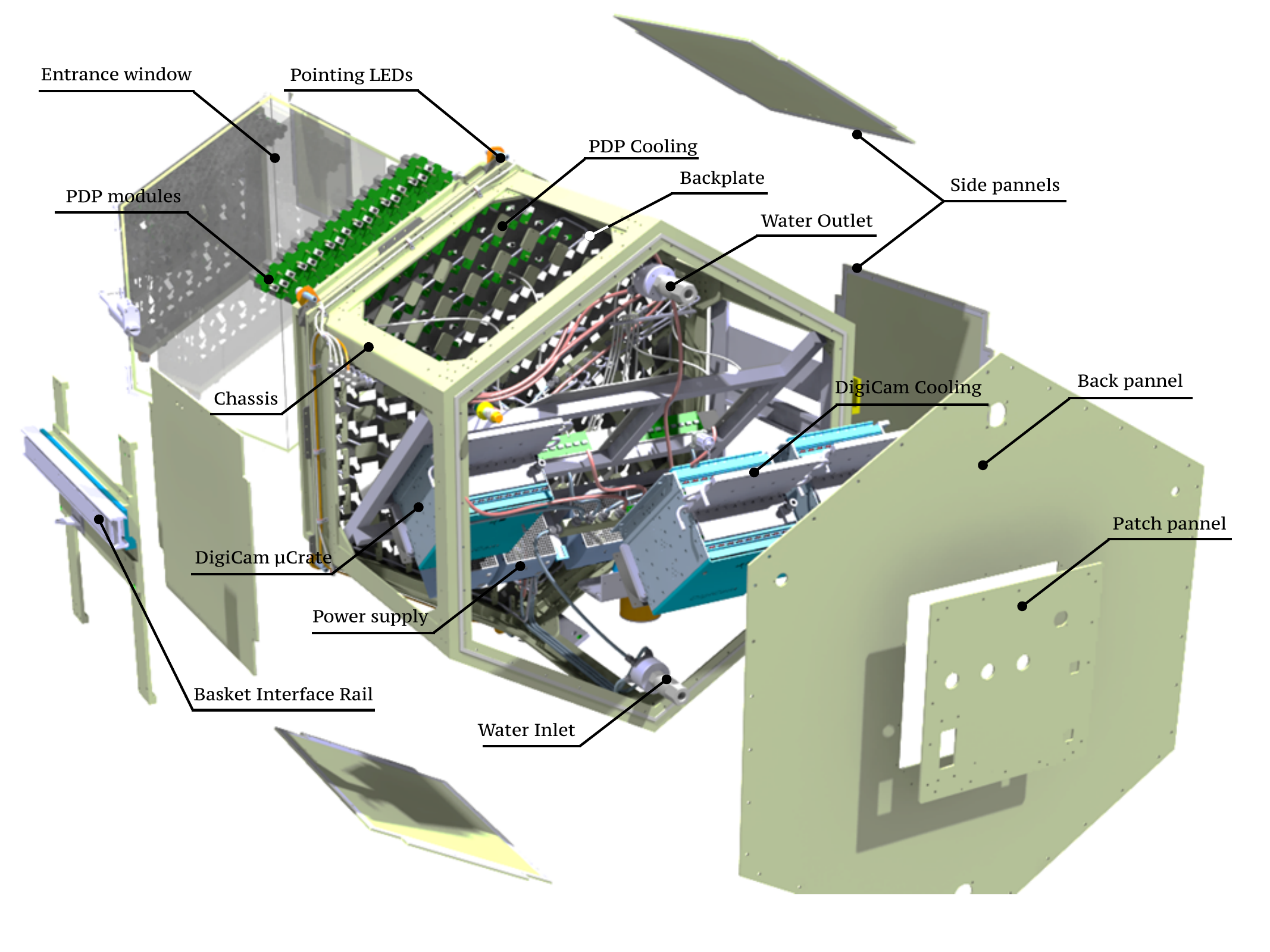}
  \caption{Exploded view of the first camera prototype. The key element of the structure discussed in this section are indicated.}
  \label{fig:CameraView}
\end{figure*}

The camera (see its picture on the field in Fig.~\ref{fig:camera} and its exploded CAD view in Fig.~\ref{fig:CameraView}) is conceived in a hexagonal chassis and its total weight is less than 200~kg, with camera structure including entrance window weighting 64~kg, readout micro-crates 41~kg (including power supplies and cables) and the PDP of about 80~kg in total.
The PDP is equipped with 1296 pixels organised in 108 modules. Each hexagonal pixel has an angular size of 0.24$^\circ$, which translates into a camera FoV of about 9$^\circ$. The \dcam{} readout is distributed in three micro-crates hosting the functional boards. Each micro-crate collects the data from a specific sector of the PDP, realising three independent sectors of the camera. 
The camera components are shown in Fig.~\ref{fig:CameraView} for the  prototype camera (the second has some mechanical differences).
The PDP, the camera mechanical structure, including the shutter, its assembly, housekeeping, and calibration boards are coordinated and developed by the University of Geneva. The \dcam{} digital signal acquisition and readout systems are developed by the team of the Jagiellonian University with contributions from the AGH University of Science and Technology, Krakow. 

\subsubsection{The photo-detection plane}
\label{sec_PDP}

The PDP is composed of several innovative elements that are illustrated in Fig.~\ref{fig:ModuleView} from the back of the camera to the front.

\begin{figure}
\includegraphics[width=\textwidth]{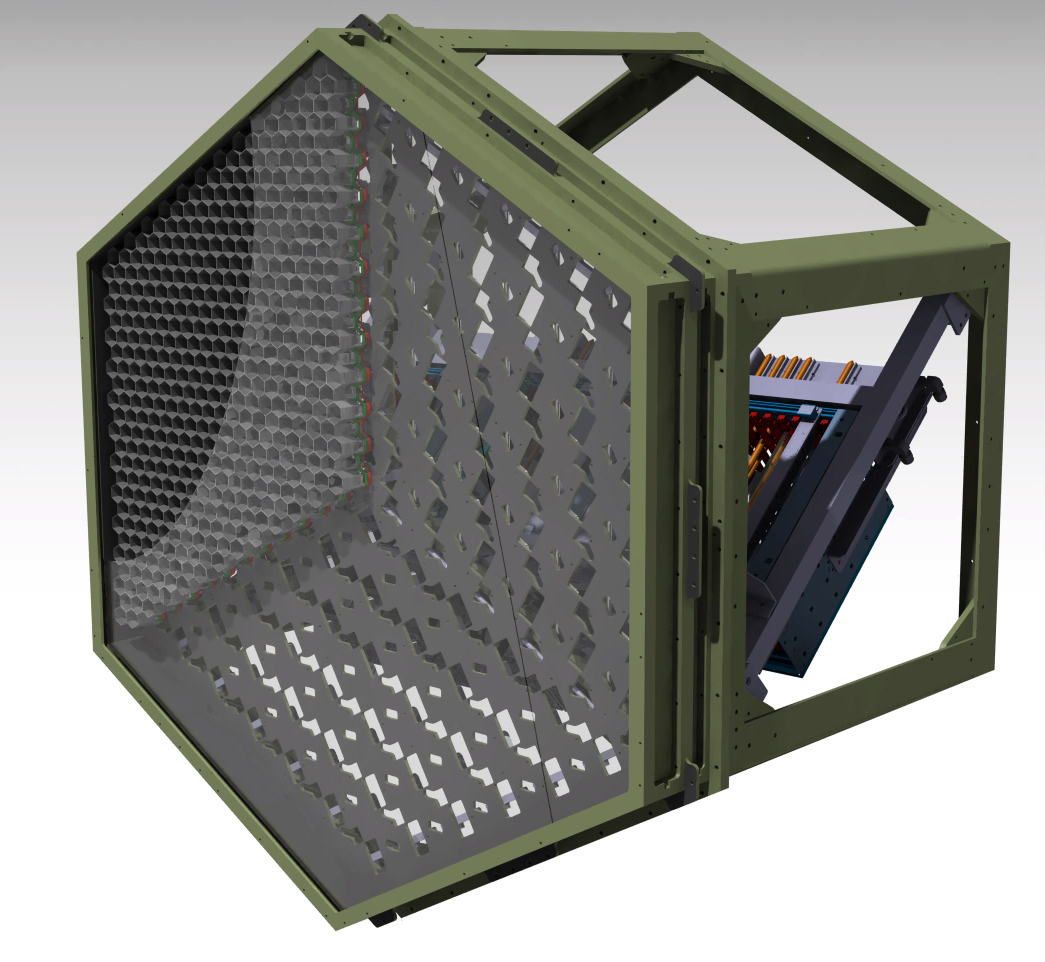}
\caption{Camera 3D CAD Model. Front view with one sector with a module installed and the front window}  
\label{fig:CameraView_front}
\end{figure}

\begin{figure}
\centering
\includegraphics[width=0.5\textwidth]{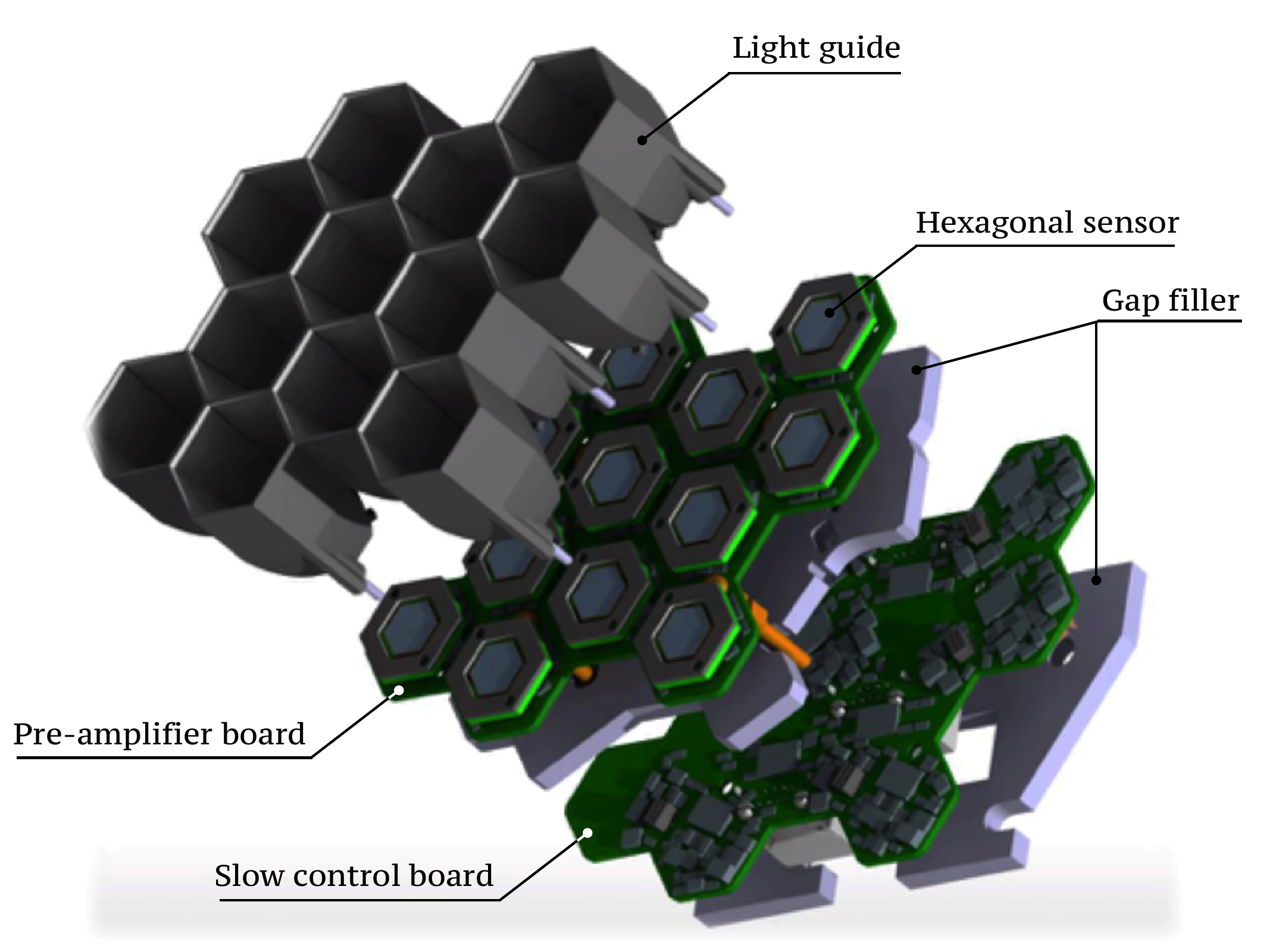}
\includegraphics[width=0.45\textwidth]{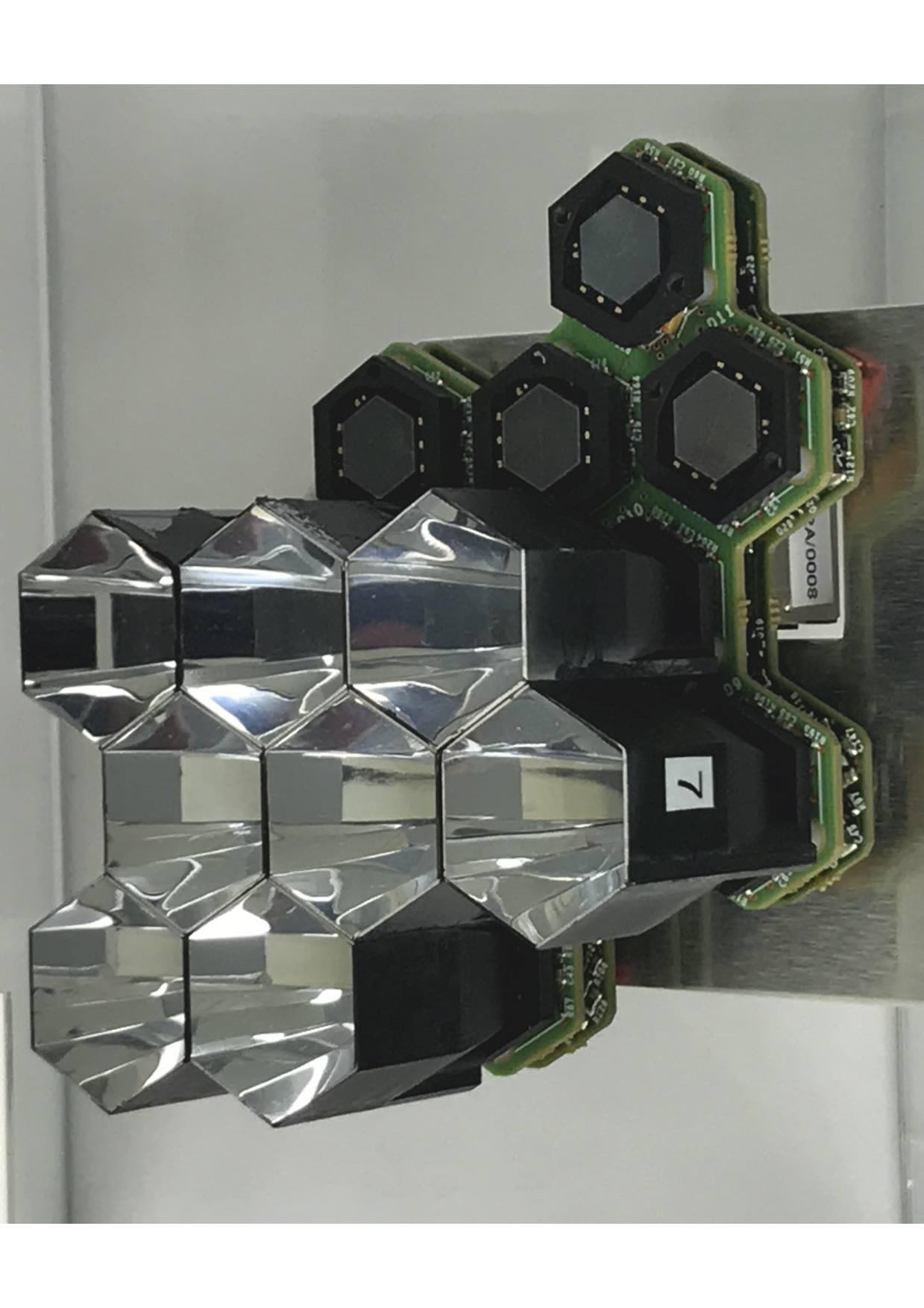}
\caption{Blow-up CAD of a PDP module (top) and its picture partially assembled (bottom). There are 108 modules in the PDP each weighting 300~g.}
\label{fig:ModuleView}
\end{figure}

\paragraph{The entrance window:}

The aperture of the Cherenkov camera is composed of a 3~mm thick Borofloat window on which a narrow band optical filter of dielectric layers and an anti-reflective coating are deposited \cite{CameraPaperHeller2017}. The window is visible in Fig.~\ref{fig:CameraView_front} reflecting sunlight at dusk in the yellow. While it allows for air shower-induced Cherenkov light transmission to the pixel sensors, it has a dielectric multi-layer coating reflecting wavelengths beyond 550~nm. The spatial uniformity of the window transmittance over the large area of the window of $\simeq 0.81~{\rm m^2}$ is a challenge to be achieved by coating companies. For the two developed windows, the coating is deposited while the about 1~m-large window rotates in the coating chamber around its centre and, therefore, the coating profile has radial symmetry and lowest transmission at the holder position in the centre. The obtained non-uniformity along the 6 diagonals of the cameras from the centre to the frame was $\lesssim 2$\%.
The camera window is effective in reducing the NSB in each pixel to about 40~MHz considering all wavelengths from $200$~nm to $1000$~nm after passage through the filter. 
With an automatic test setup of the window transmission we measured the light transmission efficiency which is shown in Fig. 11 (right) in \cite{Alispach:2020}. The overall reduction of the NSB thanks to the filter is a factor of 3.75.

\paragraph{The photosensors and funnels:}
\label{sec:des_sensors}

SiPMs have become the preferred photosensors for many applications in high-energy particle and astroparticle physics, due to their low input voltage, lightweight, good time resolution, ability to work in the presence of continuous light, and high PDE. Consequently, they are a good choice for IACT cameras, as the pioneering work of FACT demonstrated (see Sec.~\ref{sec:features}). 
To achieve the desired performance with the chosen optics, the telescope camera is composed of 1296 pixels, each of 2.32~cm linear size and angular opening of $\sim0.24^{\circ}$, providing in total about $9^{\circ}$ FoV. More details on the camera, its design and performance can be found in \cite{CameraPaperHeller2017}.

\begin{figure}
\centering
\includegraphics[width=0.45\textwidth]{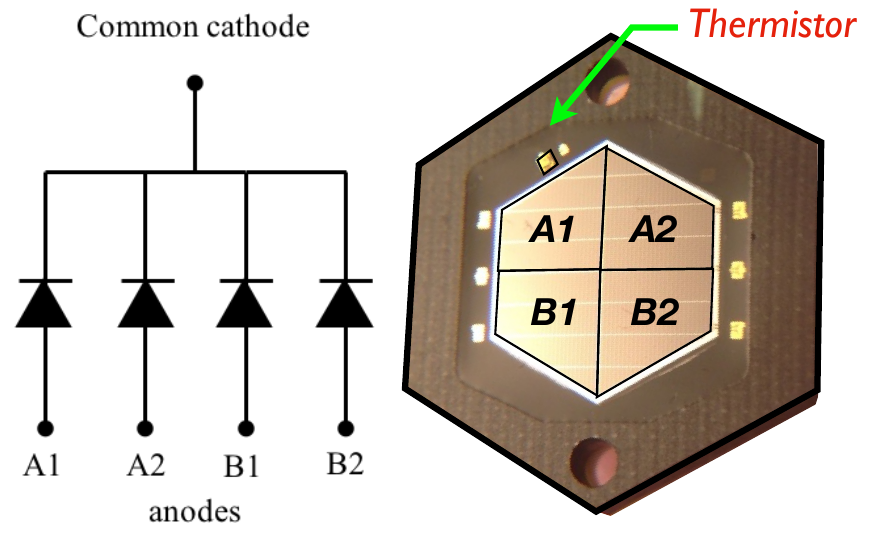}
\includegraphics[width=0.49\textwidth]{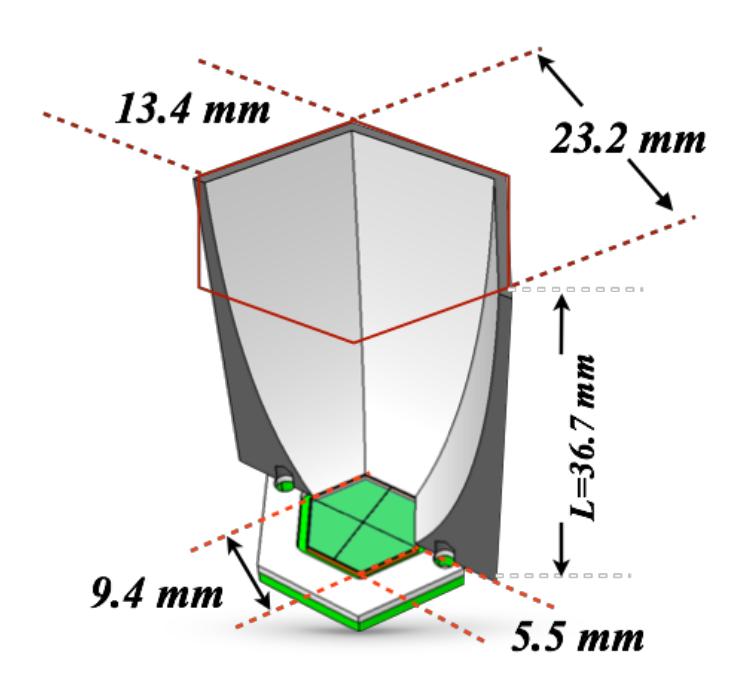}
\caption{Hamamatsu S10943-2832(X) picture and its electric equivalent model (left) and drawing of a pixel with SiPM and lightguide (right).}  
\label{fig:HexSiPM}
\end{figure}

To have a spatial uniform response of the camera, the pixels should have a circular shape to ensure the same distance between their centres in every direction. Nonetheless, a circular shape would introduce large dead spaces, and so the selected pixels are hexagonal. A hexagonal SiPM device, named S10943-2832(X), has been designed in collaboration with Hamamatsu and is shown in Fig.~\ref{fig:HexSiPM}. 
This device is based on the so-called low cross-talk technology LCT2, which was available when the camera design was done. Currently, the next generation of devices is on market such as LCT5 (S13360) and LVR (S13360). The LCT5 offers better performance than LCT2 in terms of PDE and cross talk. The sensor area is around 93.6~mm$^2$ with linear size of 9.3~mm. The lightguide design, shown on the right of Fig.~\ref{fig:HexSiPM}, is also described in \cite{Aguilar:WinstonCones_2014}, where their reflectance is also reported on. 
The sensor capacitance is directly related to its active area and this has an impact on the signal decay time and consequently on the read-out electronics noise. 

The sensor has a common cathode, but a tight requirement of 0.2~mV on the spread of over-voltage between the four channels ensure a small gain disparity across the sensor surface.
The signal is readout on four independent anodes, grouped by two and amplified before being summed, as shown in Fig.~\ref{fig:HexSiPM}~(left). 

 On each sensor package, an NTC thermistor is present, which is used to monitor the temperature variations affecting the operational parameters of the sensor, such as the dark count rate (DCR) but also all parameters depending on the value of the V$_{breakdown}$, namely the gain and PDE, for which a real-time correction can be applied to keep the working point stable \cite{SST1Melectronics}.

\paragraph{The front-end electronics}
\label{sec:des_fee}

The front-end boards read out the sensor and amplify the signals, which are input to the \dcam{} FADCs. It is named pre-amplifier board (PAB) and its design and performance are detailed in \cite{SST1Melectronics}.
From each sensor, the sum of the four channels is passed to \dcam{} via differential signals. 

The design of the pre-amplifier chain was based on the following specifications, aiming at achieving high performance even in the presence of moonlight:
\begin{itemize}\itemsep0.em
    \item The SiPMs are  DC coupled to monitor the current and correct for the NSB change effects, unlike other SiPM-based IACT cameras, such as the FACT \cite{FACT} and the ASTRI ones \cite{10.1117/12.2232464}. The DC coupling also improves the ability to determine the signal developing on top of the NSB level. The DC level is measured in the \dcam{} FPGA so that it can be monitored in real time and the trigger threshold can be dynamically adapted to variable levels of NSB.
    \item The dynamic range should be 1 to 2000 p.e. per sensor. 
    \item An amplification of 3-4~mV per p.e has to be ensured to make the camera immune to noise and cope with the limited input swing of the \dcam{} ADC  of $\sim 1.4$~V).
    \item The electronic noise has to be lower than the baseline of 0.3~mV RMS. This corresponds to the LSB value of the 12-bit ADC (1400~mV/4096).
    \item The signal rise time $\tau_{rise}\approx 2~$ns and the fall time $\tau_{fall}\approx 30~$ns. This requirement complies with the expected rate of photons of hundredths of MHz per pixel in the case of moonlight.
    \item The signal-to-noise ratio (SNR) has to be higher than 20~dB.
 \end{itemize}

The adopted solution is a trans-impedance amplifier for its high bandwidth.
Given the small size of the 12-pixel-module PCB and other constraints in terms of power and cost, it is not possible to have an amplifier for each of the 4 channels of the sensors.
The solution is summing two channel signals at the input of one low noise amplifier, so that two low noise amplifiers are needed per pixel.
The sum of the output signals is made in a differential stage which guarantees a good single photo-electron resolution with the LCT2.
In order to achieve the required dynamic range, the pre-amplifying stage saturates at around 600 p.e. Beyond this number of detected photons, the response is not linear, the amplitude saturates while the pulse integral continues to grow. A transfer function has been determined to recover from the pulse integral the number of photo-electrons.

The PAB is connected to the Slow Control Board (SCB) via six connectors conveying the power to the pre-amplifying stage, the pixel signals, the bias voltage, the temperature sensors and the ground.
The SCB is mounted in the shadow of the PAB. The board has been developed at the University of Geneva and it is by far more complex than the PAB because it has to provide more functionalities: 1) ship the signal to the \dcam{}; 2) read and write independently the bias voltage of the 12 sensors; 3) read the temperature of the NTC encapsulated in the sensor; 4) adjust the bias voltage as a function of the temperature. The first point (1) does not require too much electronics but is demanding only in terms of routing.
The remaining three points are related to the correct set of  working points.
The gain of the sensor depends on the applied over-voltage $V_{Ov}=V_{HV}-V_{breakdown}$ and then the $V_{HV}$ needs to be corrected for the $V_{breakdown}$ variation with the temperature. 
The board has a microcontroller on-board which reads the temperature of each NTC and corrects the $V_{HV}$ to compensate the temperature variation according to a look-up table. Warning flags are set whenever the difference between the read and the set temperatures exceeds a defined threshold. The same is done when the set value is out of the allowed bias voltage range. The bias voltage can be enabled or disabled for all channels independently.
The microcontroller can be also used to read via CANBus the temperature for monitoring purposes. The microcontroller is the core of the board and is also used to store information about the board (Board ID, calibration constant etc.) which can be retrieved and cross-checked with what is stored in the production database. 
The firmware can be reloaded or updated remotely via CANBus to be capable of reprogramming all the boards of all telescopes simply and efficiently in case this is needed (bug/ failure, etc.).
The SCB functionalities have been tested and validated with electronic tests and temperature variation tests in a climate chamber. 

\subsubsection{The \dcam{} readout electronics}

\begin{figure}
\includegraphics[width=\textwidth]{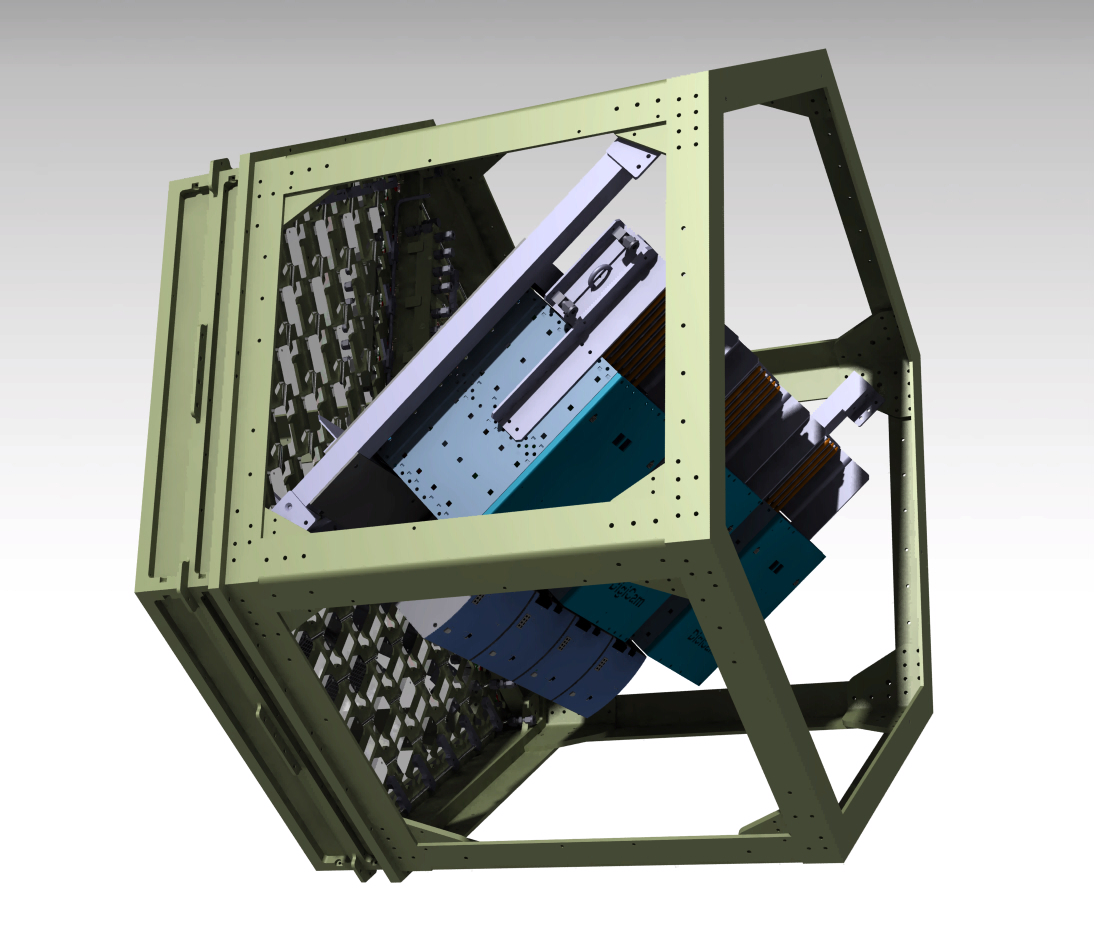}
\caption{Camera 3D CAD Model. Side view showing the micro-crate  and the power supplies below.}  
\label{fig:CameraView_side}
\end{figure}

The digital subsystem of the camera comprises three micro-crates visible on Fig.~\ref{fig:CameraView_side}, each managing a third of the PDP area. Each crate includes nine ADC boards and one trigger board, processing signals from 432 pixels through standard CAT5/6 cables. The ADC boards digitise the PDP analog signals, store them in digital ring buffers, and calculate the first-level trigger (L0) signals that are sent to the trigger board. 
The ADC boards use high-speed A/D converters and an FPGA for efficient data handling. 
The chosen converters, such as the AD9239 from Analogue Devices, offer high-speed data transfer and low power consumption. The FPGA processes and stores data in buffers, allowing for extensive continuous waveform recording. 
Data transfer to the trigger board is facilitated by a 1~Gb/s link, supporting the transmission of up to 20'000 events/s.

The L0 trigger relies on triplets of 3-neighbour pixels for triggering, and the trigger board processes these signals to generate the L1 trigger signal. 
The trigger decision is distributed across the entire digital system. 
The trigger board also gathers event data from ADC boards and sends them to the central acquisition system via a master trigger card. Equipped with an FPGA and DDR3 memory, the trigger board handles high-speed data processing and synchronisation tasks. It ensures synchronised trigger signals across all crates and manages the data flow to the central acquisition system.

The camera internal connections include a custom-designed backplane within each crate and high-throughput cables connecting all crates. The backplane connects ADC boards to the trigger board, while the cables facilitate data exchange and synchronisation between crates. 
To enhance reliability and manageability, the system includes an FPGA reconfiguration management solution, allowing for multiple configuration revisions stored in Flash memory. 
The camera uses the White Rabbit solution for clock synchronisation and time reference, ensuring all components work in unison. The master trigger card synchronises the local clock with the White Rabbit network and manages time synchronisation across all cards.
In Sec.~\ref{sec:val_cam} we discuss relevant measurements to validate the camera performance and in Sec.~\ref{sec:ondrejov} some aspects of camera operation.

\subsubsection{The mechanics and cooling}

The mechanics of the chassis of the prototype has been designed at first to minimise production cost with patch panels. As visible in Fig.~\ref{fig:CameraView}, the chassis is built out of several aluminium parts assembled using nuts and bolts. Six side panels and one patch panel mounted on the back plate provide access to the core component of the camera for assembly and maintenance activities.
Despite having sealed the all panels and nuts, the mechanics was not fully water tight when operating at IFJ-PAN first testing site. Currently, it is operated using an watertight cover.
The chassis of the second camera has been entirely redesigned to avoid, as much as possible, water and dust ingress; the frame of the chassis is machined in a single block of aluminium while being assembled; the patch panels and the six side panels are equipped with O-rings; there are no through-holes.
Now the chassis satisfies the requirement of IP65 to stand hard weather conditions and watering on the camera. 

\begin{figure}
    \centering
    \includegraphics[width=\textwidth]{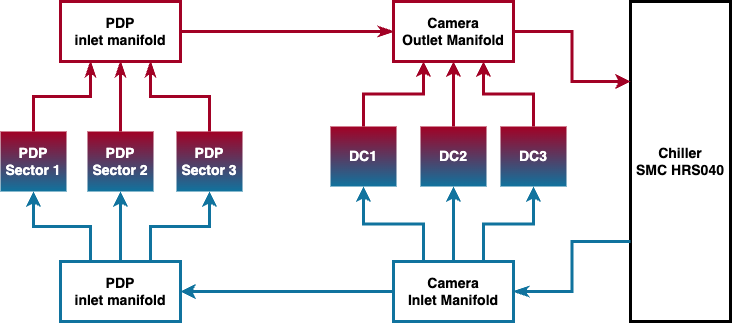}
    \caption{SST-1M camera cooling flow chart.}
    \label{fig:CamCool}
\end{figure}

The camera cooling is divided into a part serving the PDP and a part for the readout electronics. Its flow chart is shown in Fig.~\ref{fig:CamCool} and its conception and validation have been described in \cite{CameraPaperHeller2017}. 
Each channel of the PDP requires a power of 0.38~W for a total of 500~W in operation. 
The PDP conduction cooling is designed to maintain its temperature between $15-20^\circ$C. The cooling system principle is based on conduction for the PDP. Both the PAB and SCB design are optimised to better in-plane conduction by increasing the copper layer thickness.  As visible in Fig.~\ref{fig:ModuleView}, to improve the out-of-plane conductivity the PAB, the SCB and the back plate are thermally connected using a thermal gap filler. The cooling of the PDP is achieved by having cold water circulating in small aluminium pipes distributed across the back plate as shown in Fig.~\ref{fig:CameraView}. The pipes are thermally connected to the back plate via small cooling blocks. These blocks are also used to mounted the four brass nuts which ensure good thermal contact between the modules and the back plate.
Because of the power density in the \dcam{} micro crate, both conduction and convection are used. Each board is equipped with an aluminium plate which makes thermal contact with the components having the highest heat dissipation (FPGA, FADCs, power stage). In the first camera prototype, the heat from the plate is extracted thanks to removable heat pipes that connect the aluminium plate to the cold block through which the water is running. The efficiency of the heat pipes is maximum when they are installed vertically. In order to ensure that independently of the pointing direction similar performance would be obtained, The crates, and therefore the heat pipes, are there oriented at 45$^{\circ}$ which ensure maximum efficiency for a zenith angle of 45$^{\circ}$. Due the small size of the camera, the heat pipes cannot be removed without extracting the crate from the chassis. This has been identified as a issue for potential corrective maintenance activities and addressed with the second camera design as described in Sec.~\ref{sec:ondrejov}.
For both camera prototypes, fans in push configuration are mounted at the bottom of the crates.

\subsubsection{The auxiliary systems}
The environmental conditions inside the camera together with a few other monitoring and controls are realised via the so-called housekeeping board.
The board can trigger an alarm to the PLC and also send the probes reading to the PLC which can then react in case of problems.
Furthermore, a few probes in critical points will be also directly connected to the PLC which gives to this system the required redundancy, especially during the daytime when the system is off and unattended.
The communication with the PLC is through an RS485 connection, which requires only 2 pairs to send the data from the different probes.
The housekeeping board reads three digital three temperature and humidity sensors from Sensirion and four additional PT100s. 
It also reads the status of the four power supplies connected to the PDP and \dcam{} is monitored.
 The six pointing LEDs mounted on the camera frame (see Fig.~\ref{fig:CameraView}) are driven by the housekeeping board as well. They can be turned on or off if the pointing calibration is performed.

The camera shutter is driven by a dedicated board which features its own state machine. Three current limits are set to manage its proper functioning, from lowest to highest: moving, closing, and over-current.
To ensure the shutter door opening or closing, the first current limit has to be overcome by the current. In order to close properly the doors, a higher force needs to be applied, hence the ``closing'' limit is slightly higher than the moving one. This ensures that the sealing joint is compressed enough for shutter water tightness. 
The last limit protects anyone from being injured by the closing or opening of the shutter. It has been set such that if a hand would be in between the door and the window, it would stop.



\section{The telescope design validation}
\label{sec:design_valid}

\subsection{The Finite Element Analysis of the telescope structure}
\label{sec:fem}

As the SST-1M was initially conceived as a solution for the SSTs of CTAO, the structural requirements were defined based on the conditions at the Paranal site in Chile. 
The SST-1M primary structure robustness has been assessed using the ANSYS code which is based on the Finite Element Method (FEM). The initial analysis focused on the dynamic characteristics and load capability of the primary structure, concluding that the telescope structure is robust enough to withstand wind speeds of 200~km/h and seismic activities typical of the Chilean site. 
These results were augmented with additional FEM calculations addressing issues not covered in the first evaluation. 
The main results of these FEM modelling will be summarised below. 

\paragraph{Pre-stress of the mast rods: \label{sec:prestress}}
To ensure the required rigidity of the telescope mast, pre-tensioned steel rods are installed in the mast frame. Finite Element (FE) model calculations of the mast's deformations due to gravity load show that the maximum displacement at the camera focal plane decreases from about 59~mm without the rods to only 13~mm with the rods installed. This result pertains to the mast's displacement relative to the global reference frame at the telescope base and is not sensitive to the exact pre-stress value. However, because steel rods cannot take compressive stresses, they must be pre-stressed to ensure they remain under tension under any load. A pre-stress of about 5100~N (101~MPa) is sufficient, resulting in slight deformations of the mast, with link elements elongated by about 1.5~mm and other elements compressed by about 0.4~mm. This pre-stress ensures that stresses in the link elements under gravity load range from 55~MPa to 147~MPa, justifying the need for pre-tensioning. The 5100~N pre-stress provides a safety margin for combined loads such as gravity and wind.

\paragraph{Modal analysis: \label{sec:modal}}
Assuming that the steel rods are
pre-stressed with 5100~N, the modal analysis of the telescope was performed. The calculated values of the five lowest frequencies are presented in Tab.~\ref{tab:eigenmodes} for the telescope in the horizontal position and for three other elevation angles. It can be noted that the eigenfrequencies are always well above the 2.5~Hz lower limit and weakly depend on the elevation angle. The shape of the modes for the three lowest frequencies is related to the camera movement in the vertical plane, the camera movement in the horizontal plane, and the mast rotation around its main axis, respectively.

\begin{table}
\centering
\begin{tabular}{lcccc}
\hline
\hline
&\multicolumn{4}{c}{Elevation  angle}\\ \hline
Modes & {0$^\circ$} & {25$^\circ$} & {45$^\circ$} & {90$^\circ$}\\
\hline 
I [Hz]   & 3.48 & 3.47 & 3.46  & 3.47    \\ 
II [Hz]  & 3.80 & 3.82 & 3.86  & 3.97    \\ 
III [Hz] & 7.50 & 8.47 & 9.44 & 7.41  \\ 
IV [Hz] & 10.20 & 10.70 & 9.81 & 10.50  \\ 
V [Hz] & 12.10 & 11.10 & 11.10 & 11.70  \\ 
\hline
\end{tabular}  
\caption{Five lowest normal frequencies of the telescope frame for different elevation angles.}
\label{tab:eigenmodes}
\end{table}

\paragraph{Forces and moments in the elevation and azimuth drives and bearings:\label{par:forces}}
For the seismic analysis, the response spectrum for Chile/Armazones with PGA=0.49~g with 2\% damping for horizontal and vertical directions was applied.
Calculations were performed for the telescope at four elevation angles: $-13^\circ$, $0^\circ$, $45^\circ$, and $90^\circ$. At the parking position ($-13^\circ$), the telescope is docked at a single locking point. 
The results show that maximum forces in the drives and bearings, under both gravity and seismic loads, are 186.6~kN and 100.4~kN for the azimuth and elevation drives, and 96.3~kN and 91.5~kN for the azimuth and elevation bearings, respectively. 
The forces are significantly lower than the allowable limits of 2184~kN for the drives and 1250~kN for the bearings. 
The maximum \textit{axial} moments are 86.6~kNm and 68.2~kNm at the azimuth and elevation drives, respectively, below the
allowable axial torque of 94.8~kNm.
The maximum \textit{radial} moments are 112.5~kNm and 49.4~kNm at the azimuth and elevation drives, respectively. While the allowable \textit{radial} torque is not specified for the drives used in SST-1M, it is estimated to be no less than 540~kNm. Radial moments in the bearings are small and read 28.1~kNm and 41.8~kNm, for the azimuth and elevation bearings, respectively. 

Displacements and the equivalent stresses from seismic loads were also calculated.
The maximum displacement is about 42~mm at the camera and the maximum equivalent stress is 165~MPa. 

\begin{figure}
	\centering
	\includegraphics[width=0.8\textwidth]{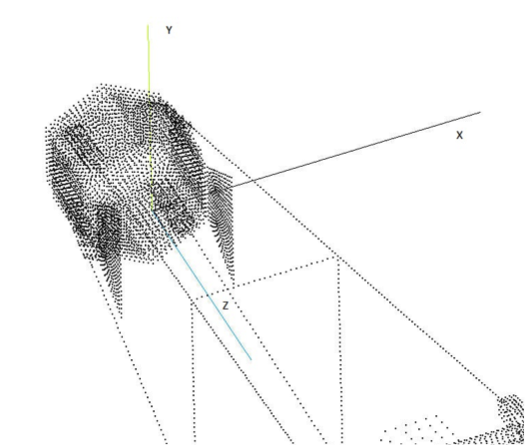}
\caption{Local coordinate system used for calculations of relative displacements and rotations between the camera and the mirror dish.}
\label{fig:local}  
\end{figure}

\paragraph{Relative displacements of the camera and mirrors: \label{par:displace}}
The optical performance of the telescope degrades under gravity, wind, and thermal loads, primarily depending on the telescope's stiffness. To quantify this degradation, calculations of telescope displacements were performed under these loads in a local coordinate system (Fig.~\ref{fig:local}). The $x$-axis and $y$-axis lie in the focal plane of the camera, with the $x$-axis always horizontal (for all elevation angles) and the $z$-axis originating at the mirror dish center. Displacements and rotations for each mirror were measured at three mirror support points. Calculations were made for mast elevation angles from $25^\circ$ to $95^\circ$,
focusing on several parameters: 
mirror displacement from the camera centre in the $xy$-plane (UW1),
change in distance between a mirror and the camera along the $z$-axis (UWZ, a change in focal length for a given mirror), 
mirror rotation around the $x$-axis (RotXm), mirror rotation around the $y$-axis (RotYm), and position change of the light ray in the camera focal plane caused by mirror displacements and rotations combined (UW).

Tab.~\ref{tab:optgrav} shows the values of these parameters values, indicating a degradation of the optical performance due to the \textit{gravity load} at selected elevation angles within the telescope observation range. The extreme values from the three points on each mirror facet are presented. 

\begin{table}
\centering
\begin{tabular}{lccccc}
\hline
\hline
&\multicolumn{5}{c}{Elevation angle} \\
 & {$35^\circ$} &  {$45^\circ$} & {$65^\circ$} & {$75^\circ$} & {$95^\circ$} \\
\hline
UW1[mm]         &  9.9 & 9.3  & 8.0 & 7.3 & 5.7 \\
UWZ [mm]       &  3.3 & 3.3  & 3.0 & 2.8 & 2.3  \\
RotX$_m$ [mrad] & -2.2 & -2.2 & -1.7 & -1.6 & -1.2\\
RotY$_m$ [mrad] & -0.2 & -0.2 & -0.2 & -0.2 & -0.2 \\
UW [mm]         &  3.3 &  3.3  & 2.5 & 2.1 & 1.8  \\
\hline
\end{tabular} 
\caption{Optical performance degradation of the telescope under the gravity load.}
\label{tab:optgrav}   
\end{table}

The contribution of wind loads to the optical performance degradation is presented in Tab.~\ref{tab:optwind36}. Here, we assumed a mean wind velocity of 36~km/h with a gust factor of 1.7 and additional drag coefficients applied. Additional calculations were conducted for a mean wind velocity of 100~km/h, under the same assumptions, Tab.~\ref{tab:optwind100}. The latter modelling aimed to assess the rate of the telescope PSF degradation with increasing wind speed.

\begin{table}
\centering
\begin{tabular}{lrrrrr}
\hline
\hline
&\multicolumn{5}{c}{Elevation angle}\\
 & $35^\circ$ & $45^\circ$ & $65^\circ$ & $75^\circ$  & $95^\circ$ \\
\hline
UW1 [mm]        &  1.0 & 1.0  & 1.0 & 0.9 & 0.7 \\
UWZ [mm]        &  0.3 & 0.3  & 0.3 & 0.3 & 0.2 \\
RotX$_m$ [mrad] &  0.2 & 0.2  & 0.2 & 0.2 & 0.1 \\
RotY$_m$ [mrad] & -0.1 & -0.1 & -0.1 & -0.1 & -0.1 \\
UW [mm]         &  0.6 & 0.6  & 0.6 & 0.3 & 0.2 \\
\hline
\end{tabular} 
\caption{Optical performance degradation of the telescope under the wind load in observation mode conditions with a mean wind velocity of 36~km/h.}
\label{tab:optwind36}   
\end{table}

\begin{table}
\centering
\begin{tabular}{lrrrrr}
\hline
\hline
 &\multicolumn{5}{c}{Elevation angle}\\
&{$35^\circ$} & {$45^\circ$} & {$65^\circ$} & {$75^\circ$}  & {$95^\circ$} \\
\hline
UW1 [mm]        &  8.1 & 8.6  & 7.9 & 7.1 & 5.7 \\
UWZ [mm]        &  2.6 & 2.7  & 2.6 & 2.0 & 1.5  \\
RotX$_m$ [mrad] & 1.9 & 1.9 & 1.7 & 1.3 & 0.8 \\
RotY$_m$ [mrad] & -1.0 & -0.9 & -0.8 & -0.7 & -0.6 \\
UW [mm]         &  4.7 &  4.6  & 3.3 & 2.4 & 1.8 \\
\hline
\end{tabular} 
\caption{Optical performance degradation of the telescope under the wind load in emergency mode conditions with a mean wind velocity of~100 km/h.}
\label{tab:optwind100}   
\end{table}

According to telescope requirements, the operational air temperature range spans from -15\degc{} to 25\degc{}, implying similar temperature changes in the telescope structure. With summertime temperatures approaching 30$^\circ$C, the total temperature difference during observations can reach $\Delta T=45^\circ$C. The optical performance degradation under this thermal load is presented in Tab.~\ref{tab:optherm}.
Note that the analysed scenario represents an extreme case and  
temperature changes occur slower than 7.5\degc/hour (as specified for CTAO sites). Such changes are expected to result in uniform temperature variations across the structure rather than strong gradients. Therefore, contributions from temperature changes are negligible.

\begin{table}
\centering
\begin{tabular}{lc}
\hline
\hline
 & Elevation angle from $25^\circ$ to $90^\circ$ \\
\hline
UW1 [mm]         &  1.1 \\
UWZ [mm]       &  3.1 \\
RotX$_m$ [mrad] & 0.1 \\
RotY$_m$ [mrad] & 0.07  \\
UW [mm]         &  0.7  \\ 
\hline
\end{tabular} 
\caption{The optical performance degradation of the telescope under thermal load of $\Delta T = - 45^\circ$C.}
\label{tab:optherm}   
\end{table}

In summary, under combined gravity, wind loads of $V=36$~km/h, and thermal load, the total  relative deflection of the camera with respect to the dish does not exceed 5~mm, thus is well within 1/3 of the physical pixel size limit (8.1~mm).
The optical performance shows only slight degradation with increasing wind speed. At an average wind velocity of 100~km/h, camera deflections are not larger than 9~mm, indicating about 10\% degradation of the PSF.
Thus, it can be assumed that the optical performance of the SST-1M would meet specifications even under wind conditions with mean velocities up to 100~km/h.

\paragraph{Deformation and stress analysis for the telescope in the parking position with one support point:} The telescope in the parking position is locked with one actuator and only side movements of the mast are fully blocked 
and displacements along the main axis of the mast are allowed.
The connection can thus accommodate the thermal, wind, and other stresses without incurring damage to the telescope. To verify this, a dedicated FEM calculation was performed, analysing different load cases (i.e., seismic, gravity, temperature gradients, snow, ice, and wind) separately and in combination.

Displacements and stresses for seismic loads alone have been already discussed above. 
The maximum displacement is about 10~mm, and the maximum equivalent stress is not higher than 127~MPa.
For the combined load of gravity, temperature ($\pm$45\degc{} from the no-stress state), wind (\textit{survival} condition with a mean wind velocity of 200~km/h and a gust factor of 1.7), snow and ice, the maximum displacement is about 18~mm, and the maximum equivalent stress is about 92~MPa. The equivalent stress limit is about 235~MPa. This indicates that the combined load case analysed would not cause plastic deformations to the structure. When combining these loads with seismic load, an unlikely real-world scenario, the maximum displacement is 28 mm, and the maximum equivalent stress is 220~MPa, still below the plastic deformation limit. In this extreme case, forces in the drives are not larger than 202~kN, and forces in the bearings are not greater than 136~kN. The axial moments reach maximum values of 64.7~kNm and 91.0~kNm at the azimuth and elevation drives, respectively. All forces and moments remain well below their limits. 
Based on these results, we can reiterate our earlier conclusions that both the drives and bearings in the telescope can carry forces and torques generated by winds and earthquakes. 

\paragraph{Deformations of the telescope caused by slewed accelerations:}
Assume a scenario where the telescope is slewing with the maximum allowable angular acceleration of 0.036~rad/s$^2$ and is suddenly halted. In this case, we can estimate the dynamic force exerted on the mast. 
Additionally, given the two lowest natural frequencies of the telescope, which correspond to horizontal and vertical movements of the camera, we can estimate the dynamic displacements of the camera in these directions due to the acceleration during the repositioning of the telescope. These estimates indicate that the maximum camera displacement caused by the slewing acceleration is below 1.2~mm.


\subsection{Validation of the optical design}
\label{sec:optval}
\subsubsection{The mirror alignment procedure}
\begin{figure*}
  \centering
  \includegraphics[width=\textwidth]{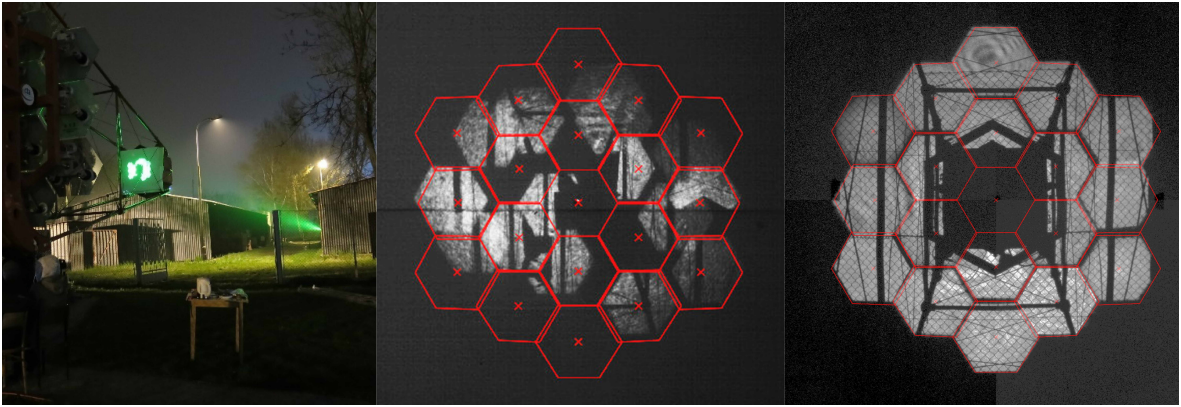}
  \caption{Setup for mirror alignment using the Bokeh method at the SST-1M prototype (left). The point-like laser source is placed at a distance of about 30~m, illuminating the screen placed in front of the telescope camera. The image of the close-distance point source is simulated and the individual mirror facet point source images are moved to a given position on the screen. The middle and right pictures show the facet images before and after alignment, respectively.}
  \label{fig:bokeh_setup}
\end{figure*}

The alignment of the mirror facets is practically done using the pre-alignment and the fine alignment methods. As a pre-alignment procedure, the FACT Bokeh Mirror Orientation Method \cite{FACT} is used after mirror installation or replacement of facets, Fig.~\ref{fig:bokeh_setup}. 
A light source is located at a far distance (tens of meters) on the optical axis of the telescope and illuminates the telescope mirrors. 
The image of the source is observed on a PSF screen covering the entire lid of the telescope camera. It consists of the combination of 18 separate images of individual mirrors. The source positions on the screen are simulated and compared with the measurements. 
The alignment procedure involves gradually adjusting the positions of individual mirror images to their ideal calculated positions. The advantage of this procedure is that it allows for the alignment of the mirrors under sunset conditions. 
The Bokeh method itself does not fulfil the D$_{80}$ specification, thus the fine alignment method needs to be used.

In the fine alignment step, a dedicated fixed PSF screen mounted on the telescope camera lid is used to optimise the PSF using a bright star in the centre of the FOV of the telescope. 
A PSF CCD camera installed in the centre of the dish acquires the images and serves for the Bokeh alignment and the PSF monitor. An automated procedure moves the individual mirror facets and overlaps the images of the star of a mirror facet to the optical axis of the telescope to minimise the D$_{80}$. 
The PSF can be periodically monitored by closing the lid and pointing to a bright star. 

\subsubsection{The optical point spread function of the telescope}
\label{sec:val_optics}
The telescope is equipped with two CCD cameras (PSF and pointing camera) located in the centre of the telescope dish oriented towards the Cherenkov camera. 
The lid of the Cherenkov camera is covered with a plastic flat screen serving as a pointing screen and a small PSF screen for PSF monitoring. 
The PSF camera observes the image of a bright star on the PSF screen during the telescope tracking and serves as the PSF control system. 

\begin{figure}
  \centering
\includegraphics[width=\textwidth]{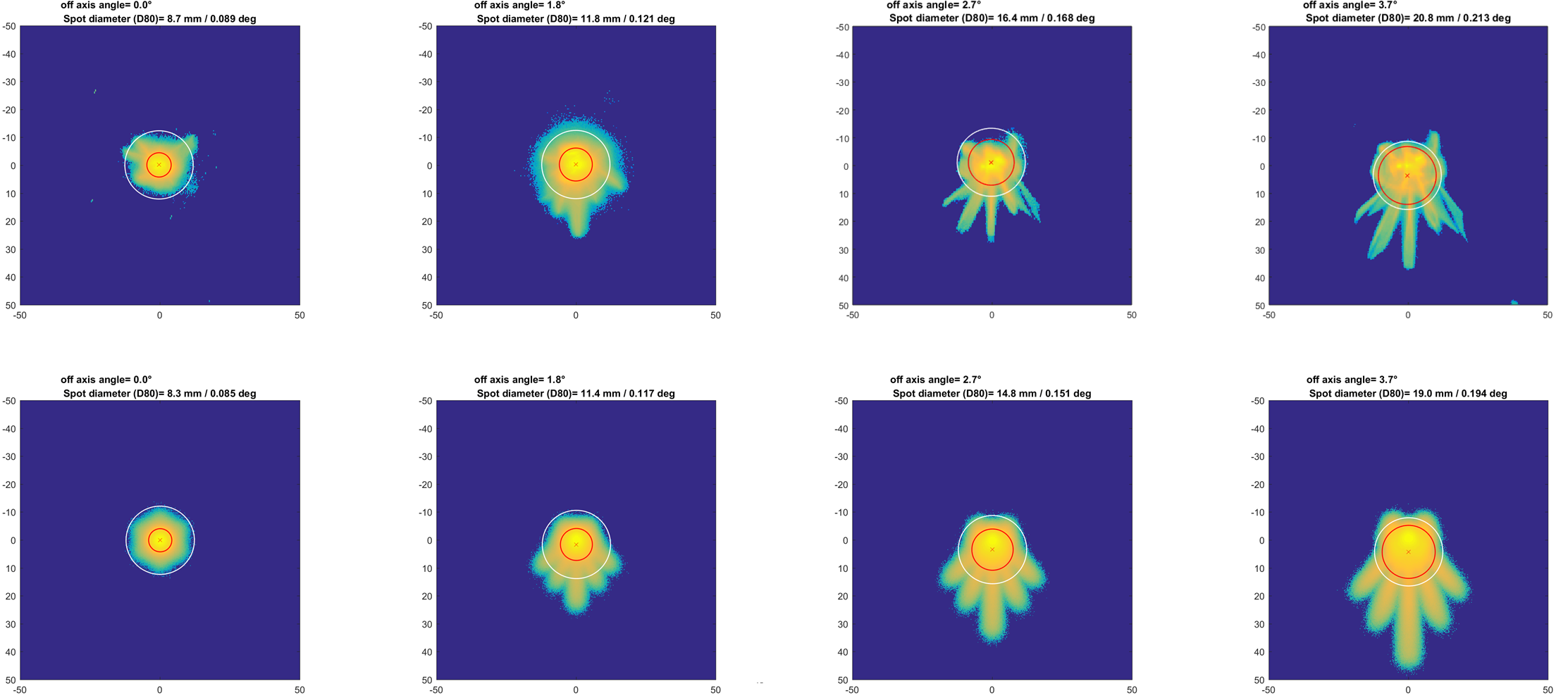}
\caption{
Measured (top) and simulated (bottom) PSF in the camera plane for different off-set angles from the optical axis. From left to right on-axis and off-set angles$= 1.8^\circ, 2.7^\circ, 3.7^\circ$. The z axis is in an arbitrary log-scale. The simulation includes real reflectivity and mirror surface shape.} 
\label{fig:PSF}
\end{figure}

The PSF represents the optical quality of the telescope. We define the PSF as the angular diameter of a circle containing 80\% of the photons emitted by a point source at infinity, denoted by D$_{80}$. The centre of the circle is chosen to be the centre of gravity of the photon distribution. Given that the spot profile at the focal plane is not necessarily symmetrical, we define D$_{80}$ by starting from the centre of gravity of the photon distribution and integrating the signal in larger and larger circles until 80\% of the total emission is reached. For the camera field of view of 9$^\circ$, the PSF needs to be determined for light rays coming from a source located at $\sqrt{0.8} \times 4.5^\circ$, i.e. at 4$^\circ$ off-axis. The PSF simulation for the various field of view of the telescope using two independent simulation tools is shown in Fig.~\ref{fig:PSF1D}. The simulation is compared with the real measured PSF of the telescope in Sec.~\ref{sec:design_principles} and it is shown in 2 dimensions in Fig.~\ref{fig:PSF}.

\subsubsection{Pointing precision of the telescope}

\begin{figure}
	\centering
		\includegraphics[width=\textwidth]{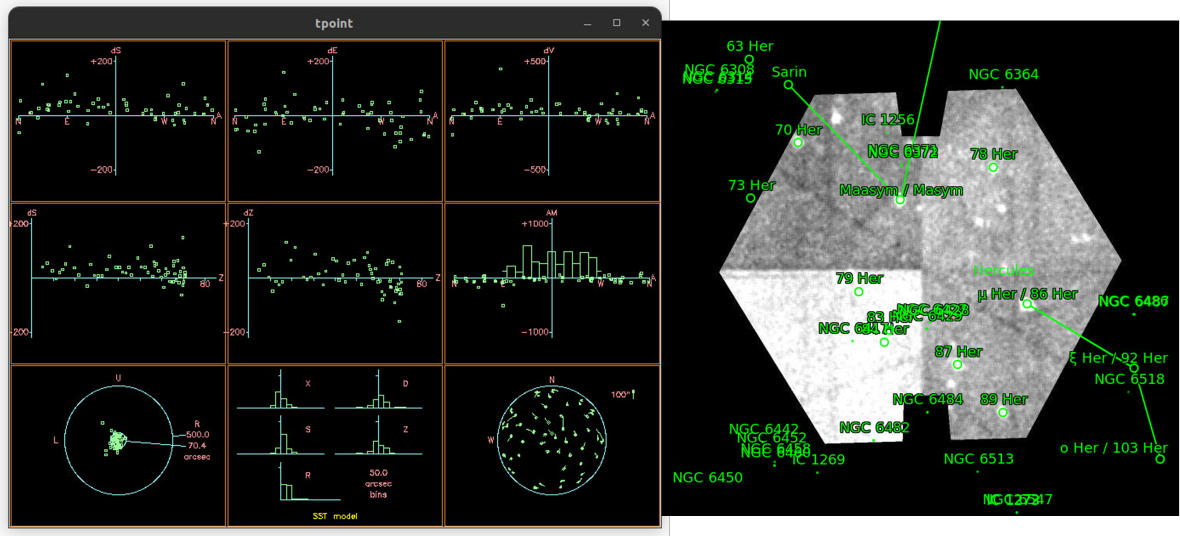}
	\caption{SST-1M TPoint result of the pointing scans. The left image shows the residuals of the pointing after implementation of the corrections. The right image shows the star field of the pointing image and the analysed star positions using astrometry.}
	\label{fig:pointing}
\end{figure}

The pointing system is based on the analysis of the star field reflected by the mirrors on the lid screen. The image is analysed by the TheSky\textsuperscript{TM} TPoint\textsuperscript{TM} software \cite{Tpoint}.
Dedicated pointing runs are performed that involve changing the pointing direction of the telescope and capturing star field images on the lid screen for different azimuth/elevation points.
These images are analysed using the  analysis scripts and TPoint software, Fig.~\ref{fig:pointing}. The resulting pointing corrections are then implemented to the drive system of the telescope. The Cherenkov camera is equipped with 6 LEDs located at the edge of the camera body. 
Once the star-filed image is acquired, the LEDs are turned on and the same pointing camera acquires the image. The star field is projected onto a special surface in front of the telescope's camera.
The actual direction of the telescope is determined by astrometry, where the centre of the telescope camera, determined by the LEDs, is pointing on the Sky field.
The Tpoint program then uses the set coordinates and the real coordinates to determine the polynomial correction factors. This polynomial correction function is implemented in to the drive system. The final pointing accuracy after the corrections is within 70 arcseconds. 

\subsection{Validation of mechanical design}

\subsubsection{Prototype structure tests}
Several measurements and tests of the first telescope prototype have been performed. The methods used and the main results of these tests are summarised below.

\paragraph{Geodesic measurements:}
The geodesic measurements were conducted by external specialised teams from AGH University of Science and Technology in Krakow.
They aimed to verify the telescope structure parameters against specifications concerning error aggregation on the focal length, relative deformation of the camera and mirror dish, relative orientation of the telescope axes, and the reproducibility of mechanical deformations. 

Structure measurements were performed using a network of 8 electronic tachymeters and rangefinders, surveying 20 symmetrical points on the telescope structure. Some of these points were in parts of the structure expected to experience minimal deformation, serving as references for determining static deformations. 
The mean accuracy of the measurements was $\pm0.5$~mm, with meaningful deformations considered to be at least three times this value ($\pm1.5$~mm). 
Measurements were taken at five elevation angles ($-13^\circ$, $0^\circ$ (twice), $45^\circ$, and $95^\circ$) for an azimuth angle of $0^\circ$, and seven azimuth positions ($0^\circ$, $\pm90^\circ$, $\pm180^\circ$, $\pm270^\circ$) for an elevation angle of $0^\circ$.

Results show that deformations affecting the focal length with elevation changes were negligible, not exceeding 1.3~mm. Relative deformations between the camera and mirror dish reached 2.5~mm, increasing with changes in the elevation angle from $95^\circ$ to the parking position ($-13^\circ$), consistent with FEM analysis estimates. 
No deformations were observed with changes in the azimuth angle, as expected. Mechanical deformations were consistently reproducible at each surveyed position. 
The relative orientation of the azimuth and elevation axes was determined, with the angle between the azimuth and  elevation planes at $90^\circ04'50''$ and the tilt angle of the azimuth axis with respect to the zenith at $0^\circ05'55''$.

\begin{figure}
	\centering
		\includegraphics[width=\textwidth]{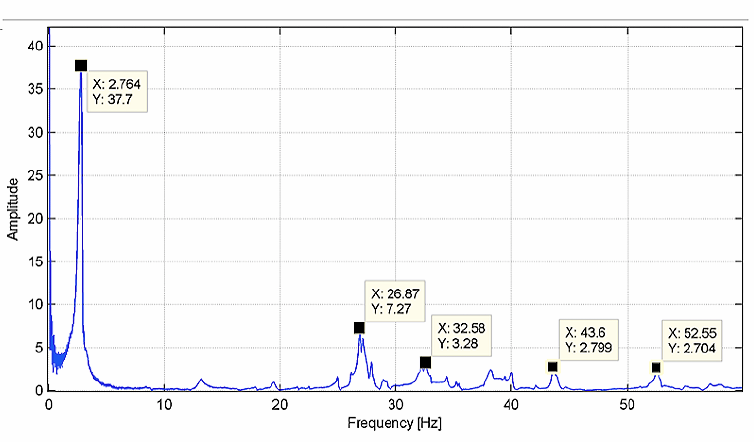}
	\caption{The amplitude spectrum for a single telescope vibration excitation measured with interferometric radar. The telescope was at $135^\circ$ azimuth and $10^\circ$ elevation angle.
}
	\label{fig:modal1}
\end{figure}

\paragraph{Modal analysis with interferometric radar:}
A limited modal analysis of the telescope structure was conducted using interferometric radar. At a telescope position with a $135^\circ$ azimuth angle and $10^\circ$ elevation angle, displacements along the radar axis were measured for all excitations. 
The sampling frequency was 197.75 Hz, providing a spectral resolution of \( \Delta f = 0.012 \) Hz. The identified modes of the telescope were observed at frequencies 2.76~Hz, 26.87~Hz, 32.58~Hz, and higher (see Fig.~\ref{fig:modal1}), showing minimal dependence on the type and strength of excitation. 
The lowest measured frequency ranged from 2.68~Hz to 2.80~Hz, consistent with full spectral analysis using accelerometers (see below) and FEM calculations (see Sec.~\ref{sec:fem}). The geodesic method did not detect frequencies predicted by our FEM model above the lowest eigenfrequency and below the second peak at 26.87 Hz.

\paragraph{Full modal analysis:}
To fully identify the spectrum of natural modes of the telescope structure, additional modal experiments were conducted using piezoelectric accelerometers by a specialised team from AGH in Krakow. 
Two accelerometers were used -- one at the bottom of the dish support structure and another on the back of the camera box. The structure was excited with a PCB 086D20 impact hammer at various points on the frame and different directions. Points of excitation were approximately the same across different elevation angles ($-10^\circ$ -- close to parking position, but telescope not locked in docking station, $-13^\circ$ -- telescope locked, $45^\circ$ and $90^\circ$), except near the camera box at $45^\circ$ and $90^\circ$ angles. Nineteen modal experiments were performed, varying the excitation points (dish support, mast, camera box, counterweight) and directions ($x$, $y$, $z$).
Some measurements were conducted with the drives activated (typical during observation) and others with the drives turned off and brakes engaged (as in the parking position). The reference frame was centred at the base of the telescope tower, with the $x$-axis aligned along the telescope mast and the $y$- and $z$-axes oriented vertically.
The analysis was based on spectral transfer functions and coherence functions.
The spectral transfer function provides information about the natural frequencies excited during an experiment, while the coherence function indicates the correlation between the excitation and the response, ranging from 0 to 1. Results are considered valid when the coherence function values exceed 0.8. Frequencies analysed ranged from 1.9~Hz to 32~Hz.

\begin{table}
\centering
\begin{tabular}{lcccccc}
\hline
\hline
exc. point & DS & M & M & C & C & C \\
\hline
direction & $z$ & $z$ & $y$ & $z$ & $x$ & $z$ \\
\hline
&\multicolumn{6}{c}{Elevation angle $-10^\circ$}\\ \hline
drive on/off & off & off & off & off & off & on \\
\hline
exp. \# & P1 & P2 & P3 & P4 & P5 & P18 \\
\hline
& \multicolumn{6}{c}{Natural frequencies}\\
\hline 
I [Hz]   & 2.98 & 2.86 &   & 2.86 & 3.00 & 2.74  \\ 
II [Hz]  & 3.47 & 3.42 & 3.46  & 3.39 & 3.46 & 3.35    \\ 
III [Hz] &  &  & 3.72 & & &  \\ 
IV [Hz] &  &  &  &  & & 5.03  \\ 
V [Hz] & 6.61 & 6.59 & 6.59 & 6.63 & 6.61 & 6.55  \\ 
\hline
\end{tabular}  
\caption{Natural frequencies of the telescope frame at $-10^\circ$ elevation angle for modal experiments that differ in the excitation point and direction. "DS", "M", and "C" in the row listing the excitation points stand for the dish support, mast, and the camera box, respectively. For experiments P1-P5 the drives were turned off. Five lowest eigenfrequencies are shown.}
\label{tab:fullmod1}
\end{table}

\begin{table}
\centering
\begin{tabular}{lcccccc}
\hline
\hline
exc. point & DS & DS & CW & CW & M & M \\
\hline
direction & $y$ & $z$ & $y$ & $z$ & $y$ & $z$ \\
\hline
&\multicolumn{6}{c}{Elevation angle $45^\circ$}\\ \hline
drive on/off & on & on & on & on & on & on \\
\hline
exp. \# & P6 & P7 & P8 & P9 & P10 & P11 \\
\hline
& \multicolumn{6}{c}{Natural frequencies}\\
\hline 
I [Hz]   &  & 2.97 & 3.16  & 2.92 &  & 3.06  \\ 
II [Hz]  & 3.46 &  & 3.45  &  & 3.48 &    \\ 
III [Hz] & 8.59 & 8.59 & 8.54 & 8.56 & 8.57 & 8.58 \\ 
IV [Hz] & 8.94 & 9.06 & 8.88 & 9.07 & 9.09 & 9.13  \\ 
V [Hz] & 9.83 & 9.91 & 9.10 & 9.91 & 10.04 &  \\ 
\hline
\end{tabular}  
\caption{Natural frequencies of the telescope frame at $45^\circ$ elevation angle. "DS", "CW", and "M" in the row listing the excitation points stand for the dish support, counterweight, and the mast, respectively. For all experiments the drives were turned on.
Compare Tab.~\ref{tab:fullmod1}.}
\label{tab:fullmod2}
\end{table}

\begin{table}
\centering
\begin{tabular}{lcccccc}
\hline
\hline
exc. point & M & M & DS & DS & CW & CW \\
\hline
direction & $z$ & $y$ & $y$ & $z$ & $y$ & $z$ \\
\hline
&\multicolumn{6}{c}{Elevation angle $90^\circ$}\\ \hline
drive on/off & on & on & on & on & on & on \\
\hline
exp. \# & P12 & P13 & P14 & P15 & P16 & P17 \\
\hline
& \multicolumn{6}{c}{Natural frequencies}\\
\hline 
I [Hz]   &  &  &   & 2.87 &  & 2.61  \\ 
II [Hz]  &  & 3.25 & 3.46  & 3.47 & 3.46 & 3.47    \\ 
III [Hz] &  & 3.82 &  & & &  \\ 
IV [Hz] &  &  & 6.58 &  & 6.68 & 6.79  \\ 
V [Hz] & 7.18 & 7.09 & 7.07 & 7.03 & 7.06 & 7.08  \\ 
\hline
\end{tabular}  
\caption{Natural frequencies of the telescope frame at $90^\circ$ elevation angle. Compare Tab.~\ref{tab:fullmod1}.}
\label{tab:fullmod3}
\end{table}

Tab.~\ref{tab:fullmod1}-\ref{tab:fullmod3} display five lowest natural frequencies for the three elevation angles and 18 measurements in which the telescope was not locked in the docking station. The lowest frequency of 2.61~Hz was detected at the $90^\circ$ elevation angle, whereas typical frequencies for the first mode were approximately 2.8~Hz. Excitation direction and point variations minimally affected these frequencies, though some modes were occasionally not visible under specific excitation conditions. The analysis indicated that frequencies with drives on were approximately 0.2~Hz lower than with brakes engaged. Note, that these modal measurement results are in good agreement with the geodesic measurements. The measured frequencies are somewhat lower than FEM simulated values (see Sec.~\ref{sec:modal}), consistent with the simplifications made in the FEM model of the telescope structure. The $19^{th}$ measurement, with the telescope locked in the docking station and the excitation at the camera box in the $z$-direction, showed increased natural frequencies at 4.03~Hz, 5.39~Hz, 6.68~Hz, 7.85~Hz, and 9.23~Hz, as expected.

\begin{figure*}
	\centering
		\includegraphics[width=0.9\textwidth]{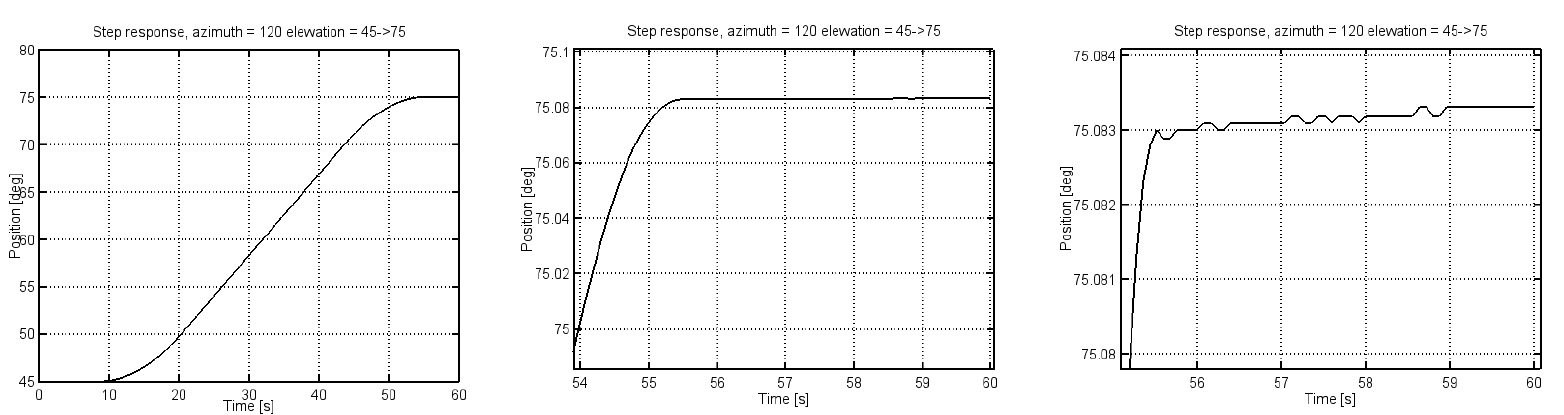}
  \includegraphics[width=0.9\textwidth]{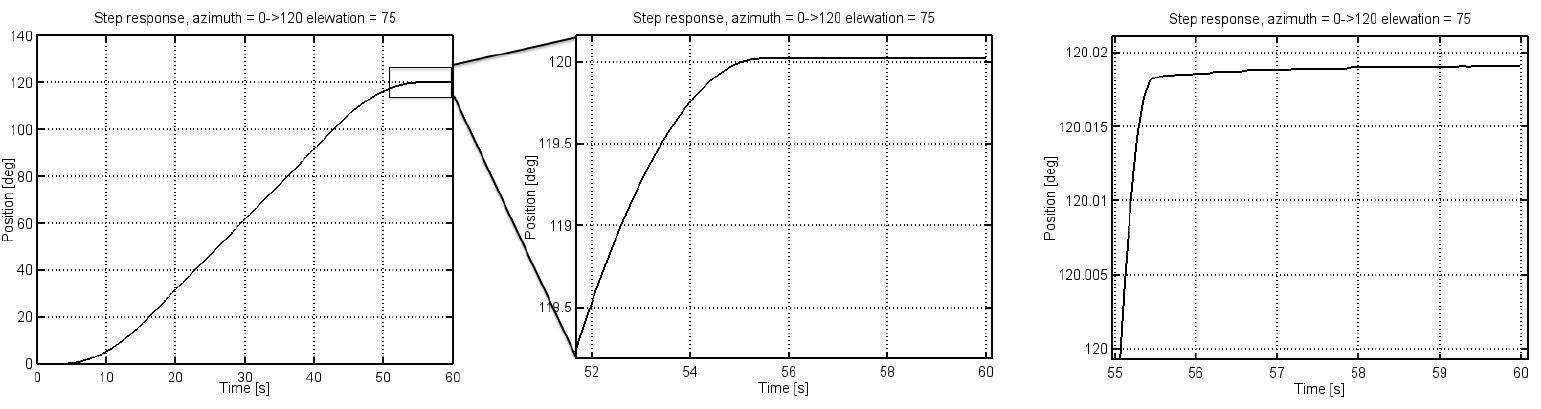}
	\caption{Step response of the telescope slewing from $45^\circ$ to $75^\circ$ in elevation and at constant azimuth angle of $120^\circ$ (top) and from $0^\circ$ to $120^\circ$ in azimuth at a constant elevation angle of $75^\circ$ (bottom). The middle and right panels show zoom-ins of the late stages of motion.}
	\label{fig:step1}
\end{figure*}

\paragraph{Drive control tests:}
To verify the positioning accuracy of the telescope and detect any structural vibrations during slewing, step response measurements were conducted for various target positions. Example results for targets in the observation range are shown in Fig.~\ref{fig:step1}.
The control software operated under default settings, including an elevation speed of 400~rpm with an acceleration of 3.5~rad/s\(^2\), and an azimuth speed of 1400~rpm with an acceleration of 12.45~rad/s\(^2\). 
Position data were recorded from external encoders at intervals of 80~ms, providing sufficient time resolution to detect oscillation modes corresponding to the natural frequencies of the telescope structure. 

As depicted in Fig.~\ref{fig:step1}, target positions were reached smoothly without vibrations with reduced acceleration and achieved required accuracy below $0.1^\circ$. 
The controlled braking mechanism implemented in the control algorithm ensured no overshoot occurred during approach to the target positions. 
After reaching the target, the telescope control automatically deactivated, with the target position maintained at a minimal drift rate during these tests. Note, that fluctuations visible in the top right panel of Fig.~\ref{fig:step1} are attributed to encoder accuracy rather than structural vibrations.

\subsection{Validation of the Cherenkov camera design}
\label{sec:val_cam}

\subsubsection{Camera environmental control}

The camera temperature is controlled all year long through the operation of a chiller unit attributed to each telescope. In summer, the coolant temperature is set at 12\degc{} 
in order to avoid long thermalisation time, while in winter it is set at 8\degc{} to warm the camera and prevent from thermal shocks when turning on the electronics. 

\begin{figure}
    \centering
    \includegraphics[width=\textwidth]{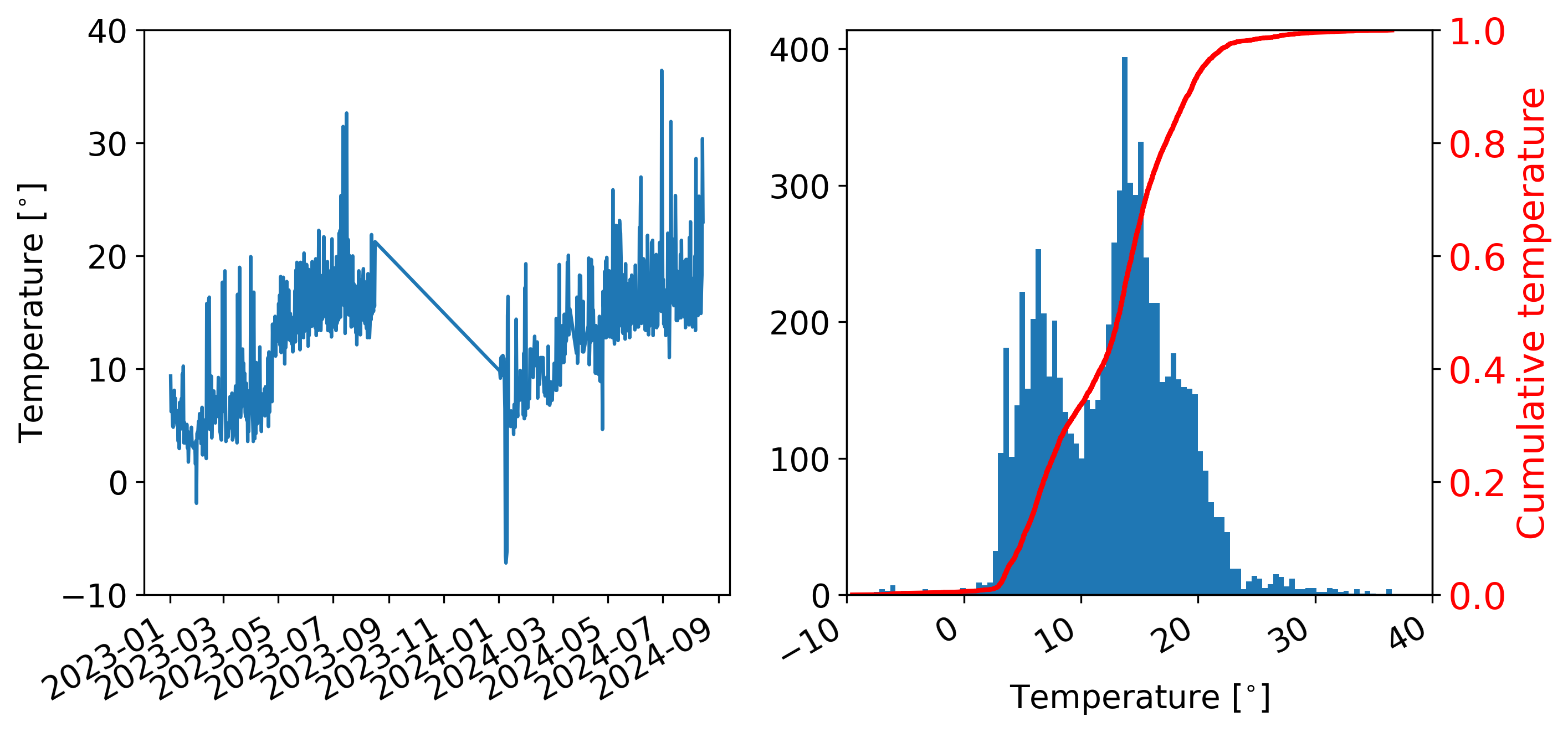}
    \includegraphics[width=\textwidth]{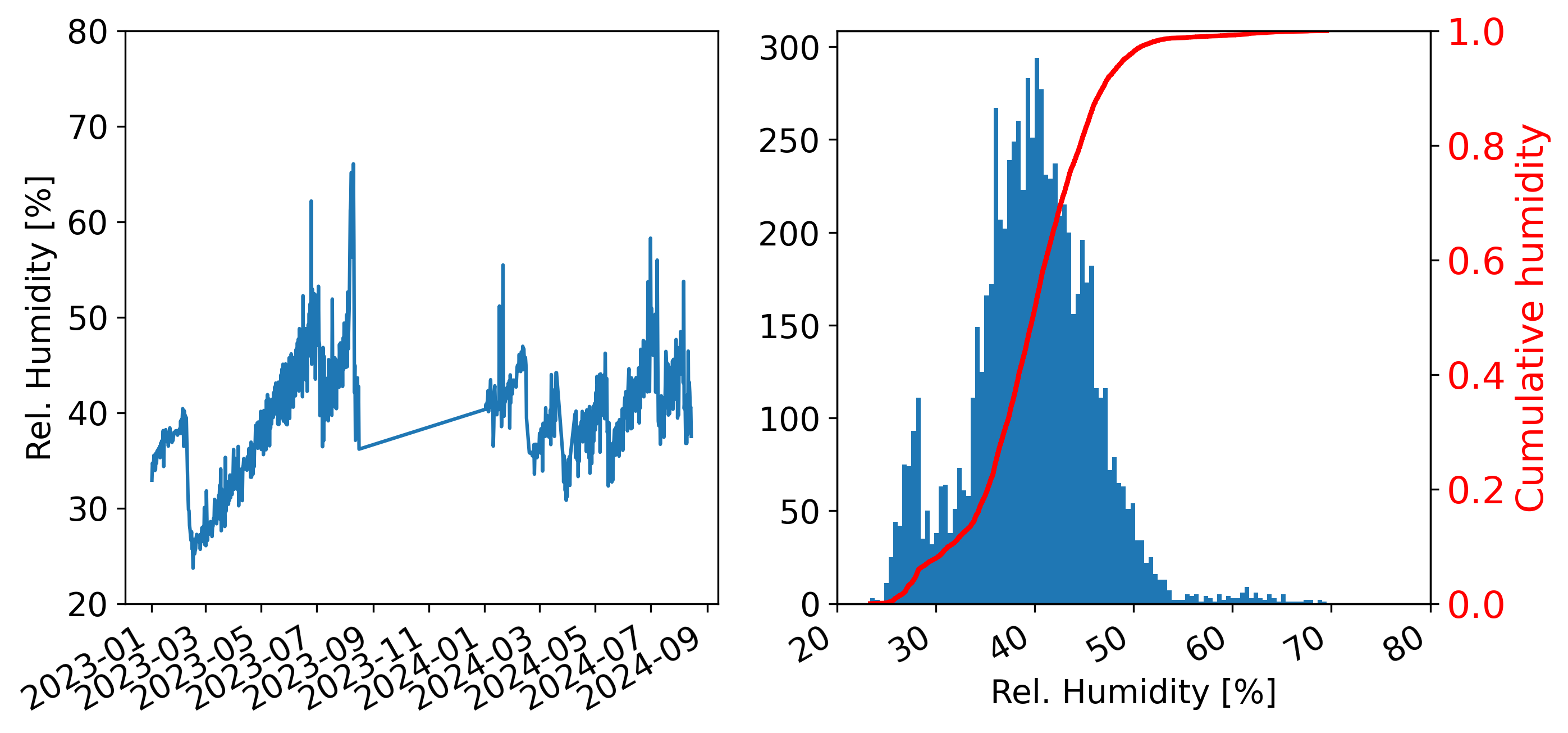}
    \caption{Top: Temperatures inside telescope 1 camera from 2023/01/01 to 2024/08/15 in Ond\v{r}ejov. Bottom: Same for relative humidity.}
    \label{fig:CamEnv}
\end{figure}

Fig.~\ref{fig:CamEnv} shows the temperature and relative humidity inside the prototype camera. These values are measured by the four SRT31-ARP sensors from Sensirion, which allow to derive the dew point and ensure that the chiller temperature is within a safe range. The second telescope shows similar behaviour and is therefore not displayed here. Fig.~\ref{fig:CamEnv} shows that the heating from the chiller always maintain the camera internal temperature above 0\degc{} in winter, except for rare occasions when both chiller and heaters inside the camera did not work.
Similarly, in summer, the chiller maintains the camera temperature below 35\degc{}.
The humidity of the first camera is below 50\% for more than 90\% of the time. However, due to humidity ingress in the prototype during heavy repeated rains, the humidity exceeded this safe value of 50\%.

\subsubsection{Telescope optical efficiency}

\begin{figure}
    \centering
\includegraphics[width=0.9\textwidth]{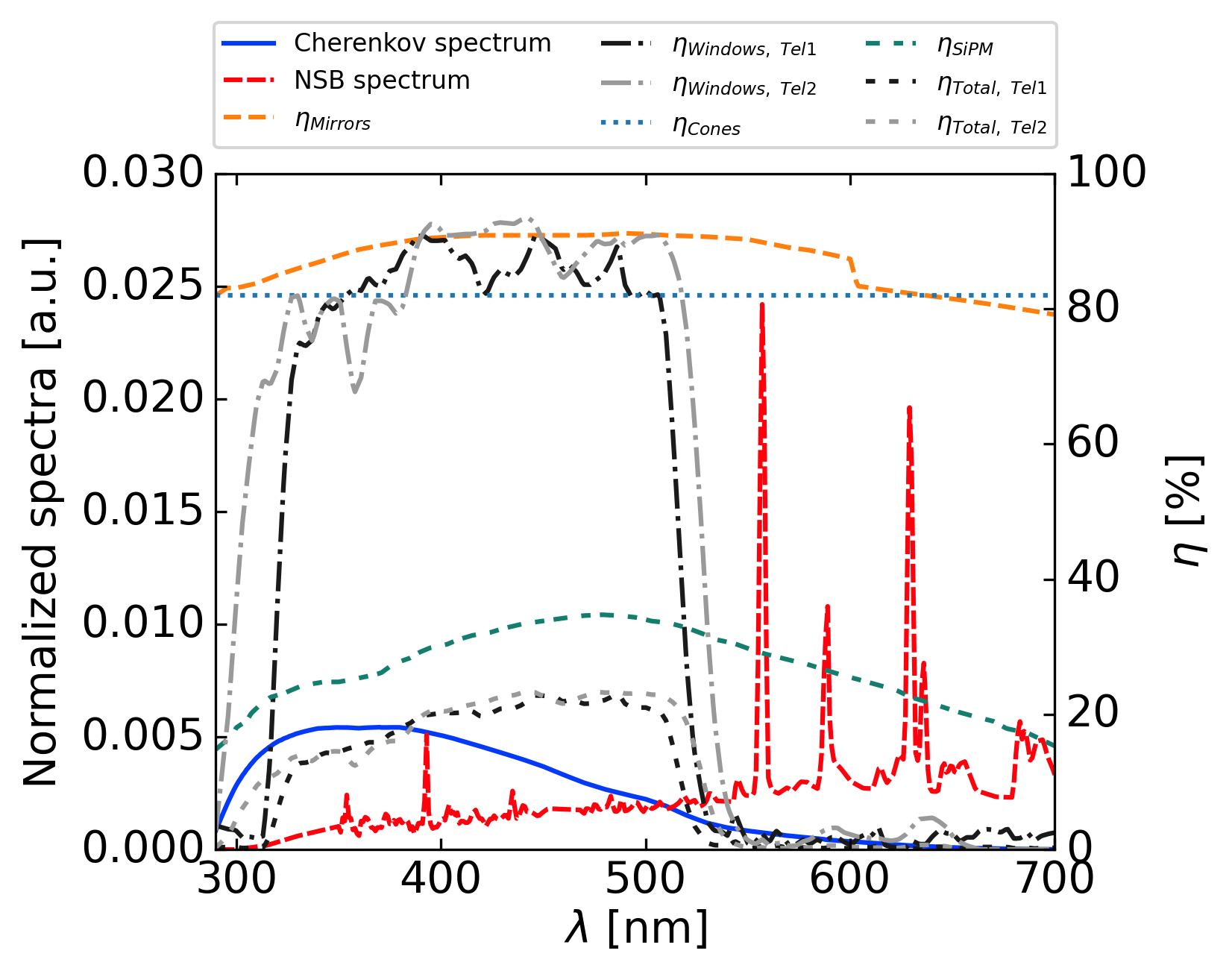}
\includegraphics[width=0.9\textwidth]{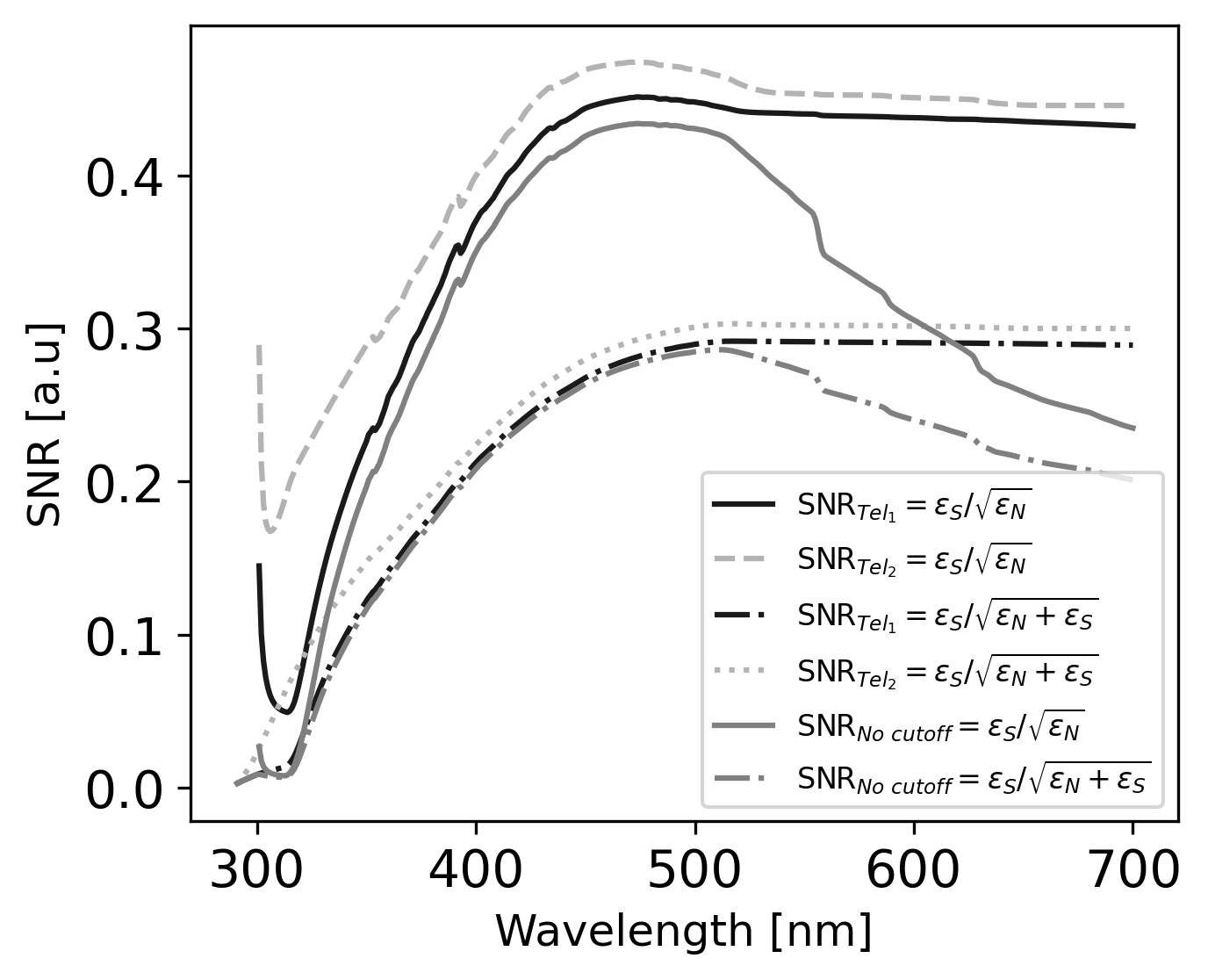}
    \caption{Top: Efficiency of the optical elements dedicated to transmission of light from the mirror dish. The total efficiencies of the 2 SST-1Ms are 15.1\% and 16.5\%. Bottom: SNRs for the two telescopes as well the one computed if the window would be identical to the one mounted on the Large-Sized Telescope that does not have any wavelength cut-off.}
    \label{fig:optical_eff}
\end{figure}

Fig.~\ref{fig:optical_eff} shows the optical efficiency measured for all the elements of the optical chain and the resulting SNR. The photons reaching the telescope aperture with an incident angle lower than 4.5$^{\circ}$ can in principle reach the camera. However, the mirror reflectivity, the camera entrance window and light guide transmission and the SiPM sensors photo-detection efficiency affect the total light collection efficiency.
As visible in Fig.~\ref{fig:optical_eff}, the only notable difference between the two telescopes comes from the entrance window. For the second telescope, the manufacturer, Thin Film Physics, managed to greatly improve the window light transmission below 320~nm. They also sharpened further the cut-off of the filter to match better the first NSB peak.

We calculate the fraction of the Cherenkov light signal that can be detected by the telescopes using the calculated total efficiency from 290~nm to 700~nm: 15.1\% and 16.5\% for the prototype and the second telescope.
The same efficiencies for the NSB are 12.2\% and 13.6\%. Two expressions of SNRs are used, the first one only considers the NSB as source of noise while the second one account for fluctuations of the signal itself as part of the noise. The resulting SNRs are shown in Fig.~\ref{fig:optical_eff} (bottom). 
For comparison, the SNRs labelled "No cut-off" displays the value computed if the entrance window was identical to the one of the Large-Sized Telescope of CTA. This window is made out of a special PMMA, called Shinkolite, with enhanced transmissivity in the UV region. 
One can clearly see the positive impact of the low-pass optical filter, which stabilises the SNR ratio above 540~nm. 
This is observed independently of the SNR expression.

\subsubsection{Sensor characteristics}
The sensor characteristics are monitored continuously before and after each night of observation by acquiring so-called dark count runs. These runs are taken with the camera shutter closed at a fixed trigger rate of 1~kHz. 
The extraction of those parameters from the charge distribution of each pixel is detailed in \cite{TavernierICRC2023}.

\begin{figure*}
    \centering
    \includegraphics[width=\textwidth]{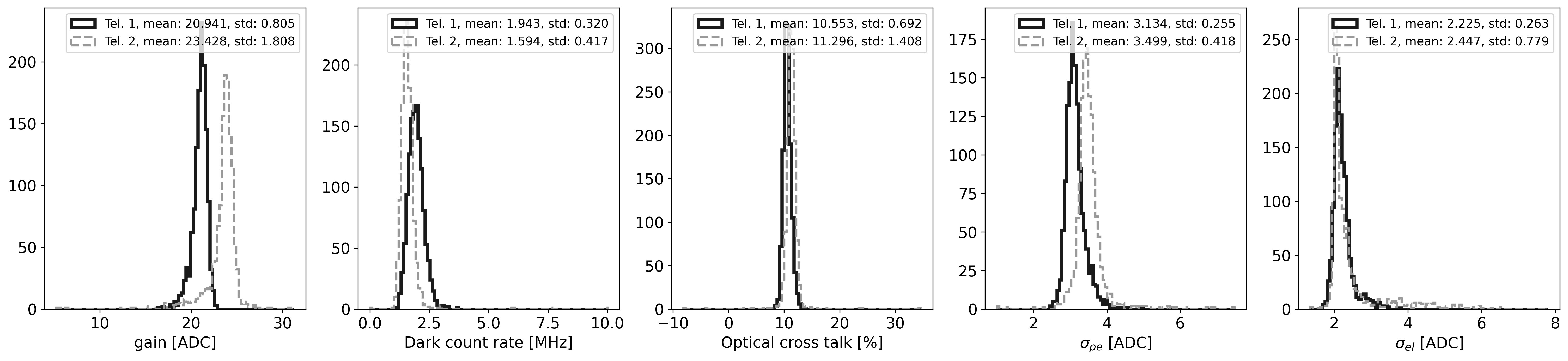}
    \caption{Main SiPM parameters extracted from dark count runs for the two telescopes.}
    \label{fig:SiPM_carac}
\end{figure*}

\begin{figure}
    \centering
    \includegraphics[width=0.8\textwidth]{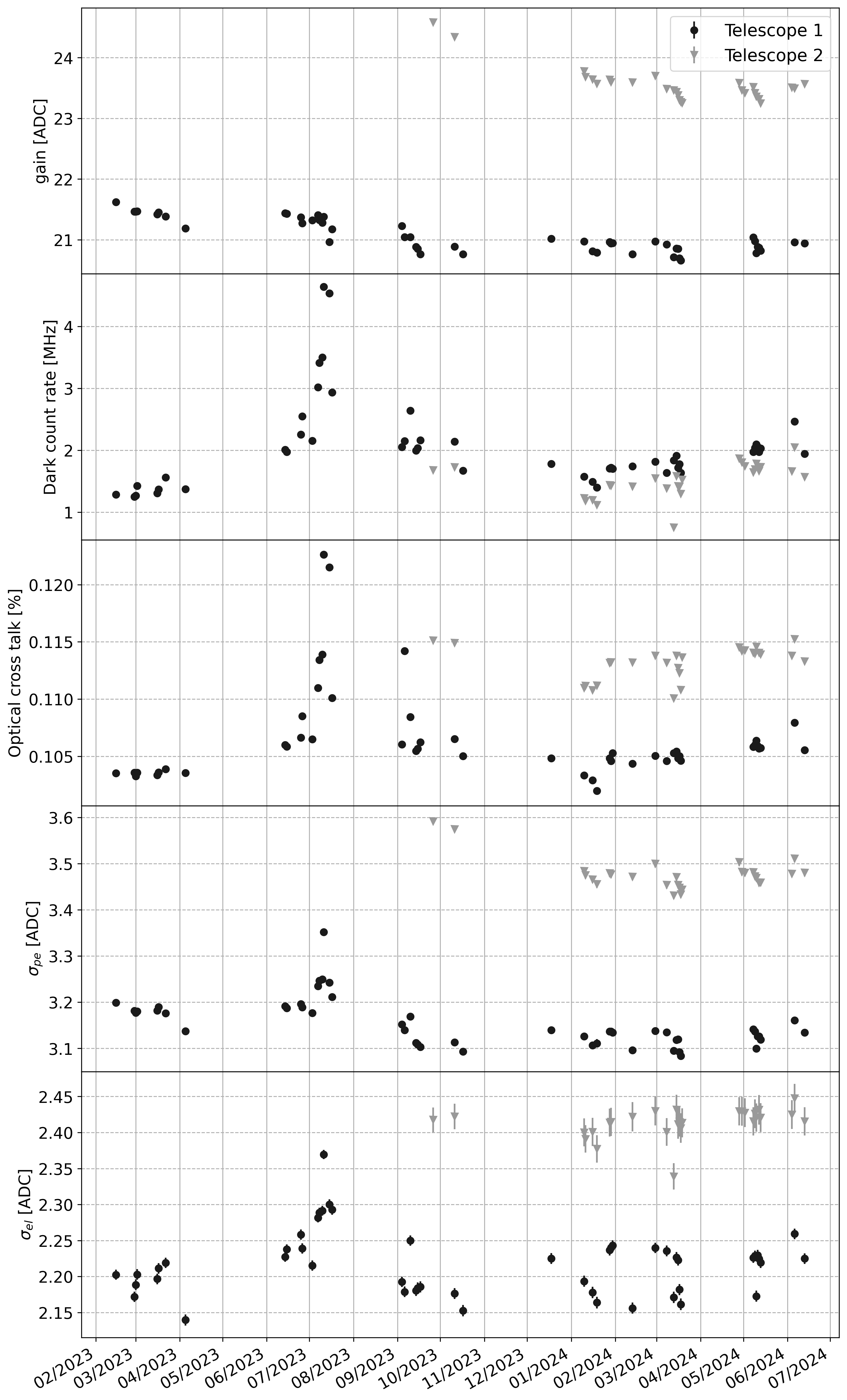}
    \caption{Evolution with time of the main SiPM parameters extracted from dark count runs for the two telescopes.}
    \label{fig:SiPM_carac_evol}
\end{figure}

The distributions of the pixel gain, dark count rate, optical cross talk, sensor noise $\sigma_{pe}$ and electronics noise $\sigma_{el}$, for a typical night are shown in Fig.~\ref{fig:SiPM_carac}. 
The characteristics are summarised in Tab.~\ref{tab:SiPM_carac}.
Their evolution from Feb.~2023 to July 2024 is shown in Fig.~\ref{fig:SiPM_carac_evol}.

\begin{table*}
    \centering
    \resizebox{\textwidth}{!}{
    \begin{tabular}{lcccccccccc}
    \hline
    \hline
    & \multicolumn{2}{c}{Gain [ADC]}  &  \multicolumn{2}{c}{Dark count rate [MHz]}  &  \multicolumn{2}{c}{Optical cross talk [\%]}  &  \multicolumn{2}{c}{$\sigma_{pe}$ [ADC]}  &  \multicolumn{2}{c}{$\sigma_{el}$ [ADC]} \\
     & Mean & Std & Mean & Std  & Mean & Std  & Mean & Std  & Mean & Std \\
                    \hline
        Telescope 1 & 20.94 & 0.80 & 1.94 & 0.32 & 10.55 & 0.69 & 3.13 & 0.25 & 2.22  & 0.26 \\
        Telescope 2 & 23.43 & 1.81 & 1.59 & 0.42 & 11.30 & 1.41 & 3.50 & 0.42 & 2.45  & 0.78\\
        \hline
    \end{tabular}
    }
    \caption{Summary of mean and standard deviation of the distributions of the main SiPM parameters}
    \label{tab:SiPM_carac}
\end{table*}

Fig.~\ref{fig:SiPM_carac_evol} shows how these parameters evolved since the cameras are regularly operated in Ondřejov. The dark count rate is not corrected for temperature and therefore increases during summer time with higher environmental temperatures.
The extraction of the optical cross talk is biased by the increased dark count rate and therefore follows the same dependency with temperature. 
In general, the parameters are stable within a few percentages.

\begin{figure}
    \centering
    \includegraphics[width=\textwidth]{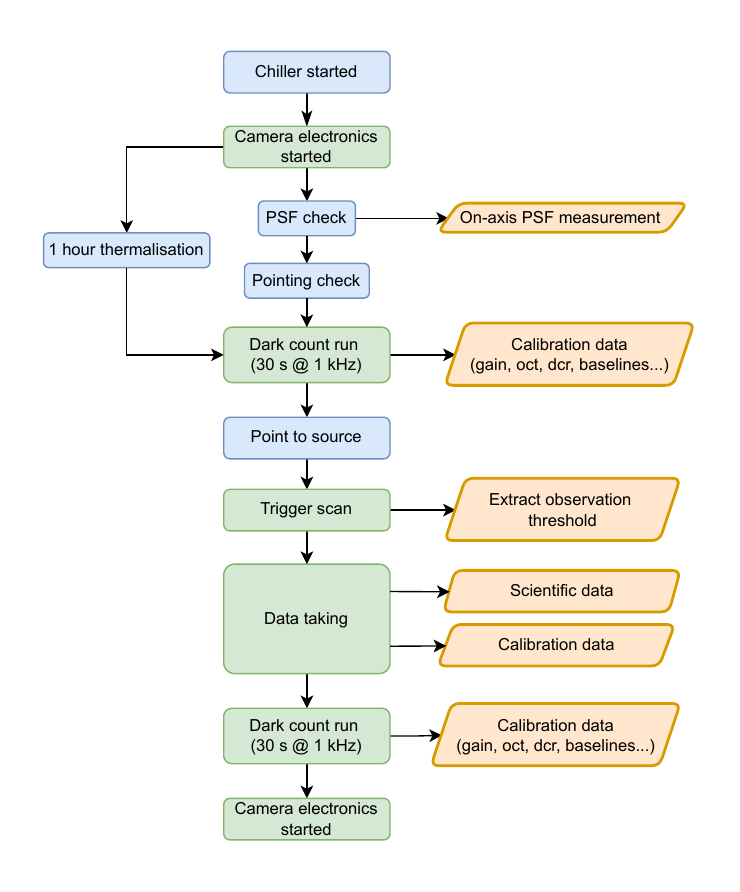}
    \caption{Operation flow for an observation night for a single telescope. A similar flow is applied for the operation of both SST-1Ms synchronised through the scheduler.}
    \label{fig:data_taking_flow}
\end{figure}

\subsubsection{Trigger scan}

As shown in Fig.~\ref{fig:data_taking_flow}, the threshold for operation of the telescope is decided after running a coarse trigger scan in the source direction. 
Independently of the source position, we also perform regularly a grid of trigger scans in order to verify the stability of environmental conditions of the simulations and deriving the trigger threshold.

\begin{figure*}
    \centering
    \includegraphics[trim = 100mm 0mm 100mm 0mm, clip, width=\textwidth]{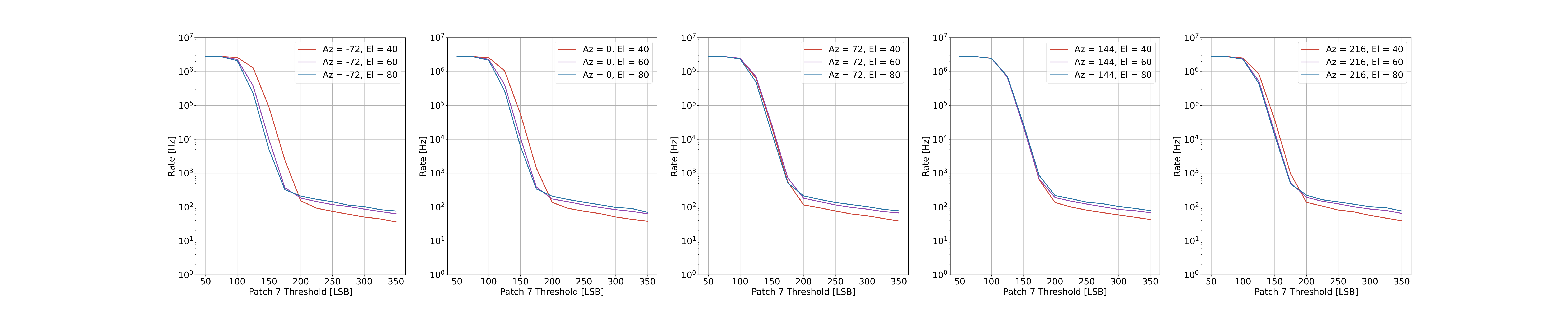}
    \caption{Trigger scan for the prototype for various (Az, El) coordinates. The known increasing proton rate with increasing elevation is observed, while the azimuth dependency is affected by the Ondřejov village lights and mostly Prague's city lights, 33~km away and in the azimuthal direction around -40$^{\circ}$.}
    \label{fig:trig_scan}
\end{figure*}


\subsection{The SST-1M prototype telescopes}
\label{sec:protos}

\begin{figure}
    \centering
    \includegraphics[width=0.9\textwidth]{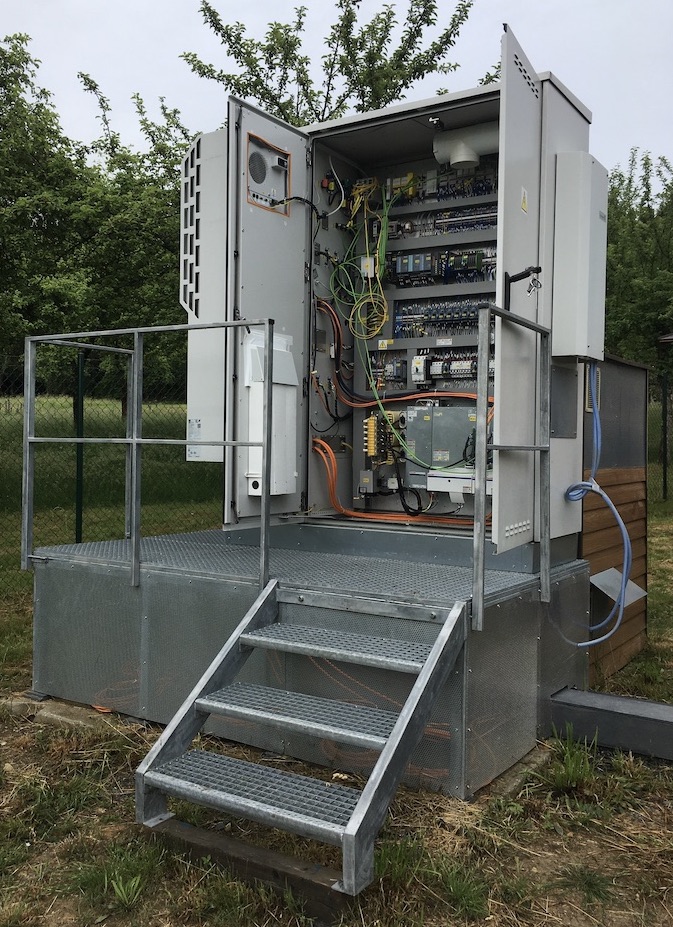}
    \caption{Control cabinet of the prototype telescope.}
    \label{fig:cabinet}
\end{figure}

The SST-1Ms are designed and built to enable remote observations, be cost-effective, and require minimal maintenance in harsh environments. The telescopes are not protected by domes. Each telescope is equipped with a stand-alone control cabinet that houses the power supply, drive, and safety system electronics, as shown in Fig.~\ref{fig:cabinet}. The control cabinet is fully waterproof and features temperature and humidity control systems. Cooling and heating are managed through fans and heat exchangers.

\begin{figure}
\centering
\includegraphics[width=\textwidth]{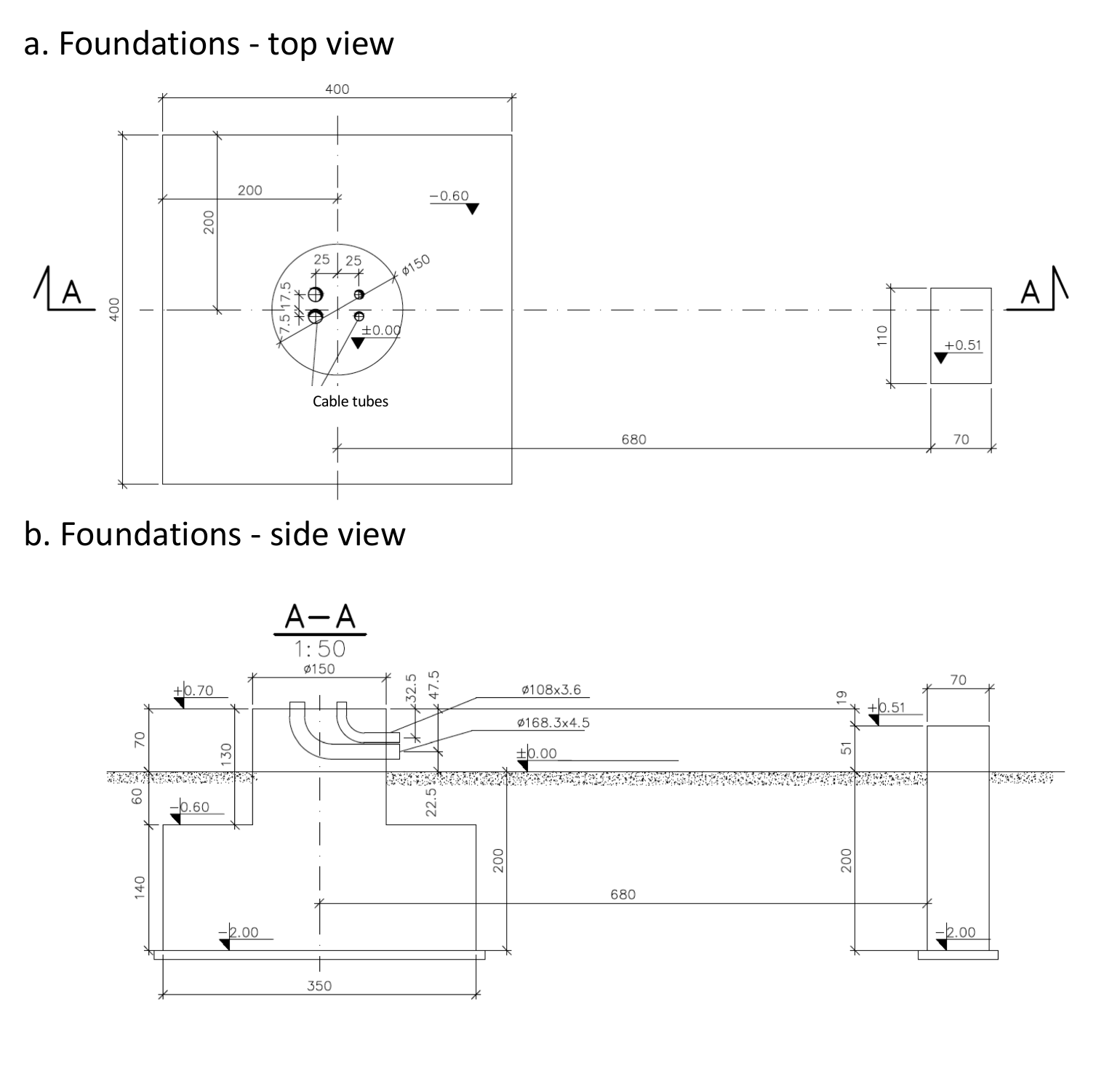}
\caption{Vertical (a) and horizontal (b) projections of the foundations of the telescope and docking station.}
\label{fig:foundation1}  
\end{figure}

\begin{figure}
\centering
\includegraphics[width=0.7\textwidth]{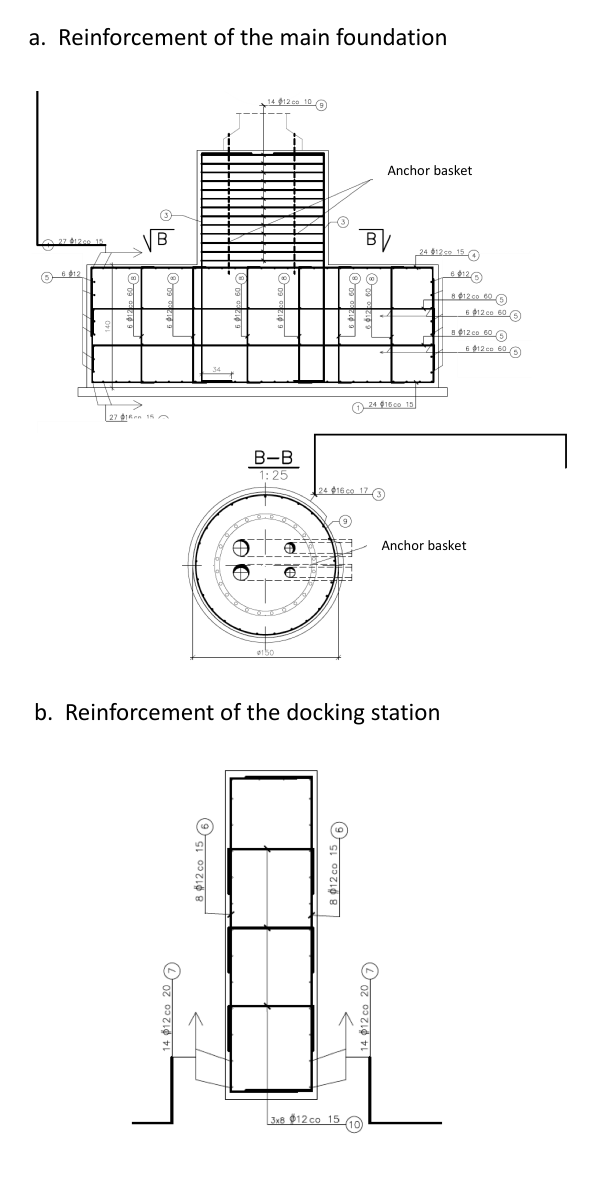}
\caption{Layout of the steel reinforcement of the telescope foundation (a) and docking station (b).}
\label{fig:foundation2}  
\end{figure}

The foundations for the prototypes installed at the Ondřejov Observatory are made as a monolithic slab, 140~cm thick, with horizontal dimensions of 4~m $\times$ 4~m. Centrally on this slab, there is a 130~cm high pedestal for mounting the telescope with the shape of a round column. The foundation is poured wet, with the slab placed at least 2.0~m below the ground level. An anchoring basket for the telescope is embedded in the pedestal (see
Sec.~\ref{sec:structure}). The construction materials include frost-resistant F100 concrete C30/37, lean concrete C8/10, ribbed reinforcing steel A-IIIN–BSt-500S, and plain reinforcing steel A-0–St0S. Dimensioned vertical and horizontal projections of the foundation and the docking station are shown in Fig.~\ref{fig:foundation1} and~\ref{fig:foundation2}.

Both telescopes, despite their similar design, have several significant differences. Compared to the prototype, the second telescope profits of some improvement, thanks to the lessons learned from the prototype testing and commissioning. The most important upgrades to mention are: 
\paragraph{Structure:} In telescope 1, cables and cooling fluid pipes are running outside the structure in the flexible tube behind the telescope and along the mast. In telescope 2, the routing was implemented inside the tower and mast using dedicated twisting system and rotary union for pipes. This required the use of larger diameter circular steel tubes in the mast structure. Only motor power cables were routed outside due to the interference with the operation of other signal cables inside the tower. Telescope 2 has also a new encoder system and an additional access door in the tower.     

\paragraph{Camera:} 
As described in Sec.~\ref{sec:val_cam}, the spatial uniformity and transmission in the 280-540~nm band of the entrance window has been improved. 
The bias stage sensitivity to variations of the NSB level has been decreased by replacing the 10~k$\rm\Omega$ bias resistor by a 2.4~k$\rm\Omega$ one.
The design of the \dcam{} crates cooling has been modified. Their new position in the chassis allows for full access to the front part of \dcam{} from the patch panel. 
Finally, the FADCs, both sampling at 250~MHz with a resolution of 12 bits, are of different manufacturer (Analog and Renegas, former Intersil).
    
\subsection{Operation at the Ond\v{r}ejov site}
\label{sec:ondrejov}

\begin{figure}
	\centering
		\includegraphics[width=\textwidth]{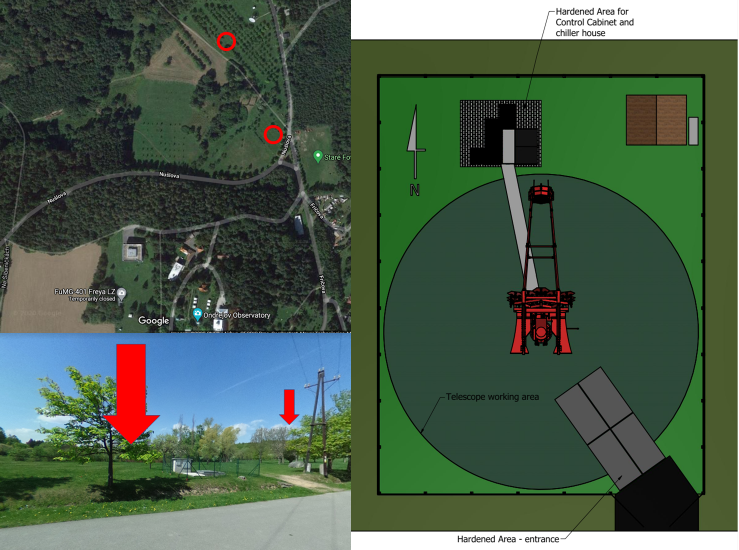}
	\caption{The location of the two SST-1Ms at the Ondřejov observatory (left). The layout of the telescope surroundings. The telescope is secured with a fence accessible via a gate. The camera servers and all necessary control electronics are located in a wooden booth at the top right of the image on the right image. The DAQ and slow control are located in the control cabinet next to the chiller. Imagery \textcopyright Airbus, CNES/ Airbus, Maxar Technologies, Maps data \textcopyright 2025}
	\label{fig:Ondrejov_site}
\end{figure}

The two SST-1Ms are located at the Ondřejov Observatory, primarily for testing purposes. This observatory, part of the Astronomical Institute of the Czech Academy of Sciences, is located near the village of Ondřejov in the Central Bohemian region, approximately 35~km South-East of Prague. The observatory operates the largest Czech optical telescope (Perek 2~m telescope) along with several radio and optical astronomical instruments. In the second half of 2020, infrastructure work started to accommodate both SST-1Ms, which are separated by about $150$~m, a distance compatible with stereoscopic observations. The exact coordinates of the first telescope are $49.9129700N, 14.7822500E$ and the altitude is $510$ m.a.s.l. The positioning of the system and its layout within the observatory area are shown in Fig.~\ref{fig:Ondrejov_site}, where it is connected to the electric power and the internet network of the observatory.

The site and telescope construction, validation and operation proceeded as follows. Between 2020 and 2021, the Ond\v{r}ejov site was prepared for the installation of the telescopes, with foundations, electrical and IT infrastructures built to meet the SST-1M requirements. In the summer of 2021, the telescope 2 structure, mirrors and camera were installed. In Nov. 2021, the prototype structure and mirrors were deployed. The first cosmic ray showers were detected with telescope 2 in Feb.~2022. In March 2022, the camera was installed on the prototype enabling the first pseudo-stereo observations with offline synchronisation in April 2022. They confirmed the ability to detect VHE gamma-ray sources using the developed software. By April 2023, time stamps in both telescopes were synchronised using White Rabbit, allowing for the first full stereo observations. The telescope 2 camera underwent maintenance in Geneva from Apr. to Sep. 2023. Throughout 2022 to 2024, intense software pipeline development occurred, resulting in the publication of the relevant software under open licence \cite{jurysek_2024_10852981}.

The observations with the SST-1M mini-array are limited by the continental climate of the site, which includes relatively high humidity, warm summers, and rainy autumns and winters. Additionally, the proximity of nearby villages and towns increases the NSB level. 
Despite these limitations, the location is still suitable for observing  standard prominent gamma ray sources, such as the Crab nebula, Mrk~421, Mrk~501, 1ES~1959+650, and more.

\begin{figure*}
	\centering
		\includegraphics[width=\textwidth]{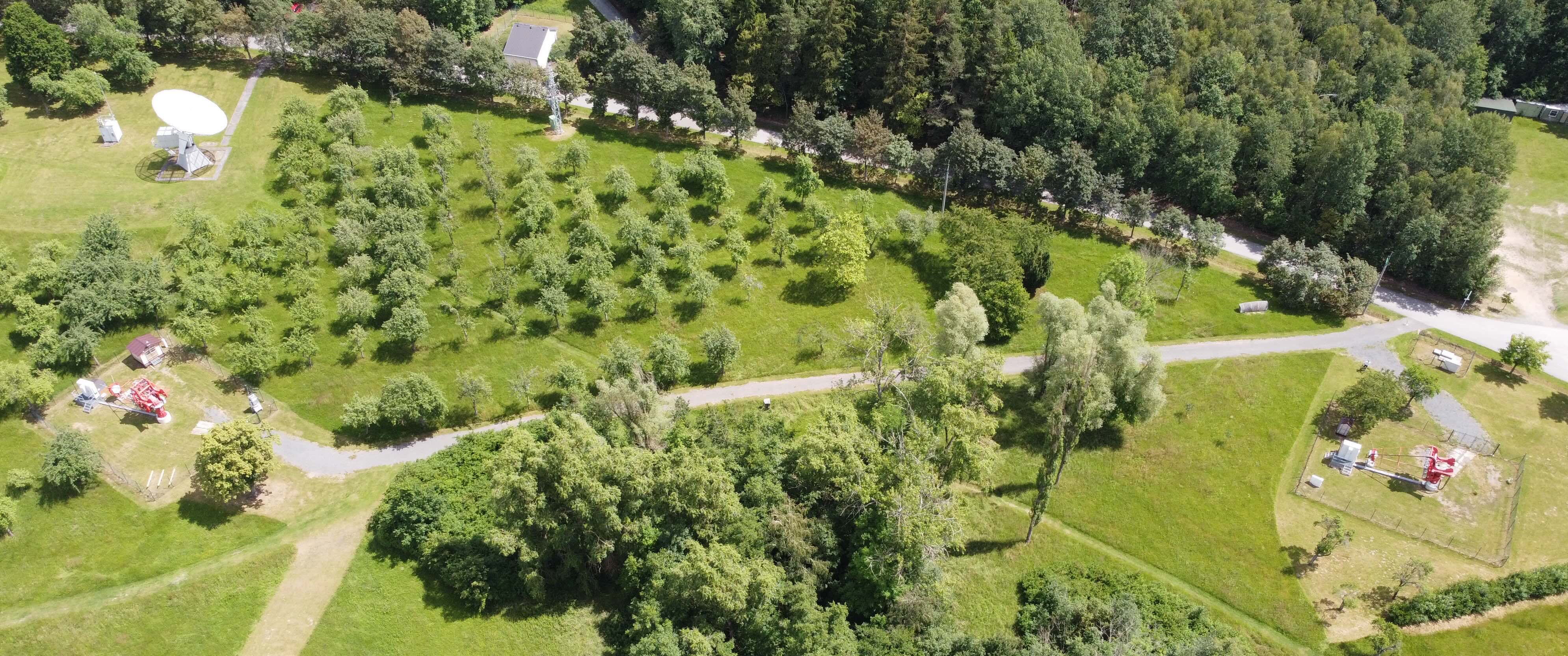}
 \caption{An aerial view of the Ondřejov site in June 2024 shows two SST-1Ms in the parked position (bottom right and left). A nearby radio dish for solar observations is also visible in the top left corner.}
	\label{fig:aerial}
\end{figure*}

\subsection{Data acquisition and stereo trigger}

The readout chain is implemented as follow: Upon reception of a trigger, each telescope sends the packets containing the event data to the Camera Server software. Each instance of the acquisition software runs on a different server. If a full event for a given telescope can be assembled, meaning that the server received all packets with a matching timestamp, trigger packets are sent to the so-called Software Array Trigger (SWAT).

The SWAT is a software component developed for use in SST-1M and CTAO.
Its main task is to accumulate trigger packets from both telescopes, merge them into a single stream with entries ordered by time, and look for subsequent triggers that took place in a narrow window of time, which could indicate coincidence. 
If such a pair is found, the SWAT assigns the same ID to events from both telescopes so that they are marked as possible stereo at the time of acquisition. The SWAT can be flexibly configured during operation: the coincidence window duration is editable during runtime and the software may be utilised to inhibit the readout
of all mono events (which is a data volume reduction measure) or only allow some mono events through as flagged by the Camera Server software (which could for example be used to store muon candidate events regardless of their coincidence with other events).

\begin{figure}
    \centering
    \includegraphics[width=\textwidth]{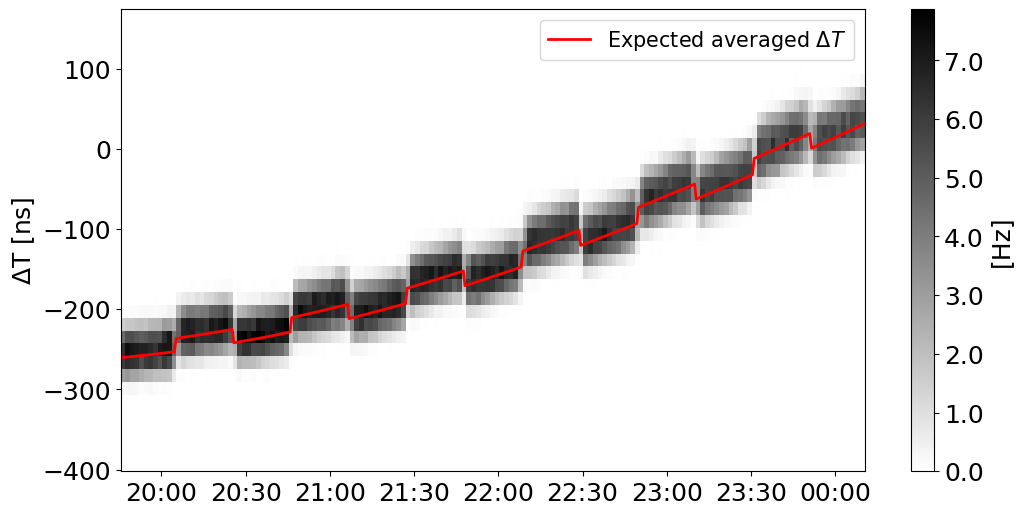}
    \includegraphics[width=\textwidth]{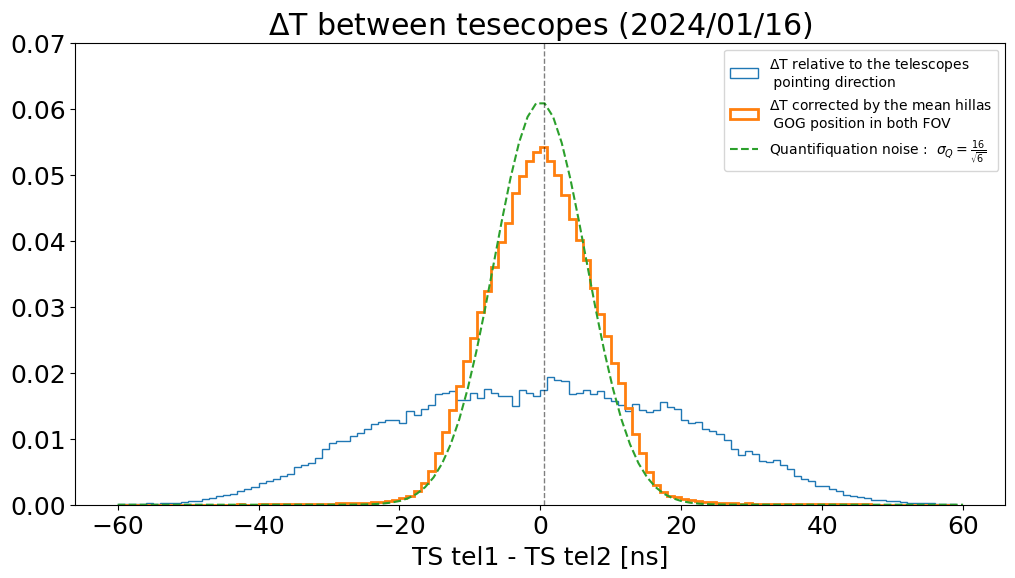}
    \caption{Top: Difference between timestamps of telescope 1 and 2 for stereo triggered and selected events as function of time. The red line shows the averaged expected time difference between the two telescopes based on the pointing direction. Bottom: Time difference between the two triggered telescope corrected for the expected averaged time difference based on the pointing direction.} 
    \label{fig:DTvsTime}
\end{figure}

Fig.~\ref{fig:DTvsTime} (top) shows the difference between the time at which the two telescopes triggered when operated in coincidence, after the standard analysis quality cuts have been applied. In this case, the coincidence window is set to $\pm$\,500~ns and does not depend on the pointing direction. 
One can see that the observed time difference matches very well the mean expected, which is derived from the pointing direction. If one subtract the mean expected time difference to the measured one, we obtain the distributions shown in Fig.~\ref{fig:DTvsTime} (bottom). It demonstrates that a dynamic coincidence window can be implemented with a width of $\pm$\,60~ns, which would include all triggered events and more efficiently reduce random coincidences. This implementation is being carried on and will be deployed in a near future. 
In Fig.~\ref{fig:DTvsTime} (bottom), the thick orange histogram shows that when the event timing is extracted from the image analysis, the coincidence window could be even further reduced. In this case, the analysis consist in extracting the time at which the signals in the pixels at the image centre of gravity reach their maximum. 

\subsubsection{Onsite infrastructures}
The telescopes are powered by the observatory grid with the flywheel and diesel generator UPS providing sufficient and stable power for the operation. The internal IT network is connected via optical fibre to the CESNET network. The gateway is the access port of the whole internal network including camera servers, switches, computing servers and auxiliary systems. The IT infrastructure is connected using optical fibres. To enable remote observations, the telescope facility is equipped with surveillance cameras for the telescope condition observation, weather station and All Sky Camera \cite{Chytka_2020} for the night cloud detection.

\section{Summary}
\label{sec:summary}

The SST-1M telescopes are now operating in stereo-mode, detecting both point-like and extended sources. Future publications will provide information on these measurements. 
This paper describes in detail the full SST-1M design and realization, which proved it a very cost-effective and performing solution for IACT arrays to cover energies beyond 1~TeV. 

The adopted SiPM technology has significantly advanced in recent years, making it  suitable for large area telescopes \cite{Heller:2023f6}.  The experience gained from the SST-1M, has been efficiently adapted and advanced for other projects.
For instance, the design and construction of the WFCTA of the LHAASO hybrid observatory were influenced by the SST-1M design \cite{Zhang:2018ygl}. The cost-effective solution of the camera based on SiPMs enabled nearly twice the number of telescopes to be built, compared to the initially planned telescopes with PMT based cameras. Additionally, the experience of the UNIGE laboratory on SiPMs was used for various collaborations with FBK from the ATTRACT project POSICS on medical imaging \cite{POSICS} to the space-based telescope Terzina on board NUSES \cite{2023EPJWC.28306006B}.

The high-performance, large-area SiPMs with dedicated slow control, their lightweight (8.6~t) and compact mechanical telescope structure and its robotic operation require minimal maintenance in harsh environment, and make the SST-1M suitable for deployment at high-altitude sites above 4000~m a.s.l. 
The SiPM market and their applications have widely expanded in the automotive, biology, medical imaging and LIDAR fields. Major companies such as Broadcom, Hamamatsu, Ketek, On-Semi, Advansid, and FBK are highly competitive and rapidly reducing costs.

\acknowledgments

This publication was created as part of the projects funded in Poland by the Minister of Science based on agreements number 2024/WK/03 and DIR/\-WK/2017/2022/12-3.

The construction, calibration, software control and support for operation of the SST-1M cameras is supported by SNF (grants CRSII2\_141877, 20FL21\_154221, CRSII2\_160830, \_166913), by the Boninchi Foundation and by the Université de Genève, Faculté de Sciences, Départment de Physique Nucléaire. The Czech partner institutions acknowledge support of the infrastructure and research projects by Ministry of Education, Youth and Sports of the Czech Republic and regional funds of the European Union, MEYS LM2023047 and EU/MEYS CZ.02.01.01/00/22 008/0004632, and Czech Science Foundation, GACR 23-05827S.

\clearpage

\appendix

\section{Telescope and infrastructure diagrams}
\label{app}

\begin{figure*}[h!]
    \centering
    \rotatebox{90}{\includegraphics[width=\textwidth]{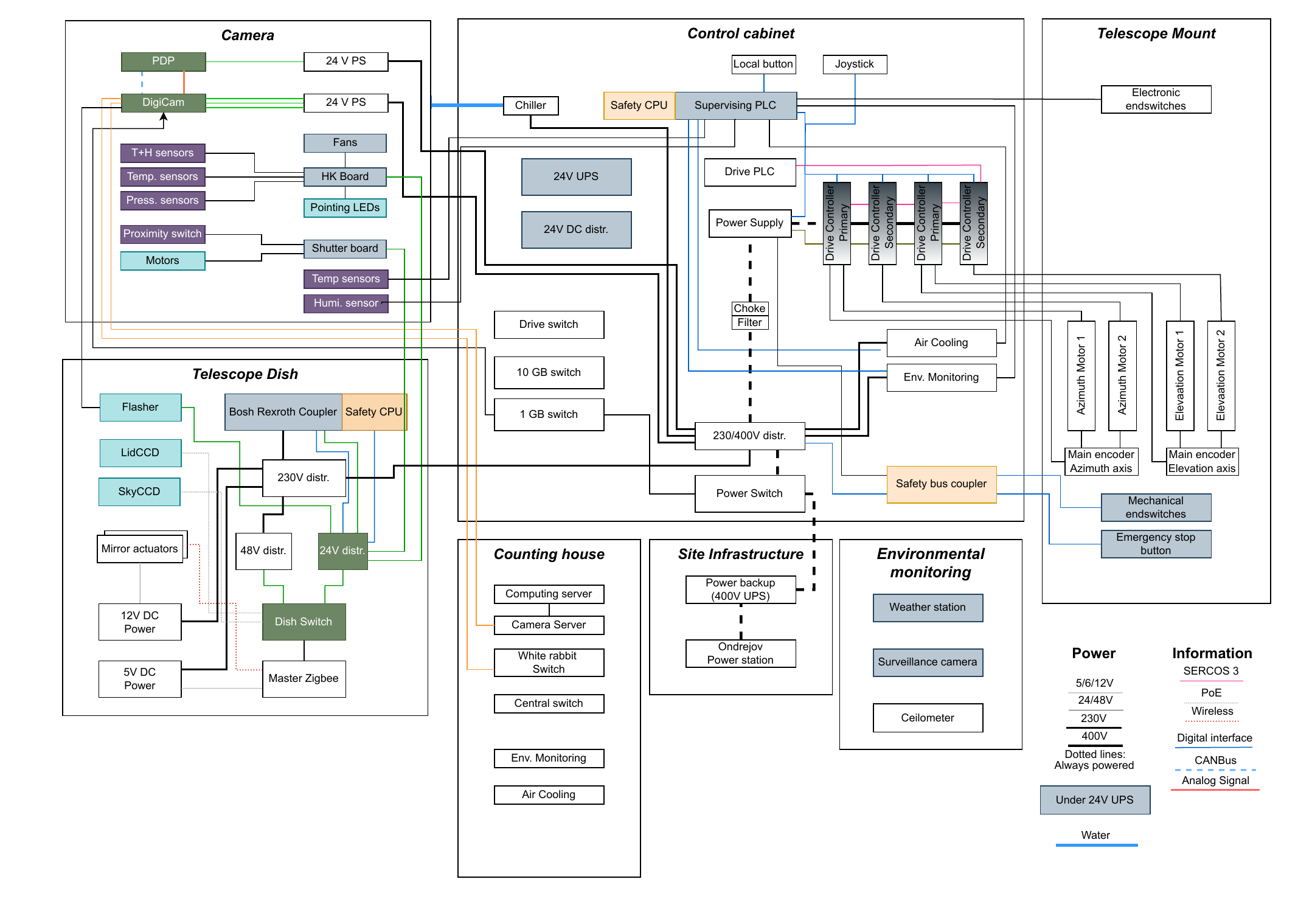}}
    \caption{The SST-1M diagram with interfaces shown only for one telescope.
    }
    \label{fig:sst1m_diagram}
\end{figure*}




\bibliographystyle{JHEP}
\bibliography{main}

\end{document}